\documentclass[%
reprint,
superscriptaddress,
nofootinbib,
amsmath,amssymb,
aps,
]{revtex4-2}

\usepackage{graphicx}
\usepackage{dcolumn}
\usepackage{bm}

\usepackage[normalem]{ulem}
\usepackage{mathtools}
\usepackage{multirow}
\usepackage{tabularx}
\usepackage{aas_macros}
\usepackage{lmodern}
\usepackage[T1]{fontenc}
\usepackage{array}
\usepackage{tabularx}
\usepackage{booktabs}
\usepackage{acronym}
\usepackage[dvipsnames]{xcolor}
\usepackage[colorlinks=true,allcolors=Maroon]{hyperref}
\usepackage{tikz}
\usetikzlibrary{shapes.geometric}
\usepackage{xspace}
\usepackage[capitalize]{cleveref}
\usepackage{varwidth}
\usepackage{enumitem}
\usepackage{boldline}
\usepackage{diagbox}
\usepackage{pifont}
\usepackage{makecell}
\usepackage[utf8]{inputenc}

\usepackage{orcidlink}
\newcommand{\orcid}[1]{\href{https://orcid.org/#1}{\includegraphics[width=10pt]{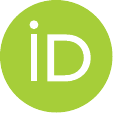}}}

\DeclareMathOperator{\erf}{erf}

\begin{document}
	
	
	\title{Systematic Biases in Estimating the Properties of Black Holes \\ Due to Inaccurate Gravitational-Wave Models} 
	
	\newcommand{\arnab}[1]{\textcolor{teal}{{#1}}}
	\newcommand{\ad}[1]{\textcolor{teal}{{AD: #1}}}
	\newcommand{\sv}[1]{\textcolor{blue}{{#1}}}
	\newcommand{\ab}[1]{\textcolor{cyan}{{#1}}}
	\newcommand{\comm}[1]{\textcolor{red}{[{#1}]}}
	\newcommand{\lp}[1]{\textcolor{purple}{{#1}}}
	\newcommand{\he}[1]{\textcolor{orange}{{#1}}}
	\newcommand{\hp}[1]{\textcolor{violet}{{#1}}}
	\newcommand{\at}[1]{\textcolor{olive}{{Alex: #1}}}
	\newcommand{\jg}[1]{\textcolor{magenta}{{#1}}}
	\newcommand{\reply}[1]{\textcolor{magenta}{{#1}}}
	
	\newcommand{\red}[1]{\textcolor{red}{{#1}}} 
	
	\newcommand{\cmark}{\ding{51}}%
	\newcommand{\xmark}{\ding{55}}%
	
	\newcommand{\eob}{\texttt{SEOBNRv5PHM}\xspace}
	\newcommand{\xphm}{\texttt{IMRPhenomXPHM}\xspace}
	\newcommand{\xoa}{\texttt{IMRPhenomXO4a}\xspace}
	\newcommand{\seob}{\texttt{SEOBNR}\xspace}
	\newcommand{\teob}{\texttt{TEOBResumS}\xspace}
	\newcommand{\bilby}{\texttt{Bilby}\xspace}
	\newcommand{\gwbench}{\texttt{GWBENCH}\xspace}
	\newcommand{\mass}{\texttt{Power Law + Peak}\xspace}
	\newcommand{\spin}{\texttt{DEFAULT}\xspace}
	\newcommand{\dynesty}{\texttt{dynesty}\xspace}
	\newcommand{\bilbypipe}{\texttt{Bilby-pipe}\xspace}
	\newcommand{\first}{\emph{Binary 1}\xspace}
	\newcommand{\second}{\emph{Binary 2}\xspace}
	\newcommand{\third}{\emph{Binary 3}\xspace}
	\newcommand{\hubble}{\ac{$H_0$}\xspace}
	\newcommand{\pesummary}{\texttt{pesummary}\xspace}
	\newcommand{\lal}{\texttt{LALSuite}\xspace}
	
	\newcommand{\aei}{\affiliation{Max Planck Institute for Gravitational Physics (Albert Einstein Institute), Am Mühlenberg 1, Potsdam 14476, Germany}}
	\newcommand{\umd}{\affiliation{Department of Physics, University of Maryland, College Park, MD 20742, USA}}
	
	\author{Arnab Dhani\,\orcid{0000-0001-9930-9101}}
	\email{arnab.dhani@aei.mpg.de}
	\aei
	
	\author{Sebastian H. V\"{o}lkel\,\orcid{0000-0002-9432-7690}}
	\aei
	
	\author{Alessandra Buonanno\,\orcid{0000-0002-5433-1409}}
	\aei
	\umd
	
	\author{Hector Estelles\,\orcid{0000-0001-6143-5532}}
	\aei
	
	\author{Jonathan Gair\,\orcid{0000-0002-1671-3668}}
	\aei
	
	\author{Harald P. Pfeiffer\,\orcid{0000-0001-9288-519X}}
	\aei
	
	\author{Lorenzo Pompili\,\orcid{0000-0002-0710-6778}}
	\aei
	
	\author{Alexandre Toubiana\,\orcid{0000-0002-2685-1538}}
	\aei

	\date{\today}
	
	\begin{abstract}
Gravitational-wave (GW) observations of binary black-hole
          (BBH) coalescences are expected to address outstanding
          questions in astrophysics, cosmology, and fundamental
          physics. Inference of BBH parameters relies on waveform
          models, and realizing the full discovery potential of
          upcoming LIGO-Virgo-KAGRA observing runs and new
          ground-based facilities (such as the Einstein Telescope and
          Cosmic Explorer) hinges on the accuracy of these waveform
          models. Using linear-signal approximation methods and Bayesian 
analysis, we start to assess our readiness for
          what lies ahead using two state-of-the-art quasi-circular,
          spin-precessing models: \texttt{SEOBNRv5PHM} and
          \texttt{IMRPhenomXPHM}.  We find that systematic biases
          increase with the spin of the BH, with parameter biases
          being approximately 6 to 8 times likelier, if the
          primary-spin magnitude exceeds 0.5 compared to when it is less than 0.5. 
Additionally, we ascertain that current
          waveforms can accurately recover the distribution of masses
          in the LVK astrophysical population, but not
          spins. Upon exploring the broader parameter space of BHs, we
          find that systematic biases increase with detector-frame
          total mass, binary asymmetry, and spin-precession, with a
          majority of such binaries incurring parameter biases,
          extending up to redshifts $\sim3$ in future
          detectors. Furthermore, we examine three ``golden'' events
          characterized by mass ratios of approximately 6 to 10,
          significant spin magnitudes ($0.6\mbox{--} 0.9$), and high precession, evaluating
          how systematic biases may affect their scientific
          outcomes. Our findings reveal that current waveforms fail to
          enable the unbiased measurement of the Hubble-Lemaître
          parameter and sky localization from loud signals, even for
          current detectors. Moreover, highly asymmetric systems
          within the lower BH mass gap exhibit biased measurements of
          the secondary-companion mass, which impacts the physics of
          both neutron stars and formation channels. Similarly, we
          deduce that the primary mass of massive binaries ($ > 60 M_\odot$) will also
          be biased, affecting supernova physics. 
Future progress in analytical calculations and 
          numerical-relativity simulations, crucial for calibrating
          the models, must target regions of the parameter space with
          significant biases to develop more accurate models. Only
          then can precision GW astronomy fulfill the
          promise it holds.
	\end{abstract}
	
	\maketitle
	

	\acrodef{BNS}{binary neutron star}
	\acrodef{BBH}{binary black hole}
	\acrodef{NSBH}{neutron-star--black-hole}
	\acrodef{NS}{neutron star}
	\acrodef{BH}{black hole}
	\acrodef{GW}{gravitational wave}
	\acrodef{CBC}{compact-binary coalescence}
	\acrodef{EM}{electromagnetic}
	\acrodef{$H_0$}{\textrm{Hubble-Lema\^{i}tre parameter}}
	\acrodef{XG}{next-generation}
	\acrodef{CE}{Cosmic Explorer}
	\acrodef{ET}{Einstein Telescope}
	\acrodef{PISN}{pair-instability supernova}
	\acrodef{PSD}{power spectral density}
	\acrodef{SNR}{signal-to-noise ratio}
	\acrodef{ADM}{Arnowitt-Deser-Misner}
	\acrodef{LSA}{linear-signal approximation}
	\acrodef{IMR}{inspiral-merger-ringdown}
	\acrodef{LVK}{LIGO-Virgo-KAGRA}
	\acrodef{EoS}{equation of state}
	\acrodef{GR}{general relativity}
	\acrodef{NR}{numerical relativity}
	\acrodef{PN}{post-Newtonian}
	\acrodef{PM}{post-Minkowskian}
	\acrodef{SFR}{star formation rate}
	\acrodef{EOB}{effective-one-body}
	\acrodef{FIM}{Fisher information matrix}
	\acrodef{GWTC}{gravitational wave transient catalog}

	\section{Introduction}
	\label{sec:introduction}
	
	Almost a decade ago, the first 
	observation of a \ac{GW} from the coalescence of two \acp{BH} marked an important milestone in the history of GW astronomy~\cite{LIGOScientific:2016aoc}. Since then, the \ac{LVK} Collaboration~\cite{LIGOScientific:2014pky,VIRGO:2014yos,KAGRA:2020tym} has detected more than 90 compact binary mergers~\cite{LIGOScientific:2021djp,LIGOScientific:2021usb,LIGOScientific:2024elc}, and independent research groups~\cite{Venumadhav:2019tad,Venumadhav:2019lyq,Zackay:2019btq,Olsen:2022pin,Mehta:2023zlk,Wadekar:2023gea} have discovered additional events. 
	Thus, GWs have become a novel tool to explore the Universe. The observed signals have been used 
		to measure the mass and spin distributions of BHs and \acp{NS}, their formation channels, and the co-evolution of their properties with that of the Universe~\cite{KAGRA:2021duu,Fishbach:2018edt}. \Ac{BNS} mergers have improved the bounds on the nuclear equation of state and the maximum allowed mass of a NS~\cite{LIGOScientific:2018cki,Radice:2017lry,De:2018uhw}. The mass distributions have been employed to constrain the observed lower and predicted upper mass gaps and 
	other features in the mass spectrum. In conjunction with the \ac{EM} counterparts observed for GW170817, or together with available galaxy catalogs, they have also been used to constrain the \ac{$H_0$}~\cite{LIGOScientific:2017adf, LIGOScientific:2021aug}. 
	In addition, \ac{GW} measurements have probed \ac{GR} as the fundamental theory of gravity~\cite{LIGOScientific:2016lio,LIGOScientific:2020tif,LIGOScientific:2021sio}.
	
	Improvements in the sensitivity of current  GW detectors and
	proposed \ac{XG} ground-based observatories like the \ac{ET} and \ac{CE}~\cite{Punturo:2010zz,Maggiore:2019uih,Reitze:2019iox} 
		will significantly increase the observational volume and, with it, the
	number of GW sources. For instance, a network of \ac{XG}
	detectors will observe \emph{every} stellar-origin 
	BBH merger and most BNS mergers across the observable
	Universe~\cite{Borhanian:2022czq}.  
	A number of studies have explored in
		detail the extent to which various science objectives can be
		accomplished 
	~\cite{Evans:2021gyd,Gupta:2023lga,Branchesi:2023mws,Bogdanov:2022faf,Iacovelli:2022bbs,Borhanian:2022czq}. 
        With $\mathcal{O}(100)$
	observations by the \ac{LVK} Collaboration, and $\mathcal{O}(10^5)$ promised 
		detections with \ac{XG} detectors, meaningful inferences on the properties of the 
		astrophysical distribution of BBHs will constrain more and more the underlying distribution
	of main sequence stars and their evolution.  EM observations in our Galaxy indicate that stellar-origin 
        BHs have masses
	above $5 \,\rm M_{\odot}$. However, these observations may be biased by properties that are unique to our Galaxy. 
	Similarly, the \ac{PISN}
	process is expected to suppress BH formation in the mass range of around 
	$50 \mbox{--}120 \,\rm M_{\odot}$~\cite{Woosley:2021xba,Belczynski:2016jno}. Confident detections of BBHs in
	these mass ranges would pose challenges to stellar-evolution models, as
	well as constrain the $\prescript{12}{}{C}(\alpha,
	\gamma)\prescript{16}{}{O}$ reaction rate that drives the \ac{PISN}
	process~\cite{Farmer:2020xne}. Gravitational-wave astronomy can also 
	determine the cosmological evolution of the Universe. 
	In particular, by combining many GW signals, it will contribute to resolving the \ac{$H_0$} tension, and provide new constraints on structure
	formation. On the other hand, loud individual events carry a lot of
	information too. Individual ``golden'' BBHs can also
	resolve the Hubble-Lema\^itre tension~\cite{Borhanian:2020vyr}. Finally, precision tests of \ac{GR} can be derived from high SNR 
	observations~\cite{Pang:2018hjb,Maggio:2022hre,Hu:2022bji,Toubiana:2023cwr,Bhat:2022amc,Saini:2023rto,Narayan:2023vhm}. 
	All of these scientific objectives are vulnerable to false positives arising from waveform inaccuracies.
	
	The source properties are estimated from the GWs via Bayesian inference using waveform models predicted by \ac{GR}.  
    Since there is no complete, closed-form analytic solution for the gravitational waveform of a compact-binary coalescence (CBC), various approximate and numerical methods have been developed to describe the GW signal faithfully. The \ac{EOB} waveforms ~\cite{Buonanno:1998gg,Buonanno:2000ef,Damour:2000we,Damour:2001tu,Buonanno:2005xu,Ramos-Buades:2021adz,Pompili:2023tna,Ramos-Buades:2023ehm,vandeMeent:2023ols,Khalil:2023kep,Nagar:2018zoe,Nagar:2019wds,Nagar:2020pcj,Gamba:2021ydi,Nagar:2023zxh} combine and resum several perturbative results, such as \ac{PN}, \ac{PM} and gravitational self-force information for the conservative and dissipative dynamics, with physically motivated ansatze for the merger, and BH perturbation theory for the ringdown. They are made highly accurate through 
		calibration to \ac{NR} simulations~\cite{Pretorius:2005gq, Campanelli:2005dd, Baker:2005vv}. The \ac{IMR} phenomenological (IMRPhenom) models~\cite{Pan:2007nw,Ajith:2007qp,Hannam:2013oca,Pratten:2020ceb,Estelles:2021gvs,Thompson:2023ase} are constructed in two steps. First, one stiches together in time domain an inspiral \ac{EOB} waveform with an \ac{IMR} \ac{NR} waveform, and Fourier-transforms it. Then, one fits the latter to a frequency-domain closed-form expression based on the
        \ac{PN} stationary-phase approximation for the inspiral and plunge, and physically motivated phenomological ansatz for the merger and ringdown. Where \ac{NR} simulations are not available, \ac{EOB} wwaveforms are used to calibrate the model. \ac{NR} simulations give the most accurate representation of a GW signal, although they are still limited by numerical truncation errors~\cite{Hannam:2009hh,Hinder:2013oqa,Boyle:2019kee}, imperfect outer boundary
        conditions~\cite{Rinne:2008vn,Buchman:2012dw,Buchman:2024zsb}, and issues with GW extraction and extrapolation~\cite{Chu:2015kft,Mitman:2020bjf}.  Moreover, \ac{NR} simulations are not available in the entire parameter space and are limited in length due to their high computational cost. 
	\ac{NR} surrogate models (NRSur)~\cite{Blackman:2015pia,Varma:2018mmi,Varma:2019csw,Yoo:2023spi} are constructed by directly interpolating \ac{NR} waveforms, where available. 

	\begin{figure*}
		\centering
		\includegraphics[width=2\columnwidth]{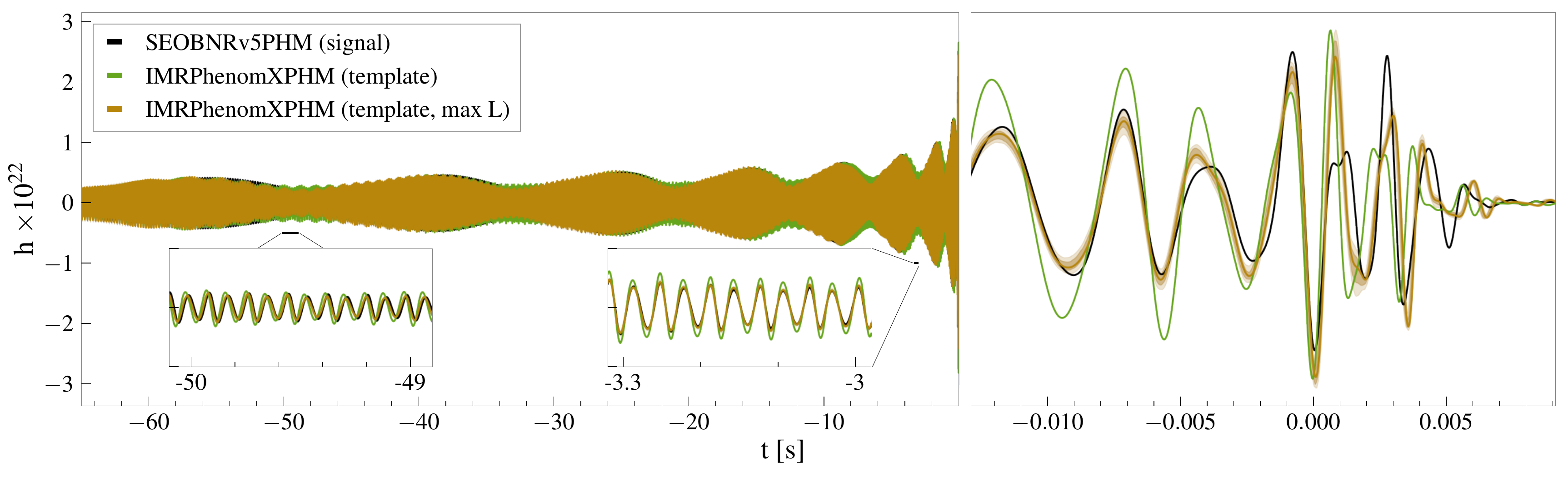}
		\caption{\Ac{GW} strains for a BBH system with parameters given in Table~\ref{tab:params} (\first) at the LIGO-Livingston detector of the \emph{O5} network. The black curve is the injected signal \eob, the green curve is the template \xphm, evaluated for the injection parameters, and time shifted and global-phase rotated to maximize their overlap with the signal; the brown curve is the template \xphm evaluated at the maximum likelihood values obtained using a Bayesian analysis. 
The reference for the time axis, $t=0$, is taken to be the peak of the \ac{GW} multipole $h_{22}$ of the signal.}
		\label{fig:S1_L}
	\end{figure*}

	Thanks to advancement in GW modeling since the discovery of GW150914~\cite{LIGOScientific:2016ebw}, 
		waveform models have been sufficiently accurate to analyze most signals 
		in the \ac{LVK} GW Transient Catalogs (GWTC)~\cite{LIGOScientific:2016ebw,Purrer:2019jcp}. In Ref.~\cite{Hu:2022rjq}, the authors 
		used the absolute value of the difference between waveform
		models to quantify the accuracy of a given pair of models, finding that a few high \ac{SNR} events in the GWTC-3 and GWTC-2.1
		fail their criterion. They also find that parameter estimation of such events shows greater inconsistencies.
		A reanalysis of the GWTC-3 by Ref.~\cite{Islam:2023zzj} finds that the \texttt{NRSur7dq4} model recovers noticeably different 
		parameters compared to LVK analyses using \xphm and \texttt{SEOBNRv4PHM} waveform models 
		for around $20 \%$ of the events where \texttt{NRSur7dq4} model can be used~\footnote{Note that some differences 
		are likely to be attributed to sampler issues rather than waveform systematics.}.
		A hypermodel approach to identify waveform systematics has also been carried out on the 
		13 heaviest \ac{GW} events from the GWTC-3. In this approach, waveform models are treated as
		parameters and directly sampled over, yielding a direct probability for each waveform model. The authors do not 
		find any waveform model to be preferred except for three events that are marred by data quality issues~\cite{Puecher:2023rxw}.
		Recently, there have also been efforts to marginalize over waveform modeling uncertainties~\cite{Jan:2020bdz,Read:2023hkv}.
	Other studies have found that even relatively low \ac{SNR}
	events could be affected by systematic biases if they lie in a region
	of the parameter space where calibration with \ac{NR} is
	sparse, which would include binaries that
	are asymmetric, eccentric, have large spin magnitudes, and/or have
	precessing orbits~\cite{Ramos-Buades:2023ehm,Hu:2022rjq}. 

	With increasing detector sensitivity and number of detections, the
	median \ac{SNR} of the observed population of binaries, as well as the
	likelihood of detecting a binary from a region of the parameter space
	where waveform inaccuracies are greater, will increase. While
	statistical uncertainties decrease with increasing \ac{SNR},
	systematic biases are independent of the signal power. Several studies
	have explored the validity of waveform models for the
	parameter estimation of quasi-circular \ac{BBH} mergers in upgraded
	and \ac{XG} detectors, mainly focusing on the biases for
	individual events~\cite{Owen:2023mid,Purrer:2019jcp,Hu:2022rjq,Kapil:2024zdn}, 
	with Ref.~\cite{Purrer:2019jcp} also showing the inferred distribution 
	of the primary mass to be biased. 
	While the negligence of subdominant modes can significantly bias the
	parameter estimation of individual
	events~\cite{Varma:2014jxa,Varma:2016dnf}, a recent study indicated that
	such biases do not affect the inference of the LVK-like astrophysical
	distribution of \acp{BBH}~\cite{Singh:2023aqh}. Other studies have
	focused on waveform systematics in the presence of
	eccentricity~\cite{Favata:2021vhw,Cho:2022cdy}, matter
	effects, and spin-precession~\cite{Samajdar:2018dcx,Samajdar:2019ulq,Gamba:2020wgg,Pratten:2021pro,Kolitsidou:2024vub}. 
	Recent studies have also explored the effect of truncation errors in \ac{NR} simulations
	employing finite-differencing methods and concluded that current
	simulations are not accurate enough for highly asymmetric binaries
	and binaries whose orbits are inclined with respect to the line of sight~\cite{Ferguson:2020xnm,Jan:2023raq}.
	However, state-of-the-art waveform models, such as \eob~\cite{Pompili:2023tna,Khalil:2023kep,vandeMeent:2023ols,Ramos-Buades:2023ehm} and \xphm~\cite{Pratten:2020fqn,Garcia-Quiros:2020qpx,Pratten:2020ceb}, are calibrated to 
	the Simulating-eXtreme-Spacetimes (SXS) Collaboration  waveforms, which employ spectral methods, 
        and the effect of truncation errors
	on these waveforms has not been explored systematically.
	An indistinguishability criterion~\cite{Flanagan:1997kp} has also been used as an easy-to-compute metric to 
	determine the accuracy requirements of waveforms~\cite{Chatziioannou:2017tdw,Purrer:2019jcp}. 
	However, this measure has been found to be very conservative. Reference~\cite{Toubiana:2024car} proposed a 
	correction to improve the reliability of the measure. 

	We illustrate the effect of waveform mismodeling in Fig.~\ref{fig:S1_L}, using a BBH with parameters given in Table~\ref{tab:params}~\footnote{We refer the reader to \cref{sec:gw_signal,sec:waveform_models,sec:alignment} for discussions on \ac{GW} parameters, waveform models, and maximization of overlaps between waveforms.}.
		We show the multipolar, spin-precessing GW strains 
		in the LIGO-Livingston detector from the \eob waveform model as the signal (black curve)
		and the \xphm waveform model as the template (brown and green curves). For the green curve, we fix 
		the polarization angle and time at coalescence, by maximizing its overlap against the \eob signal. 
		We employ the LIGO-Virgo detectors assuming the sensitivity of the upcoming fifth observing (\emph{O5}) run. 
		If the green GW strain faithfully represented the signal (black), they would perfectly match throughout the 
		coalescence. However, this is not the case, the amplitude modulations are different during the 
		long inspiral and, in particular, during the late inspiral, merger and ringdown. 
		Furthermore, while the signal and the template phases match during the early inspiral, there is significant
		dephasing near the late inspiral, merger and ringdown. In Fig.~\ref{fig:S1_L}, we also show the \xphm 
		template (brown curve) evaluated at the maximum-likelihood parameters (obtained through a 
		Bayesian analysis). It has a much better match to the \eob signal 
		even during the late inspiral, merger, and ringdown. This best match is obtained at the expense of 
		introducing a bias in the parameters; notably the total mass, mass ratio, and the spin-precession parameter 
		are biased by about $3\%$, $6\%$ and 
		$13\%$, respectively. 
		The brown  curve also has an
	associated brown band representing the measurement errors at a 90\%
	credible interval in the GW parameters, but it is barely visible to
	the naked eye, illustrating that this uncertainty, which represents the estimated statistical uncertainty from instrumental noise, is much smaller 
	than the waveform difference between the signal and the
	template evaluated at the best-fit parameters. As previously stated, this inconsistency
	manifests itself as biased parameter estimation, which could affect the 
	various science objectives. 

In this work, we start to quantify the systematic biases that can be
	expected in future observing runs with current facilities and 
		\ac{XG} detectors using the \eob and
	\xphm waveform models, which are employed for 
	parameter-estimation studies of BBHs by the \ac{LVK} Collaboration. Both models are valid for
	quasi-circular binaries and incorporate subdominant spherical
	harmonics and spin-precession effects. While it would be ideal to
	quantify the biases of each of these models against the true \ac{GR}
	signal, it is infeasible to do so everywhere in the parameter space
	 since \ac{NR} waveforms 
	are not available. 
	We leave to a future study the use of NR waveforms as synthetic signals where available. 
	Throughout this paper, we instead generate GW signals using
	\eob, considering these to represent the true signal, and
	analyze them using \xphm. However, there is a drawback to this
	approach. If both the waveform models deviate in a similar way from
	the true \ac{GR} signal, the present analysis would predict small
	biases even when the true bias is large. This outcome is especially true
	since the two waveform models are not completely independent. \xphm
	uses the \texttt{SEOBNR} waveforms (although from a previous version, 
		i.e., \texttt{SEOBNRv4})
	for calibration in parts of the parameter space
	where there is a dearth of \ac{NR} simulations---precisely
	the regions where systematic biases are expected to be more common.
	In this sense, our analysis is a conservative assessment of the prevalence of systematic biases.

	To quantify the systematics of the aforementioned waveform models in a wide range of applications, we utilize Bayesian analysis as well as the \ac{LSA}. The former is the most reliable tool to obtain the posterior distribution for a \ac{GW} signal, but it is computationally expensive. The latter allows for computational efficiency but approximates the predictions for the posterior properties, including systematic biases, and should become a good approximation only at large
    \ac{SNR}~\cite{Finn:1992xs,Cutler:2007mi,Vallisneri:2007ev,Cho:2012ed,Harry:2021hls}. We use the \ac{LSA} to study biases for \ac{BBH} populations, and a wider parameter space, which is not feasible with conventional Bayesian methods. 
	We consider three detector networks comprised of the current LIGO-Virgo network at design sensitivity (\emph{O5}), a planned network where the current LIGO detectors are upgraded to improved sensitivity (\emph{A\#}), and a \ac{XG} network comprised of two \ac{CE} and an \ac{ET}. 
	The \ac{BBH} populations we consider follow the \ac{LVK}-like distributions where the binary masses are distributed as determined by \ac{LVK} while the spins are assumed to be isotropically oriented and distributed uniformly in magnitude. We use this approach to allow for a wider range of spins. The binaries extend up to a redshift of 3 following the Madau-Dickinson \ac{SFR}~\cite{Madau:2014bja}. 
	Next, we embark on a parameter-exploration study where we consider large redshifted total masses of $200\rm\,M_{\odot}$, asymmetric systems with inverse mass ratios going up to 30, and highly spin-precessing systems. We study these as-yet unobserved regions of the parameter space in anticipation of future observations.
	We also consider three distinct prototypes of \ac{BBH} mergers, which hold great potential for various science objectives but are nontrivial to model due to precession or large mass ratios. 
The details of these three ``golden'' binary systems can be found in Table~\ref{tab:params}.

	The paper is organized as follows. 
In Sec.~\ref{sec:foundation}, we introduce 
        the main characteristics of the \ac{GW} signal and its parameters, the waveform
        models that we use, and the detector networks in which signals
        are simulated. In Sec.~\ref{sec:methods}, we describe our
        methodology comprised of Bayesian analysis and \ac{LSA}. We
        point out the importance of having consistent
        parameter definitions across waveform models and the impact on
        the systematic bias, where we show a comparison of a
        Bayesian analysis with the estimates from \ac{LSA}. 
        We also discuss the limitations of the \ac{LSA} for parameter 
       estimation (notably, the Fisher information matrix, or FIM) and biases. The study
        of systematic biases in the \ac{LVK}-like \ac{BBH} population
        and a hierarchical Bayesian inference on parameter
        distributions, reweighted to the \ac{LVK} population, is
        reported in Sec.~\ref{sec:LVK_bias}. A much broader study
        across the binary parameter space with particular focus on
        massive, highly asymmetric and spin-precessing binaries
        is reported in Sec.~\ref{sec:parameterspace}. A ramification
        on the different science applications for \acp{GW} can be
        found in Sec.~\ref{sec:science_case} where we study selected 
        \ac{GW} events or ``golden'' binaries. The discussion and conclusion can be found in
        Sec.~\ref{sec:discussion_conclusion}. 
        In Appendix~\ref{sec:toy_model}, we illustrate the effect of 
        nonuniform-parameter definitions across waveform models on the 
        estimates of the systematic bias through a toy model.
        In Appendix~\ref{sec:eob_flo}, we discuss the effect of different 
        harmonics of the \ac{EOB} model starting at different frequencies. 
        In Appendix~\ref{sec:flo_pe}, we discuss the effect of the starting frequency
        of the analysis on parameter estimation and systematic biases.
        In Appendix~\ref{sec:ratio_snr}, we provide a complementary plot to Fig.~\ref{fig:ratio} 
        by reporting the dependence of the ratio of systematic bias to statistical error 
        as a function of the \ac{SNR}. 
        In Appendix~\ref{sec:population_distribution}, we show the effect of the \ac{SNR} threshold 
        on the distribution of the population parameters. 
        In Appendix~\ref{sec:exploratory_bias}, we report the bias horizon for the $\chi_1$
        parameter of the exploratory binaries of Sec.~\ref{sec:parameterspace}.
	
	\section{Gravitational-wave parameters, models and detectors}
	\label{sec:foundation}
	
	\subsection{Gravitational-wave parameters}
	\label{sec:gw_signal}
	
	We are interested in estimating the properties of quasi-circular, spin-precessing \acp{BBH}
	observed with current and future ground-based detector networks. The
	\ac{GW} strain emitted by such binaries is characterized by
	15 parameters. The parameters intrinsic to the source are the
	component masses, $m_{i}$~\footnote{We adopt the convention
		$m_1 \geq m_2$.} and the dimensionless spin vectors, $\bm{\chi}_{i} = \bm{S}_{i}/m_{i}^2$ $(i = 1,2)$. 
	The position of the binary is described
	by its luminosity distance, $D_L$, and the coordinates on the plane of
	the sky, $(\alpha,\delta)$. The orientation of the binary 
	is described by the polar angle, $\iota$, and the azimuthal angle, 
	$\varphi$, to the observer in the source frame~\cite{Schmidt:2017btt} at the reference frequency, $f_{\rm ref}$,
	which we set to $f_{\rm ref}=20\rm\, Hz$ throughout this paper. 
	Finally, the relative
	contribution of the two gravitational polarizations, $h_+(t)$ and
	$h_{\times}(t)$, is described by the polarization angle, $\psi$, while 
	the reference for the time is given by the coalescence time,
	$t_c$. With these definitions, the \ac{GW} strain can be expressed as
	\begin{equation}
	\begin{split}
	h(t) =& h_{+}(t;m_{z,i},\bm{\chi}_{1,2},D_L,\iota,\varphi,t_c) F_{+}(\alpha,\delta,\psi) \\ 
	&+ h_{\times}(t;m_{z,i},\bm{\chi}_{1,2},D_L,\iota,\varphi,t_c) F_{\times}(\alpha,\delta,\psi)\,,
	\end{split}
	\end{equation}
	where $F_{+,\times}(\alpha,\delta,\psi)$ are the antenna pattern functions~\cite{Finn:1992xs,Sathyaprakash:1991mt}. The detector- and source-frame masses, $m_{z,i}$ and $m_{i}$, respectively, are related by $m_{z,i}=m_{i}(1+z)$ with $z$ being the redshift of the source. A superscript on any mass parameter indicates that it is the detector frame while its absence indicates that it is the source frame. The parameters $D_L$ and $z$ are related for a given cosmological model, which we take 
	to be the one from \texttt{Planck18}~\cite{Planck:2018vyg}. 
	The two \ac{GW} polarizations can be decomposed in the basis of $-2$ spin-weighted spherical harmonics, $\prescript{}{-2}{Y}_{lm}$, as
	\begin{equation}
	h_+(t) - ih_{\times}(t) = \sum_{l=2}^{\infty} \sum_{m=-l}^{+l} \prescript{}{-2}{Y}_{lm}(\iota,\varphi) h_{lm}(t)
	\end{equation}
	where $h_{lm}(t)$ are the \ac{GW} multipoles and $\varphi=\pi/2-\phi_{\rm ref}$. 
	
	It is often helpful to express the \ac{GW} signal in terms of parameters that are combinations of the component masses and spins, either because they appear in such combinations in \ac{PN} expressions or because they are conserved up to certain \ac{PN} orders. 
	In particular, the chirp mass, $\mathcal{M}_c$, and the symmetric mass ratio, $\nu$, are defined by $\mathcal{M}_c = (m_1 m_2)^{3/5}/M^{1/5}$ and $\nu = (m_1 m_2)/M^2$, respectively, where $M = m_1 + m_2$ is the total mass. The effective spin, $\chi_{\rm eff}$ \cite{Damour:2001tu,Ajith:2009bn,Santamaria:2010yb}, and spin-precession, $\chi_{\rm p}$ \cite{Schmidt:2014iyl}, parameters are given by 
	\begin{subequations}
		\begin{align}
		\label{eq:chieff}		
		\chi_{\rm eff} &= \frac{m_1 \chi_{1z} + m_2 \chi_{2z}}{m_1 + m_2}\,, \\
		\chi_{\rm p} &= \frac{1}{B_1 m_1^2} \max{(B_1m_1^2\chi_{1,\perp}, B_2m_2^2\chi_{2,\perp})}\,,
		\label{eq:chip}		
		\end{align}
	\end{subequations}
	where $B_{1,2}=2+3 \,m_{2,1}/m_{1,2}$, and $\chi_{i\perp}$ and $\chi_{iz}$ are the magnitudes of the projection of $\bm{\chi}_{i}$ onto the orbital plane and perpendicular to it, respectively. While alternative definitions of $\chi_{\rm p}$ have been proposed \cite{Thomas:2020uqj, Gerosa:2020aiw}, the \ac{GW}--Bayesian-analysis package \bilby~\cite{Ashton:2018jfp,Romero-Shaw:2020owr}, 
	which we use to analyze simulated signals employs Eq.~(\ref{eq:chip}).
	
	When transforming to a spherical coordinate system with the $z$ axis perpendicular to the instantaneous orbital angular momentum, the tilts of the two spin vectors with respect to the $z$-axis, $\theta_{i}$, are given by
	\begin{equation}
	\cos\theta_{i} = \frac{\chi_{iz}}{\chi_{i}}\,,
	\end{equation}
	where $\chi_{i}\equiv|\bm{\chi}_{i}|$ are the magnitudes of the dimensionless spin vectors. The relative angle between them in the orbital plane is parametrized by $\phi_{12}=\phi_1-\phi_2$ where $\phi_{1,2}$ are the azimuthal angles of the two spin vectors in the spherical coordinates. 
	Finally, the direction of the total angular momentum, $\bm{J}$, in the plane perpendicular to the orbital angular momentum, $\bm{L}$, at some reference time is given by the parameter $\phi_{\rm JL}$. Since $\bm{J}=\bm{L} + \bm{S}_1 + \bm{S}_2$, $\phi_{\rm JL}$ also defines the direction of the total spin vector in the orbital plane.
	The total angular momentum also defines the angle, $\theta_{\rm JN}$, which gives the orientation of the total angular momentum vector relative to the line of sight, $\bm{N}$, of the observer. 
	The angle $\theta_{\rm JN}$ can be expressed in terms of the inclination angle, $\iota$, at the reference frequency, $f_{\rm ref}$, e.g., through Eq.~(C9) of \textcite{Pratten:2020ceb}.
	In summary, the waveform depends on the following 15 parameters:
	\begin{equation}
		\bm{\vartheta} = \{ \mathcal{M}_c, \nu,\chi_i, \cos \theta_i, \phi_{\text{JL}}, \phi_{12}, 
		\alpha, \delta, D_{\text{L}}, t_{\text{c}}, \theta_{\text{JN}}, \psi, \phi_{\rm ref} \}. 
	\end{equation}
		In the following, we use bold font, like $\bm{\vartheta}$, to describe a set of parameters and regular font, like $\vartheta$, to describe a particular parameter in the set.

	\subsection{Waveform models}
	\label{sec:waveform_models}
	
	We consider two state-of-the-art, quasi-circular spin-precessing
	waveform models incorporating subdominant spherical harmonics---\eob
	and \xphm. The \ac{GW} modes $(l,m) \neq (2,2)$ are important both for
	detection~\cite{Brown:2012nn, Capano:2013raa, Harry:2017weg}, where their noninclusion leads to a loss of signal power 
	for asymmetric binaries and inclined orbits, and
	parameter estimation, where these modes can break degeneracies between
	various parameters and improve the measurement
	accuracy~\cite{Varma:2014jxa, Varma:2016dnf, Cotesta:2018fcv, Kalaghatgi:2019log}. 
		
	The \eob waveforms contain the spherical harmonics $(l,|m|) = (2,2), (2,1), (3,3), (3,2)$, and $(4,4), (4,3), (5,5)$ in the coprecessing frame. However, in this paper we do not include the $(l,m)=(5,5)$ mode. The \xphm waveforms include the $(l,|m|) = (2,2), (2,1), (3,3), (3,2), (4,4)$ modes in the coprecessing frame~\footnote{We note that, since our work started, there have been a few important updates on phenomenological models~\cite{Thompson:2023ase, Colleoni:2023czp}, which included NR calibration to the precessing sector, a more faithful ringdown model, and improvements to 
			the spin-precessing equations. However, we do not expect that our results would change substantially, if we used those new waveform models.}.
	The coprecessing frame is a non-inertial frame that tracks the instantaneous motion of the orbital plane, in which the GW radiation resembles that of an aligned-spin binary~\cite{Buonanno:2002fy, Schmidt:2010it, Boyle:2011gg, OShaughnessy:2011pmr, Schmidt:2012rh}.
	Both waveform models can be used for a wide range of mass ratios, as well as \ac{BH}-spin magnitudes up to the maximal values. However, only the aligned-spin sectors of both waveform models were calibrated to \ac{NR} simulations, and their accuracy has been assessed only in regions 
		of parameter space where NR is available.
	
	In this work, we consider a signal generated using the \eob model to be the true \ac{GW} signal and analyze it using \xphm as the template model. For the Bayesian analyses of this paper, we use this approach because \xphm is quicker to evaluate due to it being a frequency-domain model while \eob is a time-domain model and thus slower. Furthermore, the computational efficiency of \xphm can be improved by utilizing the multibanding approach~\cite{Garcia-Quiros:2020qlt} while no such analogous methods exist for time-domain models. 
	For the Fisher information matrix analysis discussed later, we find instabilities in the numerical derivatives of the \eob waveform with respect to the \ac{GW} parameters, for some regions of the parameter space, and hence restrict ourselves to computing derivatives of the \xphm model. We expect to address this issue in the future.

	
	\subsection{Detector networks}\label{sec:det_net}
	\label{sec:detnet}
	\begin{figure}
		\centering
		\includegraphics[width=\columnwidth]{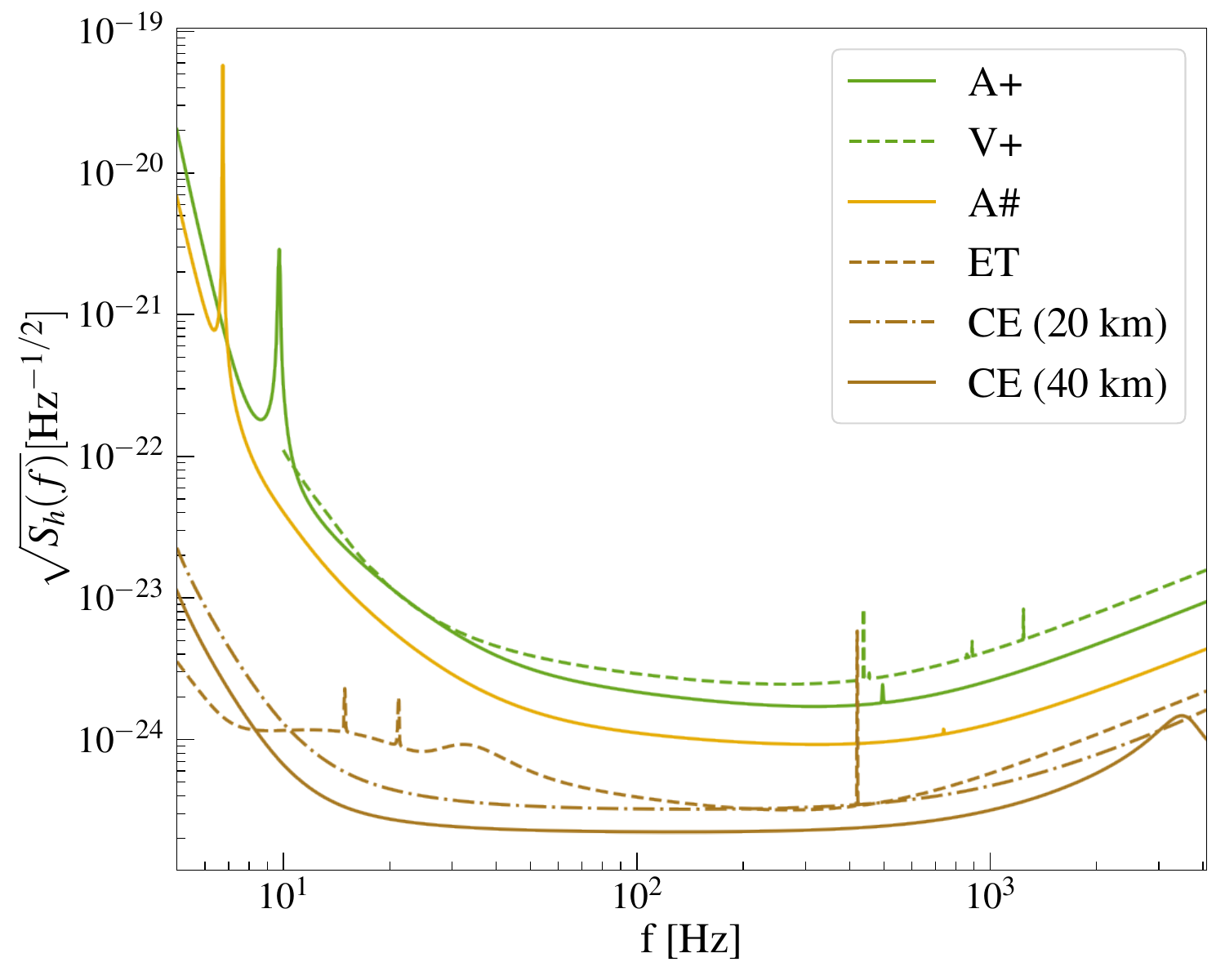}
		\caption{Amplitude--spectral-density curves of the various detectors used in this paper (see footnote \ref{curves}). The curves labeled by A+ and V+ denote the design sensitivity of the LIGO and Virgo detectors, respectively, which form part of the network of the fifth observing run (O5), while A\# refers to the LIGO detectors at upgraded sensitivity. The next-generation observatories are the \ac{ET} and \ac{CE} with the baseline network for the latter consisting of a 20-km and a 40-km detector.}
		\label{fig:psd}
	\end{figure}
	
	The current detectors are expected to achieve design sensitivity in
	the next few years, during the fifth observing (O5) run, and continue operating till the end of the
	decade~\cite{KAGRA:2013rdx}. It is anticipated that the detectors would undergo major
	upgrades thereafter and operate until next-generation detectors come online or
	even in tandem with them. 
	Since plans for future detector networks have not yet been finalized, a number of studies have 
	explored the capabilities of different combinations of detector configurations to understand what the optimal
	design is for various science goals~\cite{Borhanian:2022czq,Gupta:2023lga,Branchesi:2023mws,Srivastava:2022slt,Evans:2021gyd}.
	In this work, three \ac{GW} detector
	networks---consisting of the current detectors at design and upgraded
	sensitivity, and proposed future detectors---are considered to emulate
	a highly probable observing scenario for the coming
	decades~\footnote{\label{curves} The A+, V+, ET, and CE sensitivity curves in this work are 
		those used in Ref.~\cite{Borhanian:2022czq}
		while the A\# sensitivity curve is taken from \url{https://dcc.ligo.org/LIGO-T2300041/public}\,.}. These networks
	are enumerated below:
	\begin{itemize}
		\item \emph{O5} network: This network is comprised of the advanced LIGO detectors located at Hanford and Livingston, and the advanced Virgo detector operating at design sensitivities, A+ and V+, respectively. 
		\item \emph{A\#} network: In this configuration, the LIGO detectors operate at upgraded A\# sensitivity while the Virgo detector continues to operate at design sensitivity, V+.
		\item \emph{XG} network: This network comprises of three proposed XG observatories consisting of the baseline 40-km and 20-km \ac{CE} in the United States, and an \ac{ET} in Europe.  	\end{itemize}
	The \ac{PSD} of the individual detectors is shown in Fig.~\ref{fig:psd}.	
	
	\section{Statistical Methods}
	\label{sec:methods}
	
	This section describes the data-analysis methods used in this paper. In Sec.~\ref{subsec:bayesian_analysis} we start 
	by describing the Bayesian framework for analyzing \ac{GW} signals
	and lay out the different choices of priors
	and frequency bands used for the different networks. Thereafter, in Sec.~\ref{subsec:lsa}, we
	introduce the \ac{LSA} for the likelihood, and recount the \ac{FIM}~\cite{Finn:1992xs}
	method for estimating measurement errors. Following that, in Sec.~\ref{subsec:compare}, we elucidate
	the computation of biases (or systematic errors) under the \ac{LSA}~\cite{Flanagan:1997kp,Cutler:2007mi}, 
	and emphasize the importance of minimizing the mismatch between (i.e., 
		aligning) the signal and
	template for reliable estimates of the bias. As an example, we compare the
	posterior distributions for a chosen binary system, as obtained from a full Bayesian
	analysis with the estimates from the \ac{LSA}. Specifically, we point
	out the differences if the bias is computed without aligning the two
	waveforms. Finally, in Sec.~\ref{sec:hba} we discuss the hierarchical Bayesian method, which we 
		employ to understand the impact of biases on the inference of the properties of the 
		BBH population.
	
	\subsection{Bayesian analysis}
	\label{subsec:bayesian_analysis}
	
	The posterior probability distribution on the parameters of the waveform model, $\bm{\vartheta}$, given the observational 
		data $d$, and the hypothesis (model description) $\mathcal{H}$, is obtained using Bayes' theorem,
	\begin{equation}\label{eq:Bayes_theorem}
	p(\bm{\vartheta}|d, \mathcal{H}) = \frac{p(d|\bm{\vartheta}, \mathcal{H})\,p(\bm{\vartheta}|\mathcal{H}) }{p(d|\mathcal{H})} \,,
	\end{equation}
	where $p(\bm{\vartheta}|\mathcal{H})$ is the prior probability distribution, $p(d|\bm{\vartheta}, \mathcal{H})$ is the likelihood function, and $p(d|\mathcal{H})$ is the evidence of the hypothesis $\mathcal{H}$. If one is interested solely in parameter estimation, and not in model selection, the latter serves as a normalization constant, and can be discarded.  
	For a detector with stationary, Gaussian noise, the likelihood function for the data given the parameters $\bm{\vartheta}$ is defined as 
	\begin{equation}
	\label{eq:likelihood}
	\ln p(d | \bm{\vartheta}) = -\frac{1}{2} \left< d-h(\bm{\vartheta}) | d-h(\bm{\vartheta}) \right>\,,
	\end{equation}
	where we define the noise-weighted inner product as
	\begin{equation}
	\label{eq:ip}
	\left< h_1 | h_2 \right> = 4 \Re \left[ \int_{f_{\text{low}}}^{f_{\text{high}}} \frac{h_1(f) \times h_2^{*}(f)}{S_h(f)} df \right]
	\end{equation}
	with $S_h(f)$ being the noise \ac{PSD}, and $f_{\text{low}}$ and $f_{\text{high}}$ are the minimum and maximum frequencies 
		in the detectors' bandwidth. This inner product also defines the optimal, matched-filtering \ac{SNR} in a detector,  $\rho_n$, by
	\begin{equation}
	\rho_n^2 = \left< h(\bm{\vartheta}) | h(\bm{\vartheta}) \right>.
	\end{equation}
	The total \ac{SNR} is $\rho^2=\sum_{n=1}^N \rho_n^2$, where $N$ is the number of detectors in the network. 
	We note that all our injections are noiseless which corresponds to 
	averaging over multiple noise realizations.
	
	While the current detectors' sensitivity is limited to a minimum
	frequency of 20 Hz, at design sensitivity and with further upgrades,
	they are expected to reach a low-frequency sensitivity of 10
	Hz. Meanwhile, XG observatories are aiming to further
	this improvement to 5 Hz. Therefore, the minimum frequency
	for the \emph{O5} and \emph{A\#} networks is assumed to be
	$f_{\text{low}}=10\,\rm Hz$, while for \emph{\ac{XG}} detectors, it is taken
	as $f_{\text{low}} = 5\, \rm Hz$. On the other hand, the maximum frequency is
	kept the same for all three networks at $f_{\text{high}} = 1024\, \rm Hz$. This feature
	does not limit the analysis whatsoever since all the \ac{BBH} systems
	considered in Sec.~\ref{sec:science_case} merge at much lower
	frequencies. 

	As we mentioned earlier, the signal is generated using the \eob model with the
	same starting frequency as the analysis---$f_{\text{low}} = 10\, \rm Hz$ for
	\emph{O5} and \emph{A\#}; $f_{\text{low}} = 5\, \rm Hz$ for
	\emph{\ac{XG}}. Since \eob is a time-domain waveform model, this
	$f_{\text{low}}$ refers to the starting frequency of the $(l,m)=(2,2)$
	mode. Subdominant harmonics with $m'\neq2$ start at higher frequencies
	given by $f_{lo}^{m'}={m'}f_{lo}/2$. For instance, in \emph{O5}
	and \emph{A\#} networks, the $m'=3$ modes start at 15 Hz while the
	$m'=4$ modes start at 20 Hz. In Appendix~\ref{sec:eob_flo}, we
	show that this choice does not affect our results because of
	the minimal additional information contained in the missing
	frequencies compared to the rest of the signal.
	
	To simulate and analyze the \ac{GW} signals in Sec.~\ref{sec:science_case}, we use the publicly available \bilby
	package~\cite{Ashton:2018jfp,Romero-Shaw:2020owr}, which incorporates
	the nested sampler \dynesty~\cite{2020MNRAS.493.3132S}, interfaced
	through the \bilbypipe wrapper. Initially, a 14-dimensional \ac{GW}
	parameter space is sampled using the \dynesty sampler with a 
	distance-marginalized likelihood. The full posterior probabilities are then reconstructed
	using semi-analytic methods~\cite{Singer:2015ema,Singer:2016eax}. 
	
	All the detectors used in this study have an L-shaped interferometer
	configuration except the \ac{ET}, which is proposed to have a
	triangular configuration. However, the \bilbypipe wrapper is limited
	to L-shaped interferometer configurations. Consequently, the \ac{ET}
	telescope is assumed to be L-shaped in Sec.~\ref{sec:science_case}. 
	Our conclusions remain unaffected as the interferometer's shape has no
	significant impact on the science cases discussed here~\cite{Branchesi:2023mws}. 	
	
	We make standard choices for the priors for all the
		parameters~\cite{LIGOScientific:2018mvr}.
	The priors for the component masses are taken to be
	uniform, and the spins are assumed to be isotropic in direction and
	uniform in magnitude. For the distance we choose the prior $\propto d_L^2$, corresponding to a uniform in comoving volume distribution at low redshift. 
	We assume that the binary's position in the sky and the inclination of its orbit in the coprecessing frame are random. 
	Therefore, we assign uniform priors on $\alpha$, $\cos\delta$, and $\cos\theta_{\rm JN}$ across
	their domains. The other extrinsic parameters---namely,
	the polarization angle, coalescence time, and coalescence phase---are also taken
	to be uniform in their respective ranges. 
	
	\subsection{Linear-signal approximation for measurement errors, systematic biases and alignment}
	\label{subsec:lsa}
	
	\subsubsection*{Measurement errors}
	\label{subsec:fisher}
	
	The evaluation of the posterior probability distribution, as described
	in the previous section, is computationally expensive which makes the
	estimation of the measurement accuracies and systematic biases for a
	large number of sources computationally prohibitive using the Bayesian 
	method. An inexpensive approximate method is the \ac{LSA}, which we now briefly introduce.
	
	To estimate the parameter-estimation errors, the waveform model is
	expanded to linear order in the parameters around the maximum
	likelihood (best-fit) values, $\bm{\vartheta}_{\text{bf}}$. This process results in a
	Gaussian likelihood distribution whose covariance, $C_{ij}$, is given
	by the inverse of the \ac{FIM}, $C_{ij}=\Gamma_{ij}^{-1}$, which takes
	the form~\cite{Finn:1992xs,Cutler:2007mi},
	\begin{equation}
	\label{eq:fisher}
	\Gamma_{i j} \equiv \left<\frac{\partial h}{\partial \bm{\vartheta}^i} \middle| \frac{\partial h}{\partial \bm{\vartheta}^j}\right> \bigg|_{\bm{\vartheta}=\bm{\vartheta}_{\text{bf}}} .
	\end{equation}
	The marginalized one-dimensional errors are then given by the diagonal elements, $\Delta \bm{\vartheta}^i = \sqrt{C_{ii}}$~\footnote{$C_{ii}$ is the i-th element of $C_{ij}$}. The approximation holds for large \ac{SNR}. 
	
	We use the publicly available package \gwbench~\cite{Borhanian:2020ypi} to calculate the measurement errors. \gwbench is an easy-to-use \ac{FIM} analysis tool for ground-based detectors that implements finite-difference derivatives to estimate the approximate measurement errors. Other recent \ac{FIM} analysis codes for \acp{CBC} are \texttt{GWFAST}~\cite{Iacovelli:2022bbs} and \texttt{GWFish}~\cite{Dupletsa:2022scg}. The waveform models are internally referenced from the \lal~\cite{lalsuite} libraries. While the \xphm model is directly present in \lal, the \eob model is interfaced through the \texttt{pySEOBNR} package~\cite{Mihaylov:2023bkc} within \lal.
	For \xphm, the default model in \lal implements a multibanding approach~\cite{Garcia-Quiros:2020qlt} for faster waveform computation. However, we turn this off in our \ac{FIM} analysis because we find that the output of the last frequency bin has some randomness associated to it. This approach is harmless in a Monte Carlo sampling of the likelihood since the amplitude in that frequency bin is subdominant and does not contribute to the integral of Eq.~\eqref{eq:likelihood}. However, a \ac{FIM} analysis involves taking waveform derivatives with respect to binary parameters and the randomness manifests as a delta-function which dominates the integral in Eq.~\eqref{eq:fisher}.

	\subsubsection*{Systematic biases}\label{sec:syst_biases_meth}
	
	A further assumption in the 
	\ac{FIM} formalism is that there are no mismodeling errors,
	that is, the signal is accurately represented by the model
	waveform and errors are only due to a measurement process using
	detectors with finite sensitivity. In reality, we do not know the true 
	\ac{GW} waveform, and we use various approximate models to faithfully
	represent the true signal. As such, there is a source of error
	arising from a difference between the signal and the waveform model
	used to represent the signal (template). 
	As a result, the parameters that
	maximize the likelihood are biased from the true parameters of the
	\ac{GW} signal by $\delta \bm{\vartheta}^i$. This mismodeling error,
	henceforth called bias $\delta \bm{\vartheta}^i$, is given
	by~\cite{Flanagan:1997kp,Cutler:2007mi}\footnote{This expression first appeared in Ref.~\cite{Flanagan:1997kp}, but it 
			is often referred to as the Cutler-Vallisneri formula after a later paper~\cite{Cutler:2007mi}, which was the first to
			explore its implications.
	},
	\begin{equation}
	\label{eq:bias}
	\delta \bm{\vartheta}^i = C^{ij}\, \left< \partial_j h | \delta h \right> \big|_{\bm{\vartheta}=\bm{\vartheta}_{\text{bf}}},
	\end{equation}
	at the leading order, where $\delta h = h_{\text{s}} - h$ with $h_{\text{s}}$  being the true 
	signal. In practice, the true signal is not known, so this formula can only be used if the true signal is replaced by some fiducial reference model, here taken to be \eob. 
	
	\subsubsection*{Waveform alignment}\label{sec:alignment}
	
	We now discuss a few subtleties in the use and applicability of
	Eq.~\eqref{eq:bias} for the estimation of biases. Note that the bias
	is directly proportional to the waveform difference, $\delta h$. In part due to
		different conventions for some extrinsic parameters, $\delta h$ can be artificially large when evaluated 
		at the same value of all parameters, but it can be significantly reduced by changing the values of certain
		extrinsic parameters, such as the global phase and time shift, while keeping the intrinsic parameters fixed. 
		In Appendix~\ref{sec:toy_model} we describe a toy model that illustrates how a simple
		time shift can cause biases in physical parameters to become large.
		Since Eq.~(\ref{eq:bias}) 
	is derived under the \ac{LSA}, large
		waveform differences stretch the formula beyond
	its domain of validity resulting in unreliable estimates. However, large uncertainties in the extrinsic parameters are 
		typically not problematic for scientific applications of \ac{GW} observations; thus, if by changing only a subset of the extrinsic parameters, we can bring the waveform difference back into the range of validity of the \ac{LSA}, we should do so in order to improve the accuracy of the inferred results. 

Waveform-accuracy studies in the literature that use $\delta h$ as
	a metric to quantify waveform differences, and estimate expected biases, 
	have typically followed this approach and minimized $\delta h$ over
	the extrinsic parameters~\cite{Damour:2010zb,Hu:2022rjq} (alignment). 
	However, to the best of our knowledge, many studies employing
	Eq.~\eqref{eq:bias} to estimate the bias either neglect this aspect
	and naively use the difference between waveform models to estimate the
	systematic bias, or they do not discuss it. The incorrect use leads to unreasonably large estimated biases, particularly for the luminosity
	distance. Therefore, we describe here how we implement the alignment in the bias formula. 
	
	Using Eq.~\eqref{eq:ip}, we define the unfaithfulness or mismatch between two waveforms $h_1$ and $h_2$ as
	\begin{equation}
	\mathcal{M} = \min_{\bm{\lambda}} \left\{ 1 - \frac{\left<h_1 | h_2 \right>}{\sqrt{\left<h_1 | h_1 \right>\left<h_2 | h_2 \right>}} \right\}
	\label{eq:mm}
	\end{equation}
	where the minimization is performed on a subset of the binary's parameters, which we denote 
		$\bm{\lambda}$. For nonprecessing waveform
	models employing only the dominant quadrupolar mode, 
	$\bm{\lambda} = \{\psi,t_c\}$. In this case, $\psi$ is degenerate with $\phi_{\rm ref}$, so we
	need to consider only one of them.
	On the other hand, since we are considering spin-precessing waveform models,
	$\bm{\lambda} = \{\psi,t_c,\phi_{\rm ref},\phi_{\rm JL}\}$ where $\phi_{\rm JL}$ 
		is a rotation of the in-plane spin angles. 
	For spin-precessing waveform models, some
	studies have chosen to minimize the mismatch over the reference
	frequency instead of in-plane spin
	rotations~\cite{Khan:2019kot,Ossokine:2020kjp}. However, in this
	study, we choose to optimize the mismatch by rotating the in-plane
	spin components~\cite{Pratten:2020ceb,Ramos-Buades:2023ehm}, 
	thus keeping the reference frequency fixed at $f_{\rm ref}=20\rm\,Hz$. 
	
	Starting with the set of parameters $\bm{\vartheta}$, we
	find the parameters $\overline{\bm{\lambda}}$ that minimize $\mathcal{M}$ in Eq.~(\ref{eq:mm}) for the detector
	network being considered. 
	The minimization over the polarization angle $\psi$ is performed analytically,
	while the coalescence time $t_c$ is optimized by convolving the two waveforms
	utilizing the convolution theorem~\cite{Sathyaprakash:1991mt,Allen:2005fk,Buonanno:2009zt}. The
	reference phase $\phi_{\rm ref}$
	and in-plane spin rotations $\phi_{\rm JL}$ are optimized numerically by
	using standard optimization algorithms. 
	Having found the 
		parameters that minimize Eq.~(\ref{eq:mm}), $\overline{\bm{\lambda}}$,
		we have a new set of parameters $\overline{\bm{\vartheta}}_{\text{bf}}$, where the 
		parameters $\bm{\lambda}= \{\psi,t_c,\phi_{\rm ref},\phi_{\rm JL}\}$ have been replaced by the values obtained 
		through Eq.~(\ref{eq:mm}). We use this set 
		of parameters to compute the \ac{FIM}, as well as, the
		$\delta h$ in Eq.~\eqref{eq:bias}. 
		Therefore, the alignment procedure modifies Eq.~\eqref{eq:bias} to
		\begin{equation} \label{eq:mod_bias}
			\delta\bm{\vartheta}^i = C^{ij}(\overline{\bm{\vartheta}}_{\text{bf}})\, \left< \partial_j h (\overline{\bm{\vartheta}}_{\text{bf}}) | h_{\text{s}} (\bm{\vartheta}) - h (\overline{\bm{\vartheta}}_{\text{bf}}) \right> \,.
		\end{equation}
		If the parameters $\bm{\lambda}$
		are uncorrelated with the other binary parameters, the bias formula
		Eq.~\eqref{eq:bias} should give $\delta \bm{\lambda}=0$. However, in general,
		that is not the case and, therefore, $\delta \bm{\lambda} \neq
		0$. Thus the total bias for $\bm{\lambda}$ is
		$\Delta \bm{\lambda} = \delta \bm{\lambda} +
		\left(\bm{\lambda}_s- \overline{\bm{\lambda}}\right)$ where
		$\bm{\lambda}_s- \overline{\bm{\lambda}}$
		is the difference
		between the parameters $\bm{\lambda}$ of the fiducial signal and those obtained
		after the optimization procedure.
	
Note that we chose to modify the
	template in Eq.~\eqref{eq:bias} following the optimization procedure Eq.~\eqref{eq:mm}. Under the
	\ac{LSA}, we are free to modify the signal evaluating the
	template at the fiducial parameters. However, we notice a
	slightly better agreement of the bias with full Bayesian 
	results when modifying the template because the
	Bayesian analyses are performed using the fiducial parameters as the
	values of the synthetic-injected signal; thus we find the systematic bias to be more sensitive
	to small changes in the injected values compared to the measurement errors.
	
	Lastly, we note that $\mathcal{M}$ could already be close to the minimum for 
	certain pairs of waveform models at a given set of parameters out of the box. In such cases, 
	the optimization procedure will have a minimal effect on the total bias and one could simply use 
	the bias formula as it is.
	However, the total bias
	$\Delta \bm{\lambda}$ would be the same regardless of
	whether one chooses to perform the initial optimization or not, even though
	the output of the bias formula will not be the same. Correspondingly, the
	net bias does not depend sensitively on the precision of the
	optimization routine as the bias formula compensates for
	it. Therefore, it is more prudent to compare the net biases rather
	than the optimized values following the initial minimization. 
	We verify our minimization procedure using a brute-force
	4D minimization algorithm and find that, while $\overline{\bm{\lambda}}$ 
	is slightly different between the two minimization routines, $\Delta\bm{\lambda}$ 
	remains the same.
	
	In the following, we calculate the systematic biases with and without the optimization procedure outlined above. 
		We compare these estimates with the full Bayesian-analysis results for a subset of the parameters, and find 
		agreement with the Bayesian analysis when using the optimization procedure in Eq.~\eqref{eq:mm}.
	
	\subsection{Comparing Bayesian and linear-signal approximation analyses}\label{subsec:compare}
	
	\begin{figure}
		\includegraphics[width=\columnwidth]{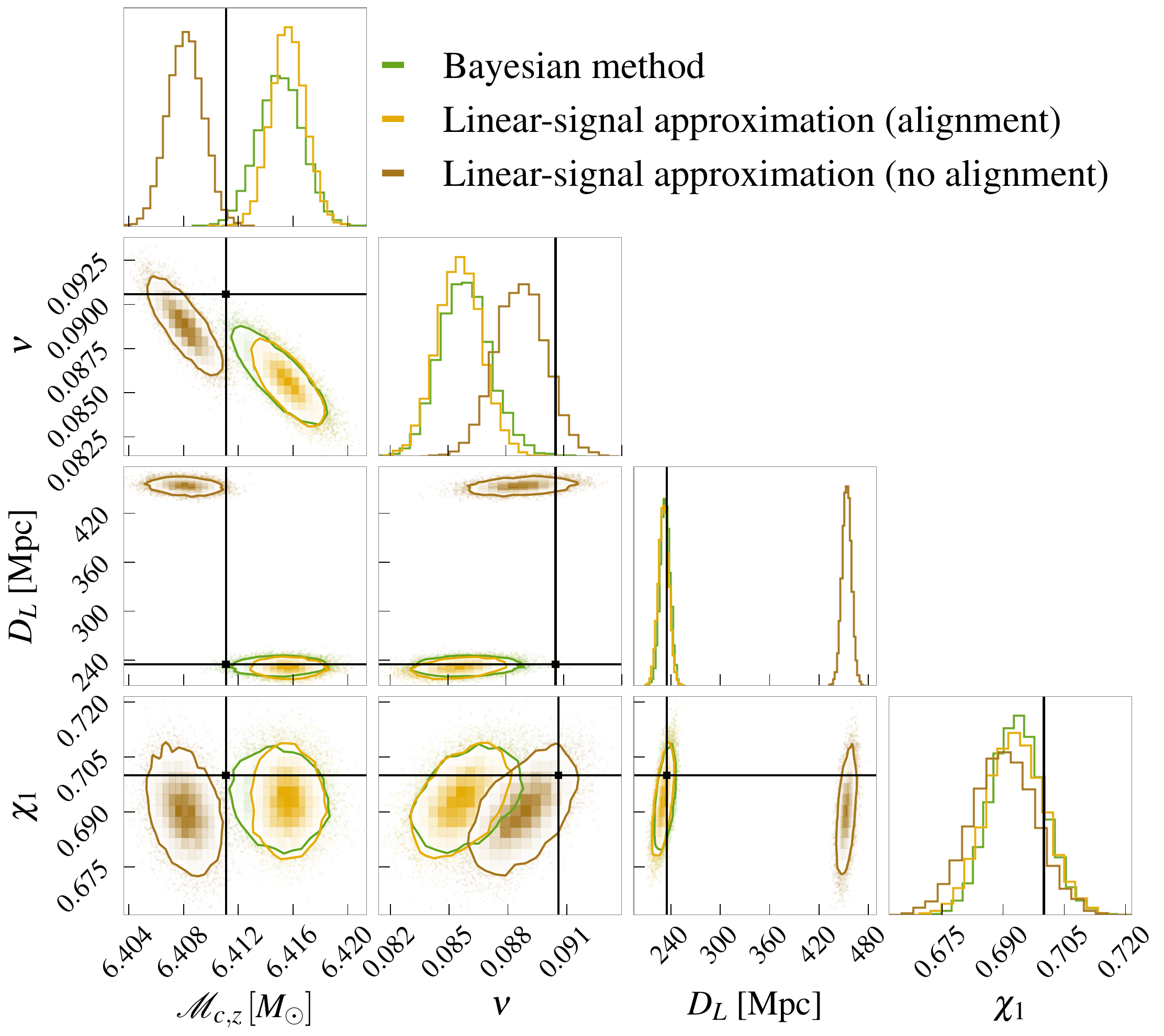}
		\caption{Comparison of the posterior distributions for the 
			chirp mass, symmetric mass ratio, luminosity
			distance, and primary spin magnitude for \first with parameters given in Table~\ref{tab:params} 
			and in the \emph{O5} detector network. 
			The distributions obtained from a Bayesian parameter
			estimation using \bilby are shown in green. The estimates
			from the \ac{LSA} with and without
			the minimization procedure (\cref{eq:mm}) (alignment) are shown in orange and brown, respectively. The
			black cross-hairs show the true injected value. The
			parameter estimation is performed by injecting a \eob signal
			and recovering it with the \xphm waveforms. The 
			Bayesian posteriors are accurately represented by the \ac{LSA} 
			when the alignment is enforced.}
		\label{fig:S1_bilby_fisher}
	\end{figure}

	As a representative case to compare results between the Bayesian
		method, and the \ac{LSA} with and without the optimization procedure
		of Eq.~(\ref{eq:mm}), we consider a BBH with parameters given in
		Table~\ref{tab:params} (see Sec.~\ref{sec:hbbh}), denoted as \first in the \emph{O5} detector network. 
		In Fig.~\ref{fig:S1_bilby_fisher}, we show the posterior
	distributions for selected parameters, namely, $\mathcal{M}_c$, $\nu$,
	$D_L$, and $\chi_1$, using the Bayesian analysis and the \ac{LSA} estimate for
	the errors and biases computed from Eq.~\eqref{eq:fisher} and
	Eq.~\eqref{eq:bias}, respectively. The full Bayesian posterior estimates
	are shown in green. The \ac{LSA} posterior distributions are multidimensional normal distributions
	centered at the biased value with the covariance matrix given by
	Eq.~\eqref{eq:fisher}. 
	The curves in orange show the distributions when the biases are estimated by minimizing the mismatch
	between the waveforms (see Eq.~(\ref{eq:mod_bias})), while the ones in brown are
	the estimates when the minimization is not performed (as it 
	is typically done in the literature; Eq.~\eqref{eq:bias}). Since the covariance matrix 
	is approximately the same in a neighborhood, the posterior widths are 
	similar. However, the predicted bias differs substantially between the
	two procedures. The effect on the estimation for the distance bias is especially
	noticeable, with the traditional method predicting an approximately $50\%$ bias
	when it is unbiased in actuality. This effect would be of particular
	importance for cosmology studies where the traditional estimates found 
		in the literature would be overly pessimistic. The contours show the 90\% 
	credible intervals of the parameters.
	
	We now briefly discuss the validity of the \ac{LSA}. Even though we find
	excellent agreement between the \ac{LSA} and the Bayesian analysis for
	this fiducial case, it is important to keep in mind that the estimates are approximate. 
	Particularly, both the \ac{FIM} and the bias formula (Eq.~(\ref{eq:mod_bias})) are derived under the assumption that
	a waveform model can be expanded linearly in its parameters. While the \ac{FIM} 
	approximation improves with increasing \ac{SNR}, with higher-order contributions 
	scaling as $\mathcal{O}(1/\ac{SNR})$, the bias is independent of the \ac{SNR} 
	both in the linear approximation and the full likelihood. For the \ac{LSA}, this case can be easily gauged
	from the bias equation,  Eq.~\eqref{eq:mod_bias}, which is independent of the distance
	and/or simple scaling of the \ac{PSD}. For the Bayesian analysis, one can conclude
	from Eq.~\eqref{eq:likelihood} that a simple scaling of the \ac{PSD} will not affect
	the stationary points. Therefore, the point in the parameter space where the likelihood
	peaks remains constant, which means that the error in the bias computation is also 
	constant. In addition, note that we are interested in the bias in units of the statistical errors. 
	Hence, while the measurement becomes better with improved sensitivity (or larger \ac{SNR}), the error 
	in the systematic bias estimated using \ac{LSA} becomes more important. 
	A priori it is difficult to know the range of the optimal place where both approximations hold. 
	However, the event shown in Fig.~\ref{fig:S1_bilby_fisher} has an $\ac{SNR}\sim75$ 
	and we also observe similar agreement in the \emph{A\#} network where the event 
	has an $\ac{SNR}\sim220$ (see Table~\ref{tab:params}) prompting us to make the 
	reasonable assertion that the \ac{LSA} is most trustworthy for such ranges of the \ac{SNR}.
	We were not able to directly compare the Bayesian results in the \ac{XG} network with 
	the \ac{LSA} estimates because as we explain in Sec.~\ref{subsec:bayesian_analysis}, 
	the former assumed an L-shaped interferometer for \ac{ET} while the latter was performed using 
	a triangular \ac{ET} configuration. 
	We would also like to stress that one would expect the \ac{LSA} to hold when the mismatch
	between two waveform models is not too large. For the case illustrated above, we find the 
	mismatch, $\mathcal{M}\sim3\%$. However, the binaries that are considered in Sec.~\ref{sec:LVK_bias}
	and Sec.~\ref{sec:parameterspace} can have much larger mismatches and a more detailed analysis
	is required to quantify the validity of the \ac{LSA} as a function of the mismatch which is beyond the scope of this study. 
	
	\subsection{Hierarchical Bayesian analysis}\label{sec:hba}
	
	We now discuss the method that we employ in Sec.~\ref{sec:LVK_bias} to understand the impact of the biases on the inference of the properties of the BBH population. Given a set of $N_{\rm obs}$ observed data $\{ d_i \}$, we can estimate the underlying distribution of parameters that generated it through a hierarchical Bayesian analysis. We denote by $\vartheta$ ($\subset \bm{\vartheta}$) the set of parameters, whose distribution we wish to infer. Assuming a form for the number density of observed events, $\frac{{\rm d} N}{{\rm d}\vartheta}  (\Lambda)$, that depends on \emph{hyperparameters} $\Lambda$, the posterior on the latter is given by \cite{Mandel:2018mve,Vitale2020} 
	\begin{equation}
	p(\Lambda|\{d_i\}) \propto \pi(\Lambda) e^{-N(\Lambda)}\prod_{i=1}^{N_{\rm obs}} \int  \frac{{\rm d} N}{{\rm d}\vartheta}  (\Lambda) \frac{p(\vartheta|d_i) }{\pi_{\text{PE}}(\vartheta) }  {\rm d} \vartheta , \label{eq:pop_posterior}
	\end{equation}
	where $p(\vartheta|d_i)$ is the single-event posterior, $\pi_{\text{PE}}(\vartheta)$ is the prior used for parameter estimation, $\pi(\Lambda)$ is the prior on the hyperparameters, and $N(\Lambda)$ is the total number of events, defined as
	\begin{equation}
	N(\Lambda) = \int \frac{{\rm d} N}{{\rm d}\vartheta}  (\Lambda) {\rm d} \vartheta.
	\end{equation}
	In the analysis of real data, the above equation must be modified to include selection effects. We interpret ${\rm d} N/{\rm d}\vartheta$ as the rate density of the full population, and modify the argument of the exponential to $p_{\rm det}(\Lambda) N(\Lambda)$, where $p_{\rm det}(\Lambda)$ is the probability of detection of a source, averaged over the population model. In this work, we do not seek to perform a full astrophysical inference. Thus, we treat the distribution of events that pass
    the cut-off in SNR as the population of interest and investigate the impact of systematic effects on the shape of this distribution. This is not strictly equivalent to performing the inference on the observed population, as this would require to "renormalize" the likelihood to the observable portion of the parameter space. Here, instead, we neglect the selection process, reconstruct ${\rm d} N/{\rm d}\vartheta$ as the distribution of events that pass the cut-off and compare how this changes
    under the effect of systematic biases. Let us mention that systematic errors might also bias our estimation of the selection function. The latter is usually evaluated performing injections with GW templates, and if the true signal that generates the observed data differs from our templates, we might estimate the probability of detecting an event wrong. In the analysis performed here, we instead approximate the selection as a hard cut on the intrinsic SNR of the source. In this model,
    selection is now defined on the source parameters, not the data, and the above equation can be used directly; however, ${\rm d} N/{\rm d}\vartheta$ must now be interpreted as the rate density in this observed portion of the population. This approach, which is common in the literature, ignores the fuzziness at the detection horizon that arises from instrumental noise, but it will give quantitatively reliable and unbiased results, provided the data are simulated from the same model.
In Eq.~(\ref{eq:pop_posterior}), we use the proportionality symbol instead of the equality one because we have omitted numerical factors that depend on the observed data $\{d_i\}$, but not on $\Lambda$, i.e., the individual event evidence and the overall model evidence. These factors are required to perform model selection, but are unimportant when the goal is to obtain the posterior distribution on $\Lambda$. 
	
	We perform a hierarchical Bayesian analysis for each source parameter separately---i.e., $\chi_1$, $q$, $\cos \theta_1$, and $\mathcal{M}_c$---so that $ {{\rm d} N}/{{\rm d}\vartheta}$ is a one-dimensional function. This approach yields optimistic measurements for the number densities as compared to the full inference, but allows us to have a quick assessment of the impact of systematic biases on population inference. 
	Adopting the approach of \textcite{Toubiana:2023egi}, we describe the number density of observed events, ${{\rm d} N}/{{\rm d}\vartheta}  (\Lambda)$, as a piecewise linear function. The extremities of the $\vartheta$ range over which we perform the inference are fixed, and determined by the minimum and maximum samples present in the data. Thus, our hyperparameters are as follows: the values of the number densities at the extremities, the number of knots, their positions and the value of the
    number density at the knots. The number density at any point is then obtained by linear interpolation. We stress that the number of knots is a free parameter of the model, and is inferred by using a reversible-jump Markov chain Monte Carlo algorithm \cite{Karnesis:2023ras}. In this way, the complexity of the model is determined by the data themselves.
	
	For a given detector network, we perform population inference on a mock catalog with systematic biases and on one without, generated as follows. 
	\begin{enumerate}
		\item We draw the parameters $\vartheta_0$ from the population model described in Sec.~\ref{sec:LVK_bias} and select those with \ac{SNR} above a given threshold.
		\item We compute the measurement error and the systematic bias for all observable events using the LSA, as described in Sec.~\ref{sec:syst_biases_meth}.
		\item For the catalog with systematic biases, we shift the true parameters by $\delta \vartheta_{bf}$ to obtain the biased parameters, $\vartheta_{bf}$. 
		\item For each event  $\vartheta_{i}$, we attribute a measurement error $\sigma_{i}$ drawn randomly among the set of computed measurement errors, allowing for replacement. 
		\item We draw a noisy measurement $\vartheta_{n,i}$ of each event from a Gaussian centered at $\vartheta_i$ ($\vartheta_{bf,i}$ for the biased catalog), with the standard deviation given by the error drawn in step 4.
	\end{enumerate}
	Under the LSA, the posterior distribution on $\vartheta$ is a truncated Gaussian:
	\begin{equation}
	p(\vartheta|\vartheta_{n,i}) = \frac{2 \exp \left [-\frac{1}{2}\frac{(\vartheta-\vartheta_{n,i})^2}{\sigma_i^2}  \right ]}{\sqrt{2\pi}\sigma_i \left [\erf \left (\frac{\vartheta_{\rm max}-\vartheta_{n,i}}{2\sigma_i} \right )+\erf \left (\frac{\vartheta_{n,i}-\vartheta_{\rm min}}{2\sigma_i} \right ) \right ]}, \label{eq:posterior}
	\end{equation} 
	where $\vartheta_{\rm min}$ and $\vartheta_{\rm max}$ are the boundaries of the prior domain on $\vartheta$.  
	The purpose of the randomization of the errors (step 4) is to remove the dependency on $\vartheta$ from the standard deviation entering the posterior distribution. If we were to use the corresponding value predicted by the \ac{FIM}  for each event, we would have to account for the complicated dependency of $\sigma$ on $\vartheta$, and the posterior on $\vartheta$ would no longer be a Gaussian, requiring us to go beyond quadratic order in the LSA. Moreover, $\sigma$ would also depend on the
    remaining parameters in $\bm{\vartheta}$, and, by performing the inference on a single parameter, we would not be accounting for this dependence correctly, making our analysis not self-consistent. However, we observe that, for $\vartheta=q$ or $\chi_1$ or $\cos \theta_1$, the amount by which the estimated uncertainty in the parameter varies over the range of our priors is small, so we expect our procedure to yield realistic results for those parameters.\footnote{In this way, we are not
    respecting the properties of the noise since we associate a random error to a given event. Crucially, this is performed for one-dimensional distributions, so, in practice, we project all the errors onto a one-dimensional space, inducing a large scatter in the value of measurement errors for a given value of $\vartheta$. Thus, for parameters that do not exhibit a strong trend between measurement errors and the true parameter---as we observed to be the case for the mass ratio and spin angles
    and magnitude---our procedure should be close enough to what we would obtain in the full case, and it allows us to also remove the dependency of the error on the other parameters.}
	Step 5 is crucial to make sure our mock catalog is self-consistent from the statistical point of view. Working in the so-called zero-noise approximation is valid for performing parameter estimation on single mock events, because it is a fair realization of the noise in the detector. On the other hand, having zero noise for all events is no longer a fair realization, and it would be valid only if all the events were perfectly measured. Note that, in steps 3 and 5, we allow the biased
    parameters and the noisy ones to be outside of the prior range. The rationale is that those steps are meant to mimic the behavior of the likelihood function in the presence of systematic biases and noise, which, as a function, does not contain information on the physically allowed range of a given parameter. The posterior, in turn, is truncated to the prior range, as given explicitly in Eq.~(\ref{eq:posterior}).
	
	In the hierarchical Bayesian analysis, we take the parameter estimation prior $\pi_{\text{PE}}(\vartheta)$ to be flat in $\vartheta$. Thus, each of the integrands in Eq.~(\ref{eq:pop_posterior}) is the product of a Gaussian with a piecewise linear function, and we can perform the integration analytically. This approach allows us to evade problems related to having an insufficient number of samples when performing Monte Carlo integration~\cite{Talbot:2023pex}, and it speeds up the analysis.

	\section{Systematic biases in the BBH population}
	\label{sec:LVK_bias}
	
	In this section, we study the effect of systematic biases on a \ac{BBH} population. We use the GWTC-3 results~\cite{LIGOScientific:2021djp,KAGRA:2021duu} only for the distribution of masses. We explore the impact of systematic biases on this LVK-like population, considering the three detector networks introduced in Sec.~\ref{sec:det_net}. We also perform a hierarchical Bayesian inference of the underlying population where we reweight our population distribution to the current \ac{LVK} distribution of astrophysical \acp{BBH}.
	
	\subsection{LVK-like population}
	\label{subsec:population}
	We simulate $10^5$ binaries in each of the detector networks described in Sec.~\ref{sec:detnet} up to a redshift $z=3$ using the \eob waveform model. Following~\textcite{Borhanian:2022czq}, this result is around the expected number of \ac{BBH} mergers per year.  
	We choose a network \ac{SNR} threshold of 12 for detection and use it to identify the subset of simulated binaries that are in the population observed by each network. 
	
	The redshift distribution for the population is drawn from a probability distribution given by
	\begin{equation}
	p(z) \propto \frac{dV_c}{dz} \frac{1}{1+z} \psi(z),
	\end{equation}
	where $ dV_c/dz$ is a comoving volume element per unit redshift and $\psi(z)$ is the \ac{SFR}, which is taken to be~\cite{Madau:2014bja}
	\begin{equation}
	\psi(z) = 0.015 \frac{(1+z)^{2.7}}{1 + [(1+z)/2.9]^{5.6}} \rm M_{\odot} yr^{-1} Mpc^{-3}.
	\end{equation}
	
\begin{figure}
		\centering
		\includegraphics[width=\columnwidth]{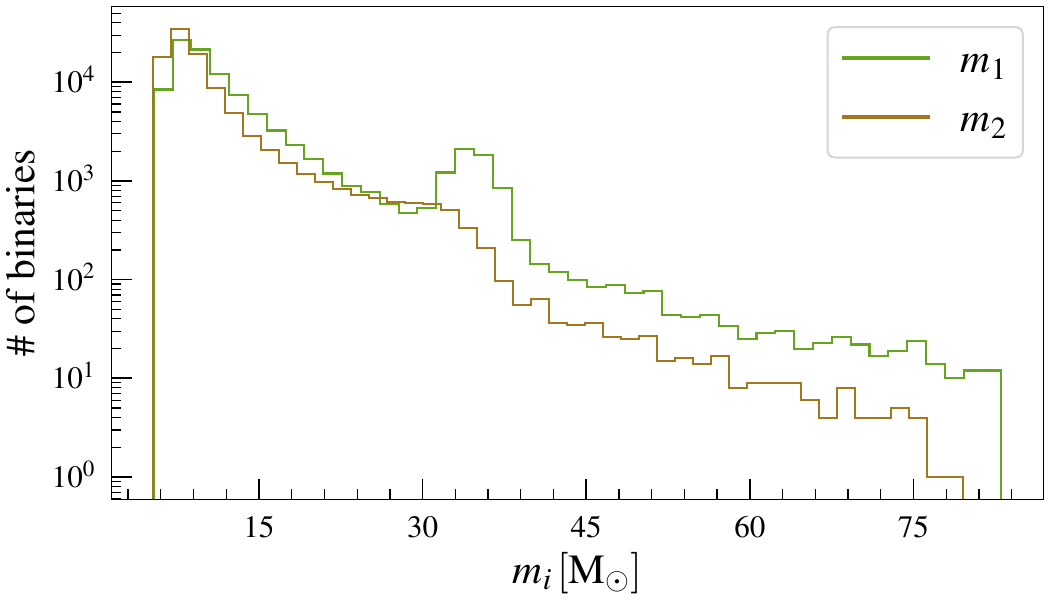}
		\caption{Distribution of component masses for a population of $10^5$ \acp{BBH} following the astrophysical distribution as determined by the \ac{LVK} Collaboration.}
		\label{fig:m1_m2}
	\end{figure}

	The distribution of the primary source-frame mass is assumed to follow the \mass model of \textcite{KAGRA:2021duu}, with the parameters fixed to their maximum likelihood values. For completeness and ease of reference, we elucidate the model  
	here. The \mass model is given by
	\begin{equation}
	\begin{split}
	p(&m_1 | \lambda_{\rm peak}, \alpha_m, m_{\rm min}, m_{\rm max}, \mu_m, \sigma_m, \delta_m) \propto \\ 
	&[(1-\lambda_{\rm peak}){\rm PL}(m_1 | \alpha_m, m_{\rm max}) + \lambda_{\rm peak}\mathcal{N}(m_1 | \mu_m, \sigma_m)] \\
	& \times S(m_1 | m_{\rm min}, \delta_m),
	\end{split}
	\end{equation}
	where $\lambda_{\rm peak}$ gives the weight of the peak component, ${\rm PL}(m_1 | \alpha_m,m_{\rm min}, m_{\rm max})$ is a normalized power-law distribution with spectral index $\alpha_m$ and truncated to the range $[m_{\rm min}, m_{\rm max}]$, $\mathcal{N}(m_1 | \mu_m, \sigma_m)$ is a normalized Gaussian distribution with mean $\mu_m$ and width $\sigma_m$
	and finally, $S(m_1 | m_{\rm min}, \delta_m)$ is a smoothing function defined by
	\begin{equation}
	S(m|m_{\rm min},\delta_m) =
	\begin{cases}
	& 0 ;  \ {\rm if} \; m < m_{\rm min},  \\
	& [f(m-m_{\rm min},\delta_m)+1]^{-1};  \\
	&\ {\rm if} \;  m_{\rm min} \leq m \leq m_{\rm min}+\delta_m  ,  \\ 
	& 1 ; \ {\rm if} \; m > m_{\rm min}+\delta_m , \label{eq:smooth}
	\end{cases}
	\end{equation} 
	with 
	\begin{equation}
	f(m, \delta_m) = \exp \left(\frac{\delta_m}{m} + \frac{\delta_m}{m-\delta_m} \right)
	\end{equation}
	where, $\delta_m$ regulates the sharpness of the smoothing function.
	The maximum likelihood values for the fit to GWTC-3~\cite{KAGRA:2021duu} are $\lambda_{\rm peak} = 0.02$, $\alpha_m = -3.5$, $m_{\rm min} = 4.8 M_{\odot}$, $m_{\rm max} = 83 M_{\odot}$, $\mu_m = 34 M_{\odot}$, $\sigma = 1.9 M_{\odot}$ and $\delta_m = 5.4 M_{\odot}$.
	The mass-ratio distribution is modeled using a power law with a smoothing function, and it takes the form
	\begin{equation}
	p(q | \beta_q, m_1, m_{\rm min}, \delta_m) \propto q^{\beta_q} S(qm_1 | m_{\rm min}, \delta_m)
	\end{equation}
	where the maximum likelihood value for the spectral index is $\beta_q = 0.76$. We show the component mass distributions in \cref{fig:m1_m2}.
	
	The analyses performed on GWTC-3~\cite{KAGRA:2021duu} suggest a broad distribution for the spin magnitude, peaking around $0.2$ and falling off to 0 for large spins. However, we are particularly interested in estimating systematic biases for large-spin systems, so we draw the magnitude uniformly between 0 and 1. The analyses on systematic effects are performed on this population; in particular, we compute the measurement uncertainties and systematic biases within the LSA for these parameters, but when performing the hierarchical Bayesian analysis, we generate the mock catalog by performing importance sampling on this flat population to obtain a spin distribution in agreement with the results of the \spin model of~\textcite{KAGRA:2021duu}. Finally, we assume the spins' orientation to be distributed isotropically, which is in qualitative agreement with the results on GWTC-3.
	
	The location and orientation of the binary in the plane of the sky is assumed to be randomly distributed. Therefore, the declination angle, $\delta$, right ascension, $\alpha$, and inclination angle, $\iota$, follow the distributions $\cos\delta \in \mathcal{U}[-1, 1]$, $\alpha \in \mathcal{U}[0, 2\pi]$, and $\cos\theta_{\rm JN} \in \mathcal{U}[-1, 1]$, respectively. The polarization angle, $\psi$, and the coalescence phase, $\phi_c$, are also drawn from a uniform distribution, $\psi, \phi_c \in \mathcal{U}[0, 2\pi]$. 
	
\begin{figure}
		\includegraphics[width=\columnwidth]{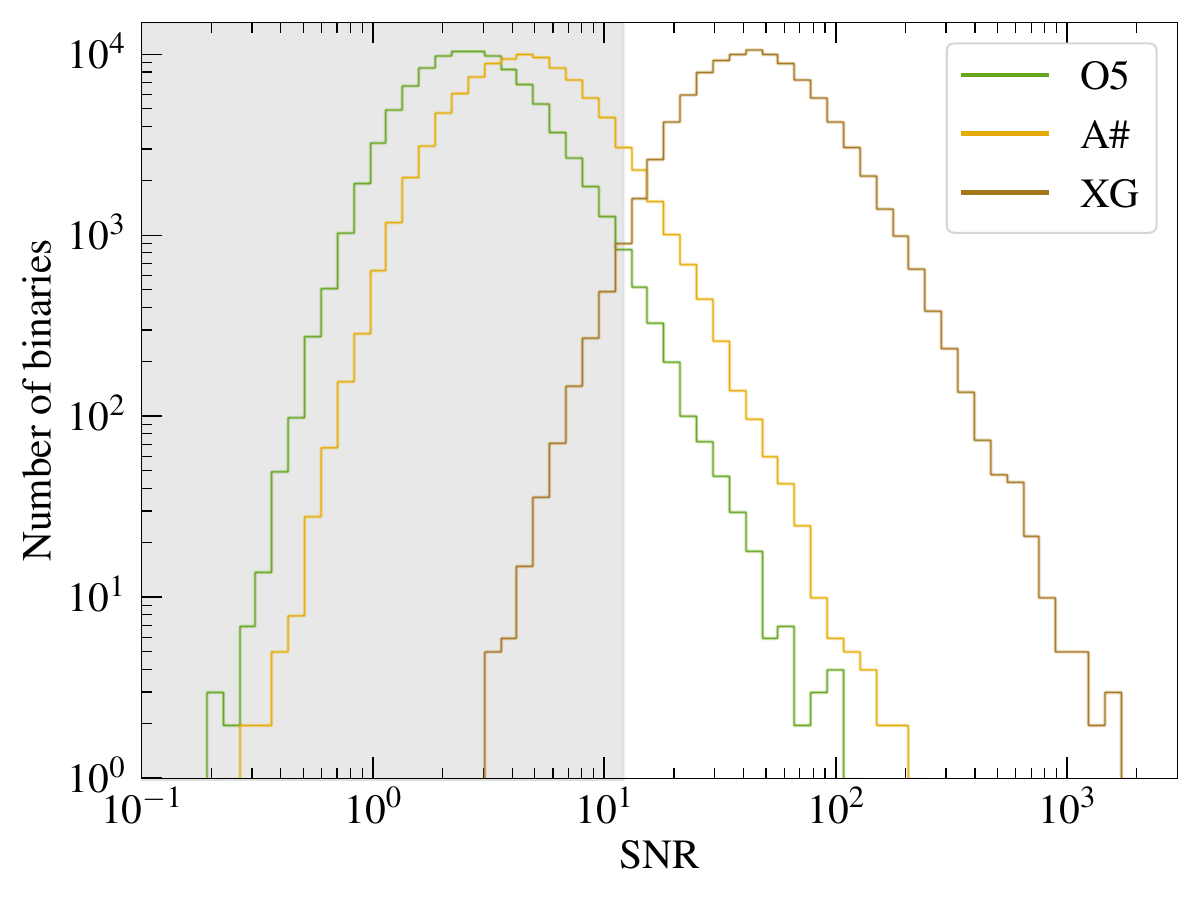}
		\caption{Distribution of the \acp{SNR} of the $10^5$ BBHs simulated in the three detector networks computed using the \xphm model. The \ac{SNR} distribution is similar when using the \eob model. In shaded gray, we indicate the region with network-\ac{SNR} threshold below 12.}
		\label{fig:snr_distribution}
	\end{figure}

	We report the \ac{SNR} distribution of the $10^5$ binaries simulated in the three detector networks in Fig.~\ref{fig:snr_distribution}. We shade the region below the \ac{SNR} threshold of 12 in gray. The tails of the three distributions exhibit the $\propto\rho^{-4}$ dependence of the rate of mergers per unit redshift in accordance with the uniform in comoving-volume distribution of sources in the nearby universe. On the other hand, the peak of the distribution correlates with the peak of the \ac{SFR}, while the initial slope depicts the first generation of stars following the dark ages. While we are limited by the sensitivity of the current detector networks and their upgrades in our ability to observe \acp{GW} from the mergers of the first stellar-origin \acp{BBH}, the \emph{XG} network will enable us to study \ac{BBH} mergers immediately following reionization.

	\begin{figure*}
		\includegraphics[width=2\columnwidth]{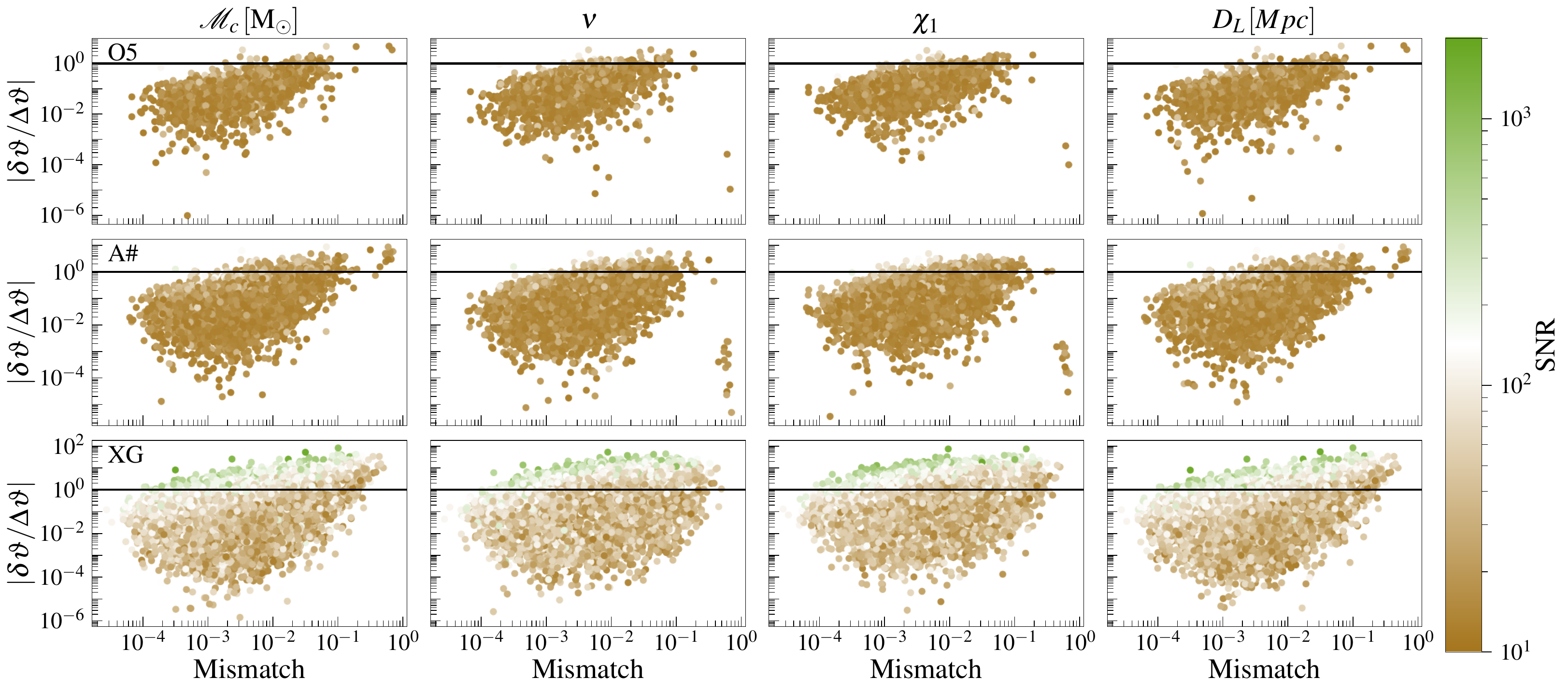}
		\caption{Ratio between the systematic bias and statistical errors, $\delta\vartheta/\Delta\vartheta$, for source-frame chirp mass, symmetric mass ratio, primary spin magnitude, and luminosity distances as a function of the mismatch between \eob and \xphm waveform models for a population of \ac{BBH} mergers as observed by the \ac{LVK}. A network \ac{SNR} threshold of 12 was imposed on the $10^5$ binaries in the population resulting in $\sim1800$, $\sim8100$, $\sim99000$ detected events in the \emph{O5} (top), \emph{A\#} (middle), and \emph{XG} (bottom) networks, respectively. The \acp{SNR} of the binaries are depicted using the different color scales. The outlier events in the top and middle panels having the largest mismatches are the heaviest and most distant events with redshifted total mass $>400M_{\odot}$ and redshift $>2$.}
		\label{fig:ratio}
	\end{figure*}
	
	\subsection{Systematic bias}
	\label{subsec:systematic_bias}
	In the following, we first discuss the systematic biases for the individual events in the \ac{LVK}-like population.  Then, we carry out a hierarchical Bayesian inference of the population distributions by reweighting the parameter distributions to the \ac{LVK} distribution. Finally, we determine the type of binaries more likely to be biased, which motivates our analysis in Sec.~\ref{sec:parameterspace}.

	\begin{figure*}
		\includegraphics[width=2\columnwidth]{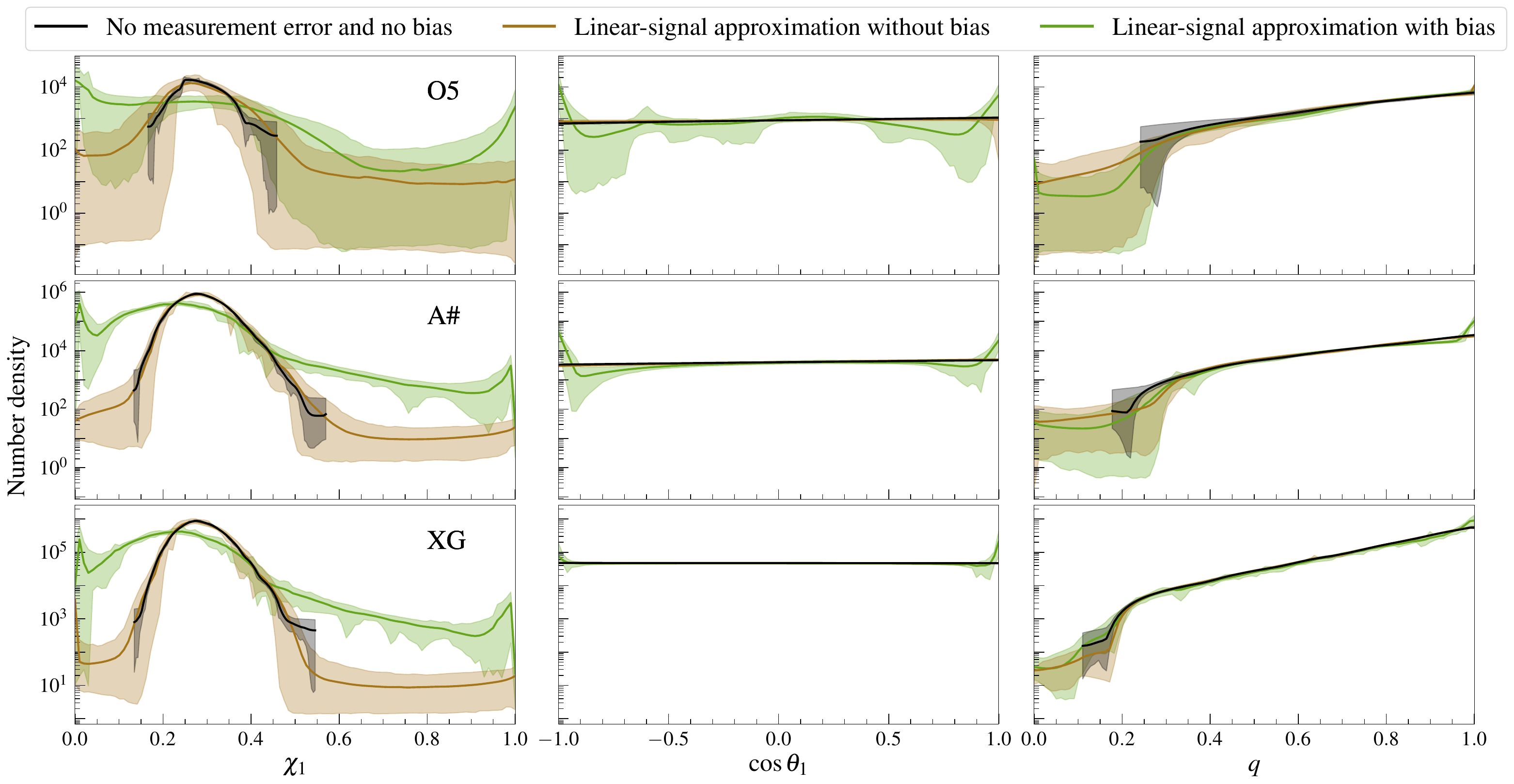}
		\caption{Inferred distribution of $\chi_1$, $\cos\theta_1$, and $q$ for different detector networks, in the case where the parameters are measured perfectly (black), where there is measurement error but no bias (brown), and where there is measurement error and bias (green). The solid lines show the mean value in a given bin and the shaded areas the $90\%$ confidence intervals. The ``no measurement error and no bias'' case does not correspond to any realistic scenario and is there to show the underlying distribution of observed events, accounting for Poisson errors. For $\chi_1$, when including biases, we observe a shift towards smaller values together with a broadening of the distribution, particularly for \emph{A\#} and \emph{XG}. For $\cos\theta_1$, we observe an excess of events at both ends and for $q$ at $\sim 1$. The distributions in the \emph{XG} case for those two parameters are less biased because the bias for individual events is smaller in \emph{XG}.}
		\label{fig:observed_pop}
	\end{figure*}

	\subsubsection*{Variation with mismatch}
	
	Armed with the biases in the \ac{GW} parameters and their measurement errors, in Fig.~\ref{fig:ratio} we report their ratio $|\delta\vartheta/\Delta\vartheta|$ as a function of the mismatch, $\mathcal{M}$, between \eob and \xphm for the chirp mass, symmetric mass ratio, primary spin magnitude, and luminosity distance.	
We recall that the biases are computed using Eq.~\eqref{eq:mod_bias}.
	The \ac{SNR} for the binaries of the population are portrayed using a color scale with lighter colors representing smaller \ac{SNR} and vice versa. A value of $|\delta\vartheta/\Delta\vartheta|>1$ indicates that systematic biases are larger 
 than the typical size of statistical errors. 
 A common feature for all the parameters is a direct correlation between $|\delta\vartheta/\Delta\vartheta|$ and the mismatch. This feature is intuitive because a larger mismatch implies a greater difference between the two waveform models and, therefore, larger biases assuming the measurement errors do not vary significantly with 
 the mismatch which we find to be broadly true for the population. On the other hand, the color scale shows that the loudness of a signal is not a guarantee for a dominant systematic effect with quieter signals exhibiting significant systematic biases particularly when the mismatch is greater, which is especially true for the \emph{O5} and \emph{A$\#$} networks. 
	We provide a complementary plot of  $|\delta\vartheta/\Delta\vartheta|$ as a function of the \ac{SNR} 
	in Fig.~\ref{fig:ratio_snr} in Appendix D for the interested reader.
	
	In this section and in other places where measurements of the spin using \ac{LSA} are discussed, we examine the primary spin magnitude instead of the popular variables $\chi_{\rm eff}$ and $\chi_{\rm p}$. This choice is made for the practical reason that the Fisher matrix is evaluated in $\chi_1$. It is not trivial to transform the likelihood to other variables since it can extend beyond the prior bounds for a parameter. In fact, imposing priors, particularly in the presence of large biases, is also nontrivial since the true posterior distribution need not simply be a truncated likelihood distribution (in the case of a uniform prior with boundaries) peaking outside the prior boundary but rather peak around the subdominant maximum of the likelihood that is within the bounds. However, we do not have any information about such features since we are working under the \ac{LSA}. Hence, in order to avoid such difficulties and keep the analysis simple, we ignore the effects of prior for individual binaries. Furthermore, since $\chi_1$ is better measured than $\chi_2$ in a majority of the cases, it is reasonable to consider that its distribution is closer to a Normal distribution and, hence, better described under the \ac{LSA}.
		
	A few events appear as outliers in the figure with large mismatches but extremely small $|\delta\vartheta/\Delta\vartheta|$ for the \emph{O5} and \emph{A$\#$} networks. These events are the heaviest and most distant, with redshifted total masses $>400M_{\odot}$ and redshifts $>2$. As such, these signals are extremely short, consisting of only the merger ringdown. The \ac{FIM} in these cases is close to singular because the signal does not contain much information, resulting in extremely
    large errors. This result suppresses the ratio for every parameter except the luminosity distance and the chirp mass. For the former, this is because the ratio is directly proportional to the mismatch and, hence, large; while for the latter, the effect of the luminosity distance dominates in the conversion from detector-frame chirp mass to source-frame chirp mass. As a result, even though the events appear as outliers for the detector-frame chirp mass, they follow the behavior of the luminosity distance for the source-frame parameter. The 
	\ac{LSA} approximation is not reliable for such cases, and we should resort to full Bayesian analyses. Nevertheless, we include these binaries 
	in the figure to show 
	their existence in the population observable by these two networks. 
	
	Figure~\ref{fig:ratio} also shows that the number of binaries with biased parameters as a fraction of the detected population increases with improved detector sensitivity. This finding can be simply understood as due to the independence of the overall scale of the \ac{PSD} in estimating the parameter bias (see Eq.~\eqref{eq:bias}) while the covariance is inversely proportional to it. While an improving detector sensitivity leads to an increase in the total number of binaries that are
    significantly biased, the increase in the biased fraction has to do with the finite number of stellar origin \ac{BBH} mergers in the universe. Note that because of different \ac{PSD} shapes, interferometer designs, and minimum frequencies, the three rows of Fig.~\ref{fig:ratio} are not simply shifted versions of one another. Nevertheless, even for the \emph{XG} network, only a minority of events are biased, with the biased fraction ranging from $10\%-25\%$ depending on the parameter. For
    detectors of the current generation, the biased fraction is even smaller with only $\sim2\%$ and $\sim2.5\%$ of binaries significantly biased for the \emph{O5} and \emph{A$\#$} networks, respectively. This finding suggests that biases in the parameter estimation will only be of importance for extraordinary individual events rather than for inferring general characteristics of the population. We check this case more carefully in the next section.
	
	It is also important to realize that a larger value of $|\delta\vartheta/\Delta\vartheta|$ does not necessarily mean a larger value of the absolute bias. For instance, as we will see in Sec.~\ref{sec:science_case}, the \emph{XG} network quite often has smaller absolute biases due to improved low-frequency sensitivity, where waveform models agree to a greater extent. However, the improved sensitivity reduces the measurement error more than the decrease in the systematic bias resulting in a
    larger value of $|\delta\vartheta/\Delta\vartheta|$. We will see the effect of this result in the next section.

	\subsubsection*{Inferred distributions}
		
Figure~\ref{fig:observed_pop} shows the inferred number density of events for $\chi_1$, $\cos\theta_1$, and $q$ for the three detector networks. Solid lines indicate the mean of the number density, and colored bands are the $90\%$ confidence intervals. The results shown in black correspond to an unphysical scenario where all observed events are perfectly measured. In this case, the only source of uncertainty is the Poisson error due to the finite number of events. The black curves serve as
guidance to indicate the underlying distribution on which we perform the inference. In brown, we show the case with measurement error and no bias, following the procedure outlined in Sec.~\ref{sec:hba}. The brown and black shaded areas overlap, with the red encompassing the blue most of the time due to the inclusion of measurement errors, indicating that our procedure yields unbiased results. Thus, differences between the cases without (in brown) and with (in green) biases are due to waveform systematics. 
	
\begin{figure}
		\includegraphics[width=\columnwidth]{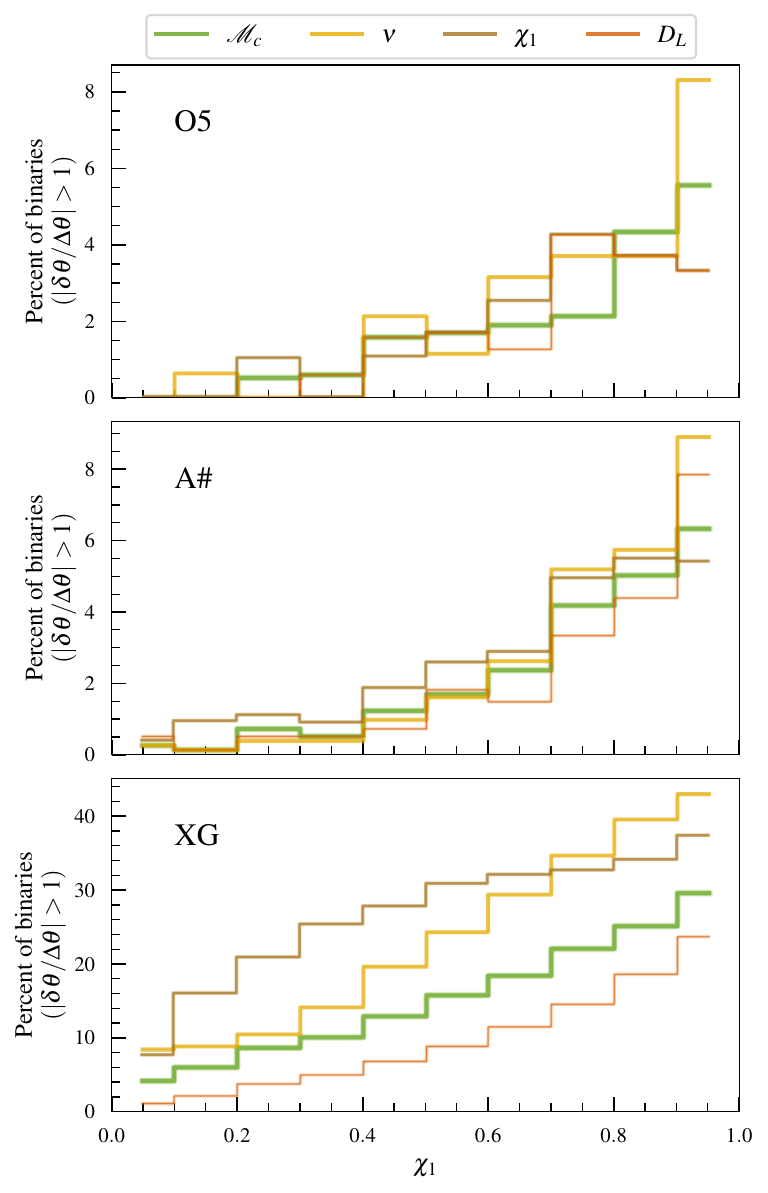}
		\caption{Percentage of binaries in each $\chi_1$ bin with $|\delta\vartheta/\Delta\vartheta|>1$ for different parameters. The top, middle, and bottom panels show the percentages for the \emph{O5}, \emph{A\#}, and \emph{XG} networks, respectively.}
		\label{fig:cdfratio_a1}
	\end{figure}

	We recall that, when performing the hierarchical Bayesian analysis, we resample the flat $\chi_1$ distribution into the distribution inferred by the \ac{LVK} Collaboration~\cite{KAGRA:2021duu} by means of importance sampling. For O5, the biased and nonbiased distributions are mostly compatible. However, for \emph{A\#} and \emph{XG}, we observe that, when including bias, the inferred $\chi_1$ distribution is broadened, with the peak being shifted to lower values. The broadening is a
    consequence of the occurrence of large biases, while the shift happens because the systematic bias typically increases with $\chi_1$ (see Fig.~\ref{fig:cdfratio_a1} and the associated discussion). Events with large spin are more shifted, with a small preference for shifts towards lower $\chi_1$, than events with small spins. From the astrophysical point of view, the shift of the peak is rather negligible, but the tail of high spin events would lead to an overestimation of the number of
    \acp{BH} with high spins, by up to 2 orders of magnitude, potentially challenging formation scenarios.   
	
	For the tilt angle, we observe an excess at the ends due to events with more precession ($\cos\theta_1\sim 0$) being more biased than those with aligned spins. We note that the \emph{XG} population is less biased than the \emph{A\#} one. This feature is a consequence of the improvement at detectors at low frequencies, which increases the proportion of inspiral signal that is observed, where waveform models agree best, and yielding a less-biased estimate of precession effects. We observe a
    similar behavior for the $q$ distribution. Asymmetric events are more biased than nearly equal-mass ones, shifting the overall distribution to $q\sim 1$. As for $\cos\theta_1$, the \emph{XG} population is less biased than the \emph{A\#} one. At first glance, this finding might seem in contradiction with Fig.~\ref{fig:ratio}, which shows that the ratio between the systematic bias and statistical error on $\nu$ tends to be larger in the \emph{XG} case. However, as further discussed in
    Sec.~\ref{sec:science_case}, when comparing the full posteriors, in many cases the result in the \emph{XG} case is closer to the true value than in the \emph{A\#} case. The ratio between the systematic bias and statistical error is larger for \emph{XG} because the measurement error decreases more than the bias (in relative terms); however both errors decrease, and 
	the fact that the absolute bias is smaller 
	ends up reducing the bias at the level of the population inference.  
	We also perform hierarchical Bayesian analysis on $\mathcal{M}_c$ and find no bias at the level of the population, as expected given that this parameter is typically less biased. 
	
	The narrowing of the error band for next
		generation detectors is more important for $\cos \theta_1$ and $q$ than for $\chi_1$, because, in the astrophysical model we use, the spin magnitude distribution is more concentrated in a narrow region than the distribution of the inclination angle and mass distribution. As a rule of thumb, the error in a bin goes as $\sqrt{N_b}$, where $N_b$ is the number of events in the bin. The expectation value of $N_b$ can be related to the total number of events $N$ as $<N_b>=N p_b$, with $p_b$ being the probability of the bin. Therefore, in the bins with low probability, the number of events needed for the error in the bin to go below a threshold is larger, explaining why the error on the spin magnitude distribution remains large in the tails of the distribution. Moreover, for the inclination angle distribution, the improvement is also driven by the reduction in parameter estimation errors. 
	
	Finally, let us stress again that those results were obtained using the LSA for the measurement error and the systematic bias. As explained in Sec.~\ref{sec:hba}, we allow the biased estimate of the parameters to be outside of the physical range, with the idea that this would mimic the likelihood behavior: It is reasonable that the likelihood of an event $\chi_1 \sim 0$ seems to peak at $\chi_1=-0.1$, and when performing parameter estimation, we would observe a truncated distribution due
    to the physical prior. However, in some cases, our formula predicts biases that are orders of magnitude outside of the physical range (e.g., $\chi_1\sim -10$), most likely indicating that the LSA should not be trusted. Indeed, the LSA relies on the quadratic approximation to the likelihood, which should hold only in a region around the peak of the likelihood, with a better agreement at high SNRs. Thus, the reason for our estimates of the bias with \emph{O5} (\emph{A\#}) being so much larger
    than with \emph{XG} might also be due to the invalidity of the LSA in some cases. However, observing more of the inspiral certainly contributes, as discussed in more detail in Sec.~\ref{sec:science_case}. Overall, we expect the results shown here for \emph{XG} to be the most reliable.

	\subsubsection*{Which binaries are likely biased?}

	Having explored the effect of waveform systematics on the full detected LVK-like population and studied the inferred population properties via hierarchical Bayesian inference, we turn our attention to the subset of binaries with significant parameter biases. In the following, we identify the properties of binaries that have a greater susceptibility to systematic biases. To accomplish this goal, we explore the dependence of the systematic bias as a direct function of the binary parameters. 
	In Fig.~\ref{fig:cdfratio_a1}, we show the percentage of binaries in each $\chi_1$ bin with $|\delta\vartheta/\Delta\vartheta|>1$ for various parameters, such as the chirp mass, symmetric mass ratio, primary spin magnitude, and luminosity distance. It is immediately clear that the number of binaries with biased parameters increases with increasing $\chi_1$. Notice that for current detectors and their upgrades, only a tiny fraction of binaries ($\lesssim1\%$) have biased parameters when
    $\chi_1<0.4$. Even for highly spinning binaries, the biased percent is less than 10\%. In contrast, we observe that a relatively large fraction ($10\%-25\%$) of the binaries have biased parameters in the \emph{XG} network even when $\chi<0.4$.
	
Before ending this section, we remark that the results for the systematic biases of the LVK-like population have been obtained by comparing 
two state-of-the-art quasi-circular, spin-precessing, multipolar waveform models. However, when assessing the accuracy 
of the waveform models, more robust and definitive results can be achieved when comparing models to NR waveforms. We plan to carry out such a 
study in the near future, although it will be limited by the number of NR waveforms and their length.

	\section{Systematic biases across binary parameter space}
	\label{sec:parameterspace}
	In Sec.~\ref{sec:LVK_bias}, the general properties of systematic bias across an LVK-like \ac{BBH} population were explored. 
	Here, we consider an agnostic BBH population in order to explore a wider region of the binary parameter space and identify the regions with greater susceptibility to systematic biases. 	
	
	\begin{figure*}
		\includegraphics[width=2\columnwidth]{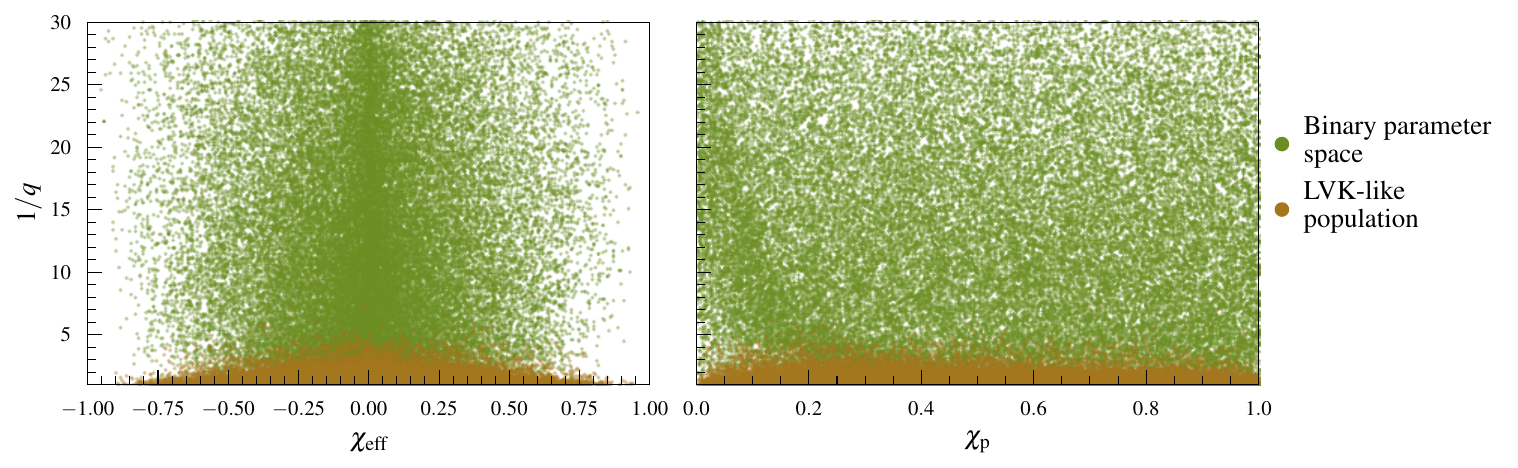}
		\caption{Comparison of the LVK-like population and distribution of exploratory binaries.}
		\label{fig:lvk_agnostic}
	\end{figure*}
	
	We sample uniformly in the total redshifted mass, $M^z \in \mathcal{U}(10, 200) [M_{\odot}]$, and inverse mass ratio, $1/q \in \mathcal{U}(1, 30)$. However, we impose a constraint on the mass of the lighter object, 
	$m_2 \geq 5 M_{\odot}$, and only select those binaries that satisfy this constraint. This approach results in a nonuniform distribution in the two masses.
	Regardless, in this section, our interest is not in any particular distribution of parameters but rather in the coverage of the parameter space.
	Exploring the region of large inverse mass ratio is interesting since the waveform models considered here are expected to differ more in this part of the parameter space due to differences in the models' calibration. For \eob, SXS NR simulations for aligned-spin systems at $1/q\geq 15$ and a nonspinning simulation at $1/q\geq 30$ were included in the calibration of the model, and second-order gravitational self-force information was incorporated, improving the reliability
    of the model at large $1/q$ \cite{vandeMeent:2023ols}. An example of the different behavior of the models for large $1/q$ is shown in Fig. 21 of \textcite{Pompili:2023tna}, which illustrates the differences in parameter recovery for a large inverse mass-ratio NR simulation as shown between the aligned-spin versions of \eob and \xphm, with \eob being more reliable in the recovery of the parameters.
	
	Because of the lack of calibration to precessing NR waveforms in both waveform models, we expect waveform models to have greater differences for large $\chi_{\rm p}$ values as evident from the mismatch plot of Fig.~10 in \textcite{Ramos-Buades:2023ehm}. Therefore, it is of interest to understand the behavior of systematic biases in these parts of parameter space. 
	Thus, we create a sample of binaries having a uniform distribution in $\chi_{\rm p}$ (the LVK-like population, in turn, disfavors large values of $\chi_{\rm p}$). For this purpose, we generate a large set of samples for the spin magnitudes and tilt angles from the precessing prior and retain a subset of these samples such that the resulting distribution in $\chi_{\rm p}$ is uniform. This selection procedure has a negligible effect on the distribution of spin magnitude and tilt of the secondary companion while giving greater weight to large and in-plane spin for the primary companion. We draw 50,000 binaries using this procedure to cover the binary parameter space. A comparison in the $1/q-\chi_{\rm eff}$ and $1/q-\chi_{\rm p}$ planes between the \ac{LVK}-like and the agnostic population is shown in Fig.~\ref{fig:lvk_agnostic}. 
		
	The distributions of all other parameters are the same as for the \ac{LVK}-like population, except for the distance, which is kept fixed at $D_L=235 \,\rm Mpc$ while computing the measurement errors and biases. However, errors have a simple scaling with the distance for given redshifted masses while the biases remain unaffected, and we use this scaling to obtain results at other distances. 
	
	\subsubsection*{Bias horizon}
	\begin{figure*}
		\includegraphics[width=\columnwidth]{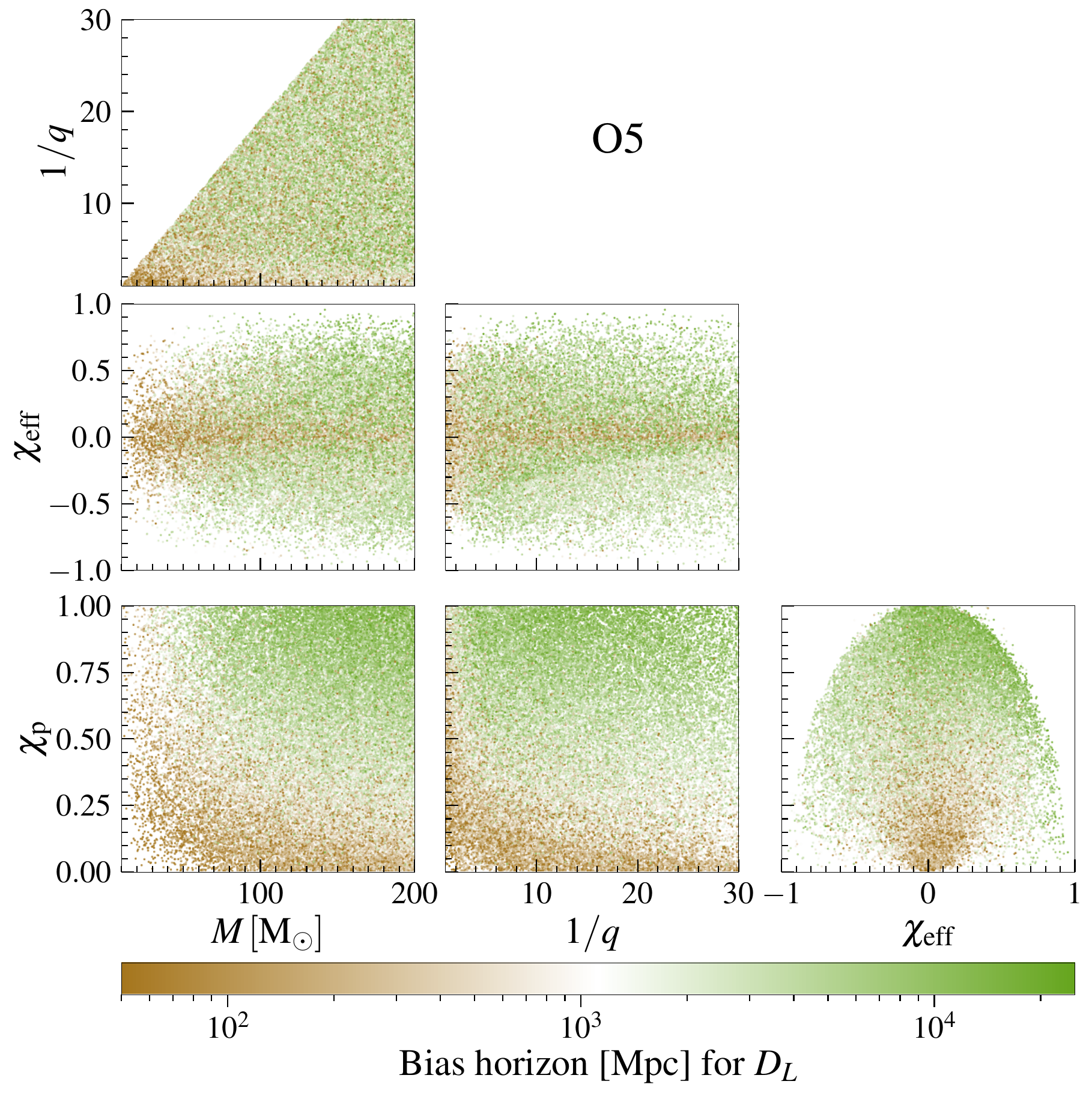}
		\includegraphics[width=\columnwidth]{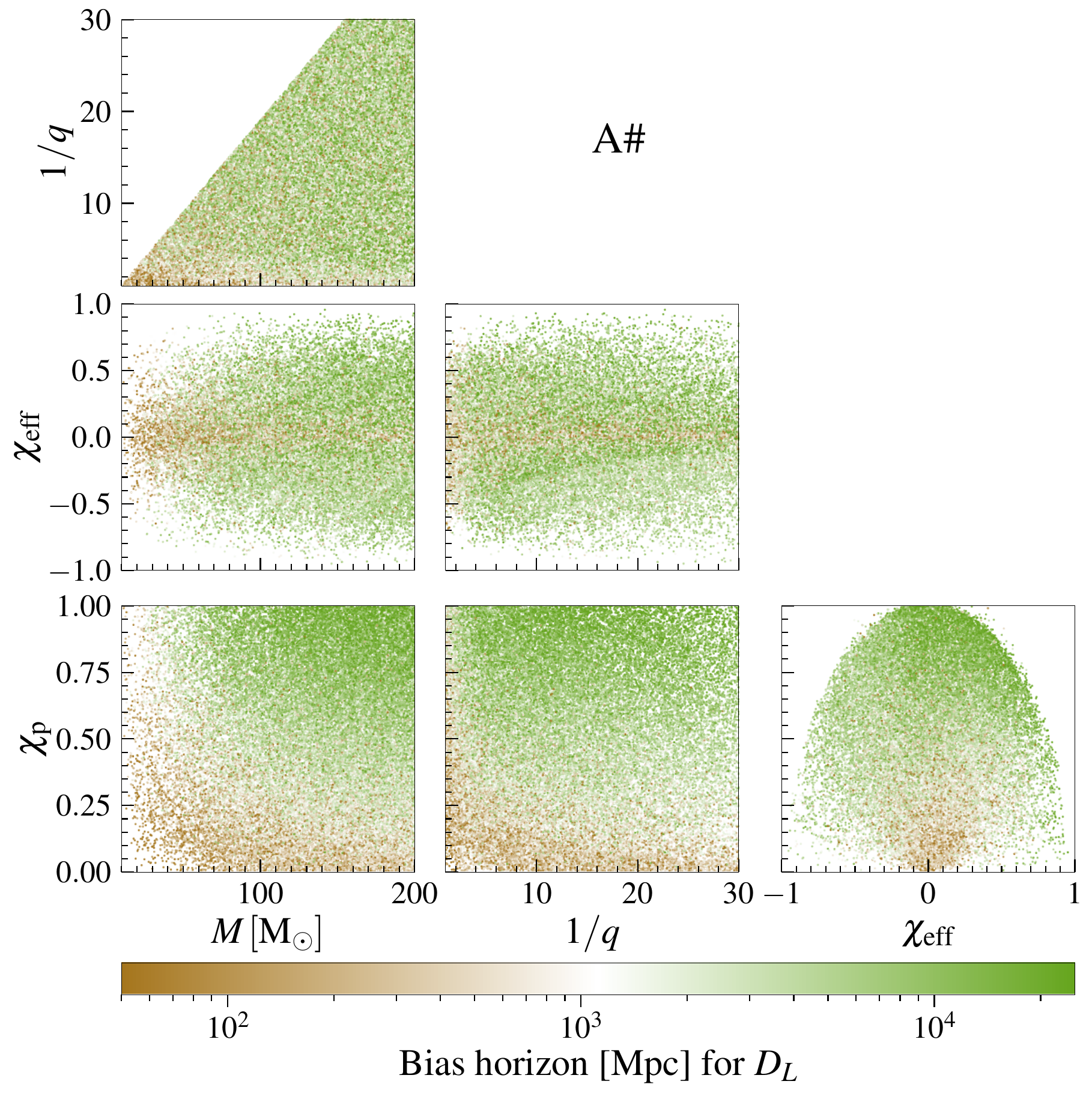}
		\caption{Distribution of the 50,000 binaries in the parameter space represented in Fig.~\ref{fig:lvk_agnostic}, with the color scale showing the distance to which the $D_L$ parameter is biased ($\delta D_L/\Delta D_L \geq 1$)
		for the \emph{O5} (left) and \emph{A\#} (right) networks. Systematic biases become less important if a binary is at a larger distance since measurement precision decreases with distance. Therefore, a large bias horizon signifies that a given parameter ($D_L$ in this case) is measured well enough for systematic biases to be important even at such large distances. Notice that the binaries are biased up to a greater distance in the \emph{A\#} network compared to the \emph{O5} network due
        to its greater sensitivity, and the resultant improvement in measurement precision.}
		\label{fig:bh_DL}
	\end{figure*}
	
	\begin{figure}
		\includegraphics[width=\columnwidth]{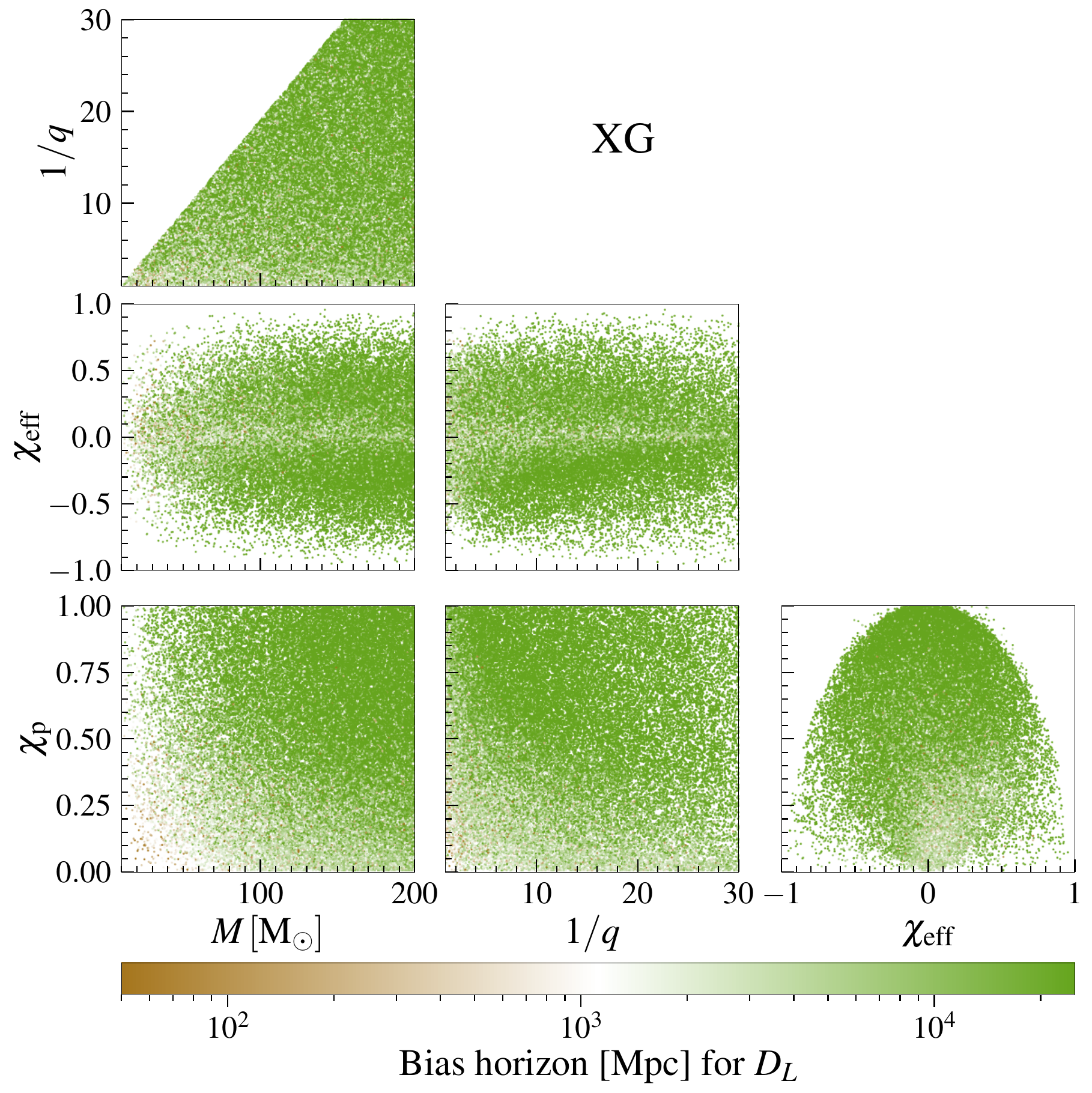}
		\caption{Same as Fig.~\ref{fig:bh_DL} for the \emph{XG} network. BBHs observed with the \emph{XG} network are biased up to a greater distance than those observed with either the \emph{O5} or \emph{A\#} network, due to its greater sensitivity, and the resultant improvement in measurement precision, with a majority of the binaries having a bias horizon $\geq 25 \, \rm Gpc$ ($z \approx 3$) beyond which stellar-origin \acp{BBH} are not expected to exist.}
		\label{fig:bh_DL_XG}
	\end{figure}
	
	We compute the biases and the measurement errors on the parameter set $\bm{\vartheta}$ for the 50,000 binaries considered in this section under the \ac{LSA}, as was done in Sec.~\ref{sec:LVK_bias}. The ratio $\delta\vartheta/\Delta\vartheta$ of systematic errors to statistical errors 
	is a function of the distance to a binary through the dependence of the statistical error $\Delta\vartheta \sim D_L$. We exploit this ratio to calculate the distance at which the ratio $\delta\vartheta/\Delta\vartheta=1$ for any given parameter. Since the measurement errors for a binary with given redshifted parameters increase with its distance, systematic biases will become less important the farther away the binary is located. Hence, the distance at which $\delta\vartheta/\Delta\vartheta=1$ is a measure of the \emph{bias horizon}, i.e., the maximum distance up to which systematic biases dominate statistical errors. Given that the biases and errors for each binary parameter are different, the bias horizons for different \ac{GW} parameters are also different. 
	
	We show the bias horizon for the $D_L$ parameter in Fig.~\ref{fig:bh_DL} for the \emph{O5} (left) and \emph{A\#} (right) networks, while that for the \emph{XG} network is shown in Fig.~\ref{fig:bh_DL_XG}. The distribution of the 50,000 binaries in the 4D space given by $\{M^z, 1/q, \chi_{\rm eff}, \chi_{\rm p}\}$ is illustrated by projecting them into 2D subspaces. 
	The bias horizon for each binary is shown by the color bar. The bias horizon increases with increasing detector sensitivity implying that systematic biases will be prevalent up to greater distances. In particular, the majority of the binaries in the \emph{XG} network have $D_L$ bias horizons exceeding $25 \,\rm Gpc$ ($z \approx 3$), the distance around which the first stars formed. 
	This finding is qualitatively different from the conclusions of Sec.~\ref{subsec:systematic_bias}, in particular, Fig.~\ref{fig:ratio}, where all the binaries have $z\leq3$ but $\sim75\%$ of them are not systematic-error dominated. This result is due to the different distributions of the parameters in this section compared to the \ac{LVK}-like population. Particularly, the majority of the binaries considered in this section are highly asymmetric with large redshifted masses. Moreover, a
    uniform distribution in $\chi_{\rm p}$ leads to a large fraction of highly spinning binaries. From Fig.~\ref{fig:cdfratio_a1}, we gather that systematic biases are more prevalent for such binaries.
	Note that the boundary in the $\chi_{\rm eff} - \chi_{\rm p}$ space is physical, and it is a result of the maximum value of the spin magnitude being 1. When comparing the biases for a given binary for different detector networks, it is important to keep in mind that a larger value of the ratio $\delta\vartheta/\Delta\vartheta$ for a better detector network need not necessarily be due to a bigger $\delta\vartheta$, but rather a much more precise measurement (see Sec.~\ref{subsec:systematic_bias} for further discussion). 
	
	We also observe that the importance of systematic biases depends on the parameter space inhabited by the binaries. For instance, from the $\chi_{\rm eff} - \chi_{\rm p}$ space, it is clear that binaries with small spins have smaller $D_L$ bias horizon compared to binaries with large $\chi_{\rm eff}$ and/or $\chi_{\rm p}$, with the binaries lying on the parameter space boundary having the largest $D_L$ bias horizon. Similarly, one can also observe that, for positive $\chi_{\rm
    eff}$, the bias horizon at large $\chi_{\rm p}$ is higher than for $\chi_{\rm eff} < 0$. This result can be intuited from Fig.~13 of \textcite{Pompili:2023tna}, which shows that the mismatch is larger for positive $\chi_{\rm eff}$ compared to negative $\chi_{\rm eff}$.
	Highly precessing binaries with aligned spins have greater $D_L$ bias horizon compared to similarly highly precessing binaries but with antialigned spins.  
	Along the same lines, we also observe that binaries with large total masses and inverse mass ratios have larger $D_L$ bias horizons. Note that we intentionally do not discuss properties relating to the distribution of binaries in the parameter space since they do not follow any physically motivated parameter distributions. We report the bias horizon for $\chi_1$ in \cref{fig:bh_a1,fig:bh_a1_XG} of Appendix~\ref{sec:exploratory_bias}. They broadly show the same dependence across the parameter
    space, although the quantitative values of the bias horizon are different for each binary parameter.

	\section{Impact of systematics on the science of individual events}
	\label{sec:science_case}
	Until now, we have discussed the effect of systematic biases for the \ac{LVK}-like population in Sec.~\ref{sec:LVK_bias}, and explored systematic biases across parameter space within the \ac{LSA} in Sec.~\ref{sec:parameterspace}. 
	In recent years a number of studies have emphasized the science objectives that can be accomplished using \acp{GW} in the near future~\cite{Evans:2023euw,Branchesi:2023mws,Gupta:2023lga}. In this section, we consider a few of those science applications. We then handpick three binaries (see Table~\ref{tab:params}) with very relevant science potential and discuss the effects of systematic biases on various science objectives. The majority of the results in this section are obtained using a
    full Bayesian analysis except where indicated otherwise.
	
	\subsection{Science objectives}
	In the following, we introduce the science applications that will be considered in this section.
	\begin{enumerate}
		\item \emph{Cosmology}: There exists a tension, at the level of $4.4\sigma$, between the value of the Hubble-Lema\^itre parameter, \hubble, measured at high redshift from the cosmic microwave background~\cite{Planck:2018vyg}, and measured using the local distance ladder comprising Cepheid variables and type-Ia supernova~\cite{Riess:2021jrx}. The emergence of \ac{GW} astronomy provides an avenue for an independent measurement of the Hubble-Lema\^itre parameter, which could importantly
            contribute to resolving this tension. 
Indeed, \acp{GW} have already provided multiple independent measurements, albeit not yet at an accuracy to resolve the tension~\cite{LIGOScientific:2017adf,LIGOScientific:2021aug,LIGOScientific:2018gmd,DES:2019ccw,DES:2020nay}. 
	
	A measurement of \hubble requires both the luminosity distance and redshift of a source to be estimated. \ac{GW} observations provide the former, but additional data or assumptions are required to provide the latter.
	For \ac{GW} observations accompanied with an \ac{EM} counterpart, primarily binaries containing \acp{NS}, the redshift information is provided by spectroscopic/photometric observations of the host galaxy
~\cite{Chen:2017rfc,Gupta:2022fwd}. 
In \ac{BNS} or \ac{NSBH} observations, \ac{NS} tides can be used to provide an independent redshift measurement~\cite{Messenger:2011gi,Messenger:2013fya,DelPozzo:2015bna} and, thereby, infer \hubble~\cite{Chatterjee:2021xrm,Ghosh:2022muc,Dhani:2022ulg,Shiralilou:2022urk}. Alternatively, features in the mass distribution of compact binaries can be exploited~\cite{Chernoff:1993th,Taylor:2012db,Farr:2019twy,Ezquiaga:2022zkx}. Finally, a statistical measurement of the redshift using galaxy catalogs~\cite{1986Natur.323..310S,DelPozzo:2011vcw,Borhanian:2020vyr,Muttoni:2023prw,Gray:2019ksv,Gray:2023wgj} or galaxy cross-correlation techniques~\cite{Oguri:2016dgk,Mukherjee:2020hyn,Ghosh:2023ksl} can be employed. 
	
	Here, the focus is on the last technique, which can be applied to \ac{BBH} systems, but relies on an accurate measurement of the distance and sky position. An indirect effect on cosmological inference due to inaccurate determination of the mass distribution will be discussed in the following subsections. 
	
	Events that have the smallest volumetric uncertainties are the most informative systems for statistical measurements using 
	galaxy catalogs~\cite{LIGOScientific:2021aug,Muttoni:2023prw}. This feature can be intuitively understood as follows: If there is a single galaxy in the localization volume of a \ac{GW} event, and one assumes that the event originated in a galaxy, then there is unit probability that the identified galaxy hosted the event and the redshift is known as well as the galaxy redshift. On the other hand, if the volumetric localization of the \ac{GW} event is poor and there are a large number of
            galaxies that are potential hosts, the redshift distribution would essentially be uniform, obtaining contributions from each possible host. All three events studied here are prime candidates for such a method due to their asymmetrical component masses, which make distance estimates more precise than analogous comparable mass mergers. This feature results from a greater contribution from subdominant harmonics that break the distance-inclination degeneracy. Spin precession also helps in breaking this degeneracy since it mixes different modes in the inertial frame.

	\item \emph{Lower mass gap}: The nature of compact objects with masses between $2 \rm \, M_{\odot}$ and $3 \rm \, M_{\odot}$ can have wide-ranging consequences in fundamental physics---from the physics of nuclear matter to primordial BH formation mechanisms---and astrophysics---from hierarchical formation probabilities to the proportion of rapidly spinning \acp{NS}. The \ac{LVK} Collaboration has observed two events where one of the components of the binary unambiguously lies in this mass range---GW190814~\cite{LIGOScientific:2020zkf} and GW200210\_092254~\cite{LIGOScientific:2021djp}. Tidal effects on \ac{GW} waveforms in highly asymmetric mergers hosting candidate \acp{NS}, such as the \first system, are minimal. Therefore, indirect constraints on the nature of the secondary and the binary's formation history are derived indirectly from their mass and spin measurements. 
	
	The existence of ultraheavy \acp{NS} has consequences for the nuclear \ac{EoS} at a few times the nuclear saturation density~\cite{Tews:2020ylw}, with the possibility of nontrivial structures in the speed-of-sound relation and related phase transition phenomena~\cite{Tan:2020ics,Tan:2021ahl} or rapidly rotating \acp{NS} stabilized against collapse by its rotation~\cite{Most:2020bba,Zhang:2020zsc,Dexheimer:2020rlp,Kruger:2023olj}. On the other hand, \acp{BH} in this mass range will inform the
            primordial \ac{BH} formation scenarios~\cite{Vattis:2020iuz,Clesse:2020ghq,Bianchi:2018ula} and hierarchical mergers in dense environments~\cite{Gerosa:2017kvu,Gupta:2019nwj,Safarzadeh:2019qkk,Gerosa:2021mno,Doctor:2019ruh}. It is also proposed that the secondary gains mass due to accretion either prior to a supernova explosion~\cite{Zevin:2020gma} or following it~\cite{Safarzadeh:2020ntc}. In all of these scenarios, accurate measurements of the mass and spin are essential to further
            the discussion.

	\item \emph{PISN mass gap}: In the mass range of about $50\mbox{--}120\rm\,M_{\odot}$, there is expected to be a dearth of stellar origin \acp{BH} because main sequence stars with masses heavier than $\sim120\rm\,M_{\odot}$ have core temperatures that facilitate electron-positron pair production, leading to a decrease in radiation pressure in the core of the star, causing explosive oxygen burning, and a resultant disruption of the entire star. This process, known as a \ac{PISN} process, does
        not leave behind a remnant, thereby producing a dearth of \acp{BH} above $\sim60\rm\,M_{\odot}$. However, if the mass of the main sequence star is greater than $\sim250\rm\,M_{\odot}$, all the heavy elements undergo photodisintegration, first to alpha particles and then further. This process reduces the radiation pressure, causing the star to implode and form a \ac{BH} with mass greater than $\sim120\rm\,M_{\odot}$. 
	
	The determination of the boundaries of this mass gap can inform the physics of \ac{PISN}, such as the $\prescript{12}{}{C}(\alpha, \gamma)\prescript{16}{}{O}$ reaction rate~\cite{Farmer:2020xne,Mehta:2022pcn}, and the role of stellar rotation~\cite{Mapelli:2019ipt}. Other astrophysical processes---such as mass reversal, mass growth due to accretion, and hierarchical mergers---can result in the formation of \acp{BH} that populate this mass gap.

	\item \emph{Spin Morphology}: The spin distribution of \acp{BH} in binaries provides crucial information on their formation channels. Upcoming observing runs of \ac{LVK} detectors and \ac{XG} observatories will measure the spins of compact binaries to ever greater precision, which will help to constrain the spin distribution of astrophysical \ac{BBH} populations and their formation channels. For instance, binaries formed via isolated evolution tend to have their spins aligned with the orbital angular momentum, while those formed dynamically are likely to have an isotropic spin distribution. Similarly, hierarchical formation is expected to produce larger spins compared to stellar collapse. However, given that the spin measurements are expected to be precise, it is crucial for them to be accurate as well to allow unbiased inference of the source properties of the underlying population. 
	According to our \ac{LVK}-like population, only 50\% of the events detected in \emph{O5} will have $\Delta\chi_1<0.8$ ($\Delta\theta_1<85^{\circ}$), while it is $\Delta\chi_1<0.1$ ($\Delta\theta_1<10^{\circ}$) in \emph{XG}.
	We discuss the indirect effects of the spin measurement on the inference of the nature of the secondary component of the \first system in Sec.~\ref{subsec:lower_mass_gap}. In the following, we discuss the systematic biases on the spin for another binary system.
	
	The origin of massive \acp{BBH}, particularly those filling the upper mass gap, can be traced using their effective spin, $\chi_{\rm eff}$, and spin-precession, $\chi_{\rm p}$, parameters. A hierarchical formation mechanism leads to large component spins since the remnant of the previous merger is expected to be spinning. 
	
	While the $\chi_{\rm eff}$ and $\chi_{\rm p}$ parameters can broadly inform and differentiate between an isolated and dynamical formation channel, an accurate measurement of the tilts of the two spin vectors with respect to the orbital angular momentum, $\theta_1$ and $\theta_2$, and their relative orientations in the orbital plane, $\phi_{12}$, provide detailed knowledge of the formation mechanisms and spin distributions of the \ac{BBH} merger population. Precessing binaries exist in
            different spin morphologies~\cite{Kesden:2014sla,Gerosa:2015tea,Gerosa:2016aus,Phukon:2019gfh} due to spin-orbit resonances~\cite{Schnittman:2004vq}. Therefore, precessing binaries can exist in subpopulations characterized by their spin morphology depending on their tilt angles at formation~\cite{Kesden:2010yp,Kesden:2010ji,Berti:2012zp,Gerosa:2013laa,Gerosa:2018wbw,Steinle:2022rhj}. Recent efforts been made to better understand the spin-precession dynamics~\cite{Gerosa:2023xsx} and
            the ability of current detectors to measure spin-precession effects on the waveform~\cite{Pratten:2020igi}. There have also been efforts to probe whether the \acp{BBH} detected by the \ac{LVK} Collaboration are associated with a particular spin morphology~\cite{Varma:2021xbh}, as well as studies on the capability of current detectors at improved sensitivities to distinguish different spin morphologies~\cite{Kulkarni:2023nes,Johnson-McDaniel:2023oea}.

	\item \emph{Remnant quantites}: The properties of the remnant \ac{BH} following a \ac{BBH} merger can be determined from its binary parameters. However, the nonlinear merger makes a fully analytical calculation intractable. As such, estimates of the final mass ($M_f$) and spin ($\chi_f$) of the remnant include information from \ac{NR}. Several studies have proposed fits for the remnant quantities for nonprecessing binaries~\cite{Healy:2016lce,Keitel:2016krm}. In the precessing case, the
        final mass is found to agree very well using a nonprecessing formula, but the same does not hold true for the final spin where the in-plane spin components are important~\cite{Johnson-McDaniel:2016,Varma:2018aht}. Some simple arguments to include in-plane contributions have also been proposed in the literature~\cite{Rezzolla:2007rz,Barausse:2009uz,Hofmann:2016yih}. Such arguments have been used to augment the nonprecessing final spin estimates~\cite{Healy:2016lce,Keitel:2016krm}. A surrogate model for the final spin using \ac{NR} has also been proposed for moderate mass ratios~\cite{Varma:2018aht}. In this paper, the reported final mass estimates are the average of the nonprecessing fits in Refs.~\cite{Healy:2016lce,Keitel:2016krm}. The final spin is computed by averaging the estimates of Refs.\cite{Healy:2016lce,Keitel:2016krm,Hofmann:2016yih} with in-plane spin augmentations for the nonprecessing fits of Refs.\cite{Healy:2016lce,Keitel:2016krm}. 
	
	Such estimates of the remnant quantities are important not only for modeling the ringdown in inspiral-merger-ringdown waveforms, but also for probing the dynamics of the merger and the nature of the remnant. The ringdown of a Kerr \ac{BH} is described by quasi-normal modes whose complex frequencies are determined from the properties of the final stationary \ac{BH}. Therefore, the ringdown signal can be used to estimate the mass and spin of the final remnant independently. The remnant
            quantities can be affected by deviations from \ac{GR} that can modify the radiated energy and angular momentum as well as an exotic remnant which will have a modified spectrum, resulting in inconsistent estimates from the ringdown signal and \ac{NR}-inspired fits. Such consistency tests have been performed for GW150914~\cite{LIGOScientific:2016lio,Isi:2019aib} and GW190521~\cite{LIGOScientific:2020ufj,Siegel:2023lxl}. With improving detector sensitivities, such tests form an important part of null hypothesis tests of \ac{GR}. 
		
	\item \emph{Maximal \ac{BH} spins}: 
	Obtaining accurate spin measurements from astrophysical \acp{BH} is a non-trivial endeavor for which different electromagnetic approaches exist. In the context of stellar-mass \acp{BH}, it is possible to apply the continuum fitting method or reflection spectroscopy to x-ray observations~\cite{Zhang:1997dy,Brenneman:2006hw} . However, modeling the astrophysical environment is non-trivial and can introduce systematic effects. \ac{GW} measurements provide a novel way to measure \ac{BH} spins,
            and because of their vacuum environment, might be easier and more robust to model. While the spin distribution of \ac{BBH} systems can provide valuable information about their formation channels, an individual event with very high spins, especially if close to extremal Kerr, would be of great scientific interest. For instance, because of the cosmic censorship hypothesis, no naked singularities should exist, which implies that \acp{BH} should not spin above $a > 1$. Moreover, quasi-normal modes of a rapidly rotating \ac{BH} become long-lived, which could lead to turbulence phenomena~\cite{Yang:2014tla}. 

	\end{enumerate}

	\subsection{Handpicked  binary black holes}\label{sec:hbbh}
	
	We pick three asymmetric and precessing systems, out of which one has low total mass and two have large total masses.	These systems have some precedence in the current \ac{LVK} GWTC, though they are not the most common events. We pick them due to their science potential and to explore the parts of the parameter space that are expected to be considerably affected by systematic biases. The parameters of these systems, their \acp{SNR} in the different detector networks, and their science objectives are listed in Table~\ref{tab:params}. We now discuss the properties of these systems and their analogs in the \ac{LVK} GWTC.
	
	\begin{itemize}
		\item{\bf \first: highly asymmetric, spin-precessing, low total-mass binary.}
		This system is modeled after GW190814 and has masses, distance, and inclination compatible with the measured values~\cite{LIGOScientific:2020zkf}. While GW190814 has no measurable spin precession, the system we consider is highly precessing. It is a reasonable choice because a dynamical capture or hierarchical merger is widely accepted as a possible formation channel for GW190814, and this channel can produce highly precessing binaries. The merger rate of such systems is estimated to be $7^{+16}_{-6} \,\rm Gpc^{-3} \,yr^{-1}$. Hence, even a pessimistic merger rate of $1 \,\rm Gpc^{-3} \,yr^{-1}$ equates to more than 400 such mergers within a redshift of 3 each year. 
		With planned and future observatories, a handful of these mergers will be observed with large SNRs. 
		
		\item{\bf \second: rather asymmetric, nonprecessing, high total-mass binary.}
		Among the \ac{LVK} observations, the candidate events GW190403\_051519~\cite{LIGOScientific:2021usb} and GW200208\_222617~\cite{LIGOScientific:2021djp} have properties similar to this system. Both the candidates are asymmetric binaries with large effective spins indicating that the spins are aligned to the orbital angular momentum. The median of the posterior distribution of the primary spin magnitude for GW190403\_051519 is $\chi_1=0.89$ while the probability that the primary spin of
            GW200208\_222617 is $\chi_1>0.8$ is 51\%. The primary components for both candidates also have a considerable probability of being in the upper mass gap. While these events were marginal detections with moderate astrophysical significance, their science potential is immense, and confident detections of similar systems in the future will be invaluable. From a modeling perspective, it is also an interesting region of the parameter space to probe, because different state-of-the-art waveform models have significant mismatches for binaries with large aligned spins.
		This is only partly due to the relatively sparse \ac{NR} coverage in this region of the parameter space. Binaries with spins aligned with the orbital angular momentum merge at higher frequencies~\cite{Campanelli:2006uy}, where the accuracy of the PN approximation --- used as a baseline in all models --- degrades. As a result, this limits the accuracy that models can achieve even after calibrating to \ac{NR}, assuming no fitting errors.
		
		\item{\bf \third: rather asymmetric, spin-precessing, high total-mass binary.}
		This system has the same parameters as \second except for the spins which are now misaligned with respect to the orbital angular momentum. The origin and associated formation channels for such systems through stellar evolution are highly uncertain~\cite{Gerosa:2015hba,Belczynski:2007xg,Zaldarriaga:2017qkw,Belczynski:2017gds,vanSon:2020zbk}. Nonetheless, the current observations provide sufficient evidence for the existence of such systems; therefore, it is essential to understand whether current waveform models have the necessary accuracy to study such mergers. 
		
	\end{itemize}
	
	\begin{table*}
		\centering
		\begin{tabular}{cc c cc}
			\multicolumn{2}{c }{} & \multicolumn{1}{c|}{\makecell{{\bf \first}: \\ highly asymmetric,\\ spin-precessing, \\ low total mass}}  & \multicolumn{1}{c|}{\makecell{{\bf \second}: \\ rather asymmetric, \\ nonprecessing, \\ high total mass}} & \makecell{{\bf \third}:\\ rather asymmetric, \\spin-precessing, \\ high total mass } \\
			\hline
			\hline
			\multirow{4}{*}{Parameters} & $m_1 [M_{\odot}]$ & 23.2 & 61.8 & 61.8 \\
			& $m_2 [M_{\odot}]$ & 2.6 & 9.5 & 9.5 \\
			& $\chi_1$ & 0.7 & 0.9 & 0.9 \\
			& $\chi_2$ & 0.4 & 0.8 & 0.3 \\
			& $\theta_1 [^\circ]$ & 40 & 0 & 140 \\
			& $\theta_2 [^\circ]$ & 40 & 0 & 120 \\
			& $\chi_{\rm eff}$ & 0.51 & 0.89 & -0.43 \\
			& $\chi_{\rm p}$ & 0.45 & 0 & 0.77 \\
			\hline \hline
			\multirow{3}{*}{Network SNR} & \emph{O5} & 75.3 & 222 & 119 \\
			& \emph{A\#} & 137 & 405 & 219 \\
			& \emph{XG} & 1040 & 3150 & 2490 \\
			\hline \hline 
			\multirow{6}{*}{Science cases} & Cosmology & \cmark & \cmark & \cmark  \\
			& Lower mass gap & \cmark & \xmark & \xmark  \\
			& PISN mass gap & \xmark & \cmark & \cmark \\
			& Spin morphology & \cmark & \xmark & \cmark \\
			& Remnant quantities & \cmark & \cmark & \cmark \\
			& Maximally spinning secondary & \xmark & \cmark & \cmark \\
			\hline\hline  
		\end{tabular}
		\caption{Properties of the three systems that we study using Bayesian analysis in Sec.~\ref{sec:science_case}. 
		The top row lists the intrinsic parameters of the systems, the middle row enumerates the \ac{SNR} of the systems in 
		the three detector networks used, and the last row illustrates the science applications associated with each system.
		} \label{tab:params}
	\end{table*}
	
	\begin{figure}
		\centering
		\includegraphics[width=1.0\columnwidth]{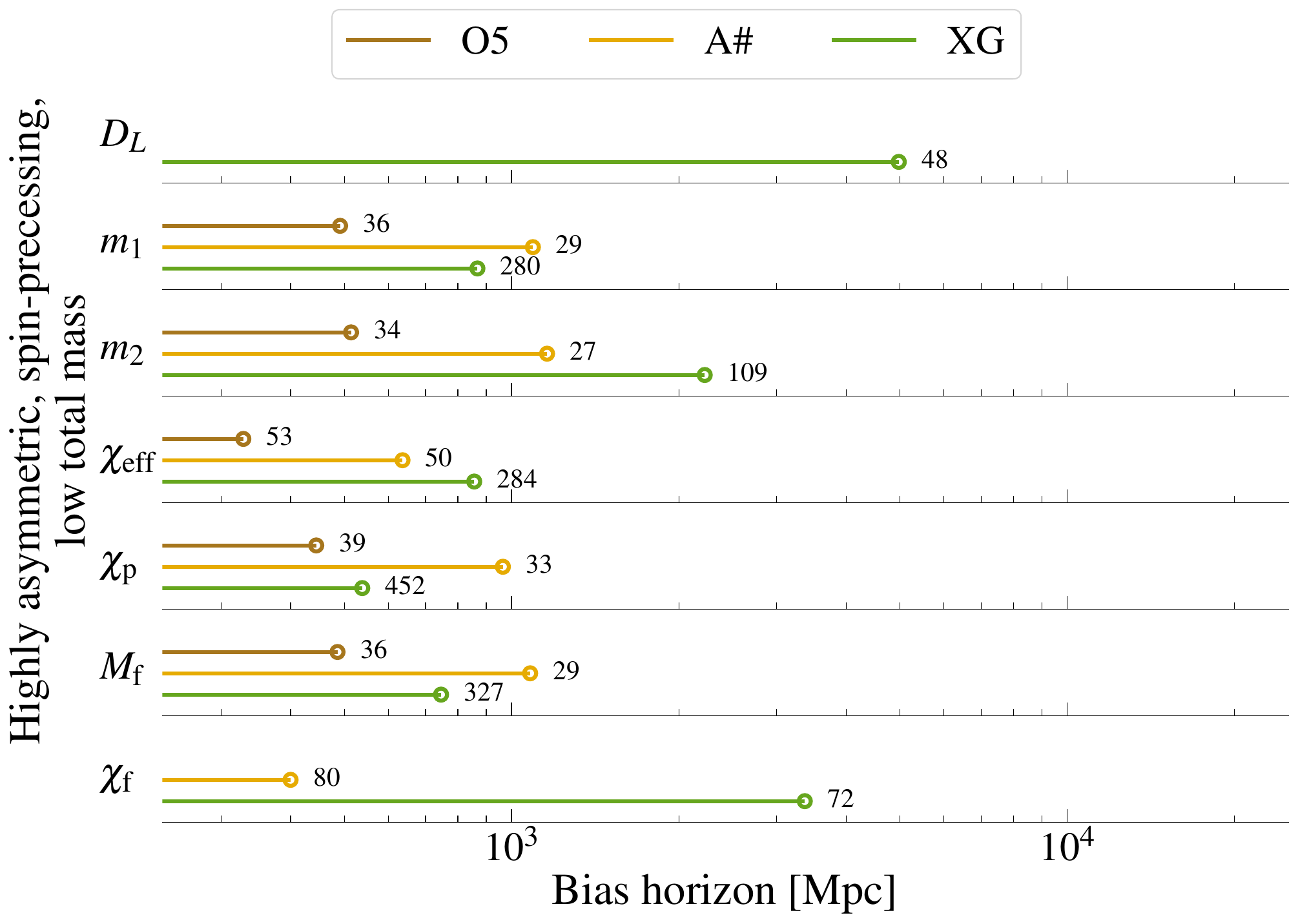}
		\includegraphics[width=1.0\columnwidth]{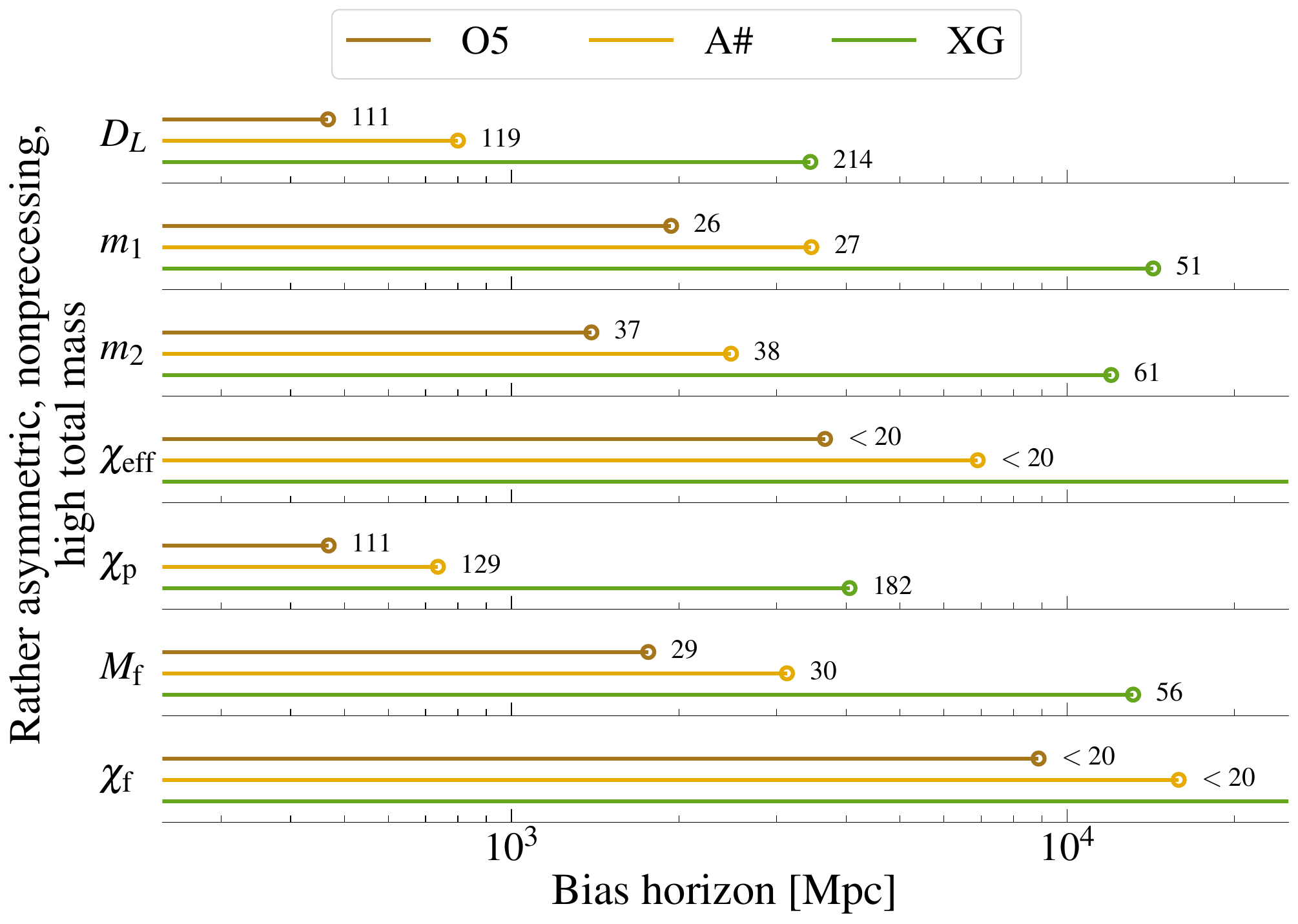}
		\includegraphics[width=1.0\columnwidth]{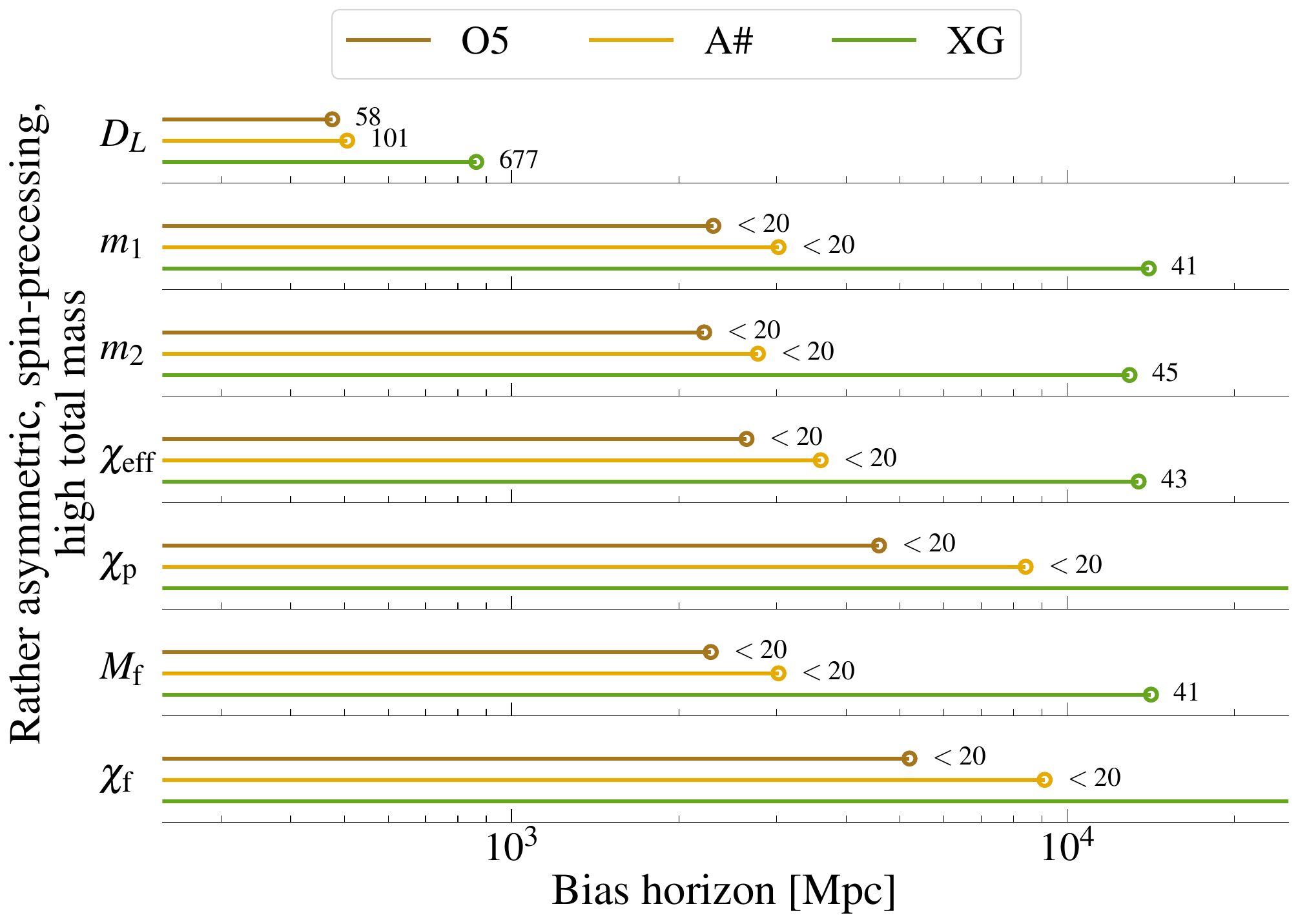}
		\caption{Bias horizon for the most relevant parameters of the three systems (from top to bottom panels) for the three different networks. The colored lines indicate up to which distance the systematic error of a parameter is outside the 90\,\% credible interval of the posterior and thus describes biased parameters. The circles at the end of each line indicate the bias horizon, which is the distance where it is at 90\,\%. The number at the end of each line reports the SNR at that distance. Note that there is a cutoff at 25 Gpc (lines without circle), and that we only explicitly show SNRs down to 20.}
		\label{fig:horizon_bias_lines}
	\end{figure}
	
	For all three events we perform a full Bayesian analysis placing them at a luminosity distance of 235\,Mpc. While \ac{BBH} mergers at this distance are certainly possible and have been observed, they are few in number and the majority of events will originate from larger distances. On the other hand, the \ac{SNR} for events that are sufficiently far away will be small and the systematic biases will be inconsequential compared to the measurement errors. It is therefore desirable to know the
    maximum distance up to which a given science objective will be affected due to biases in parameter estimation. However, it is infeasible to repeat the full parameter estimation calculation for multiple distances. For that reason, we model the posterior distribution using its median and covariance, which is equivalent to the \ac{LSA} approximation in the large \ac{SNR} limit. In this limit, the bias is independent of the strength of the signal while the covariance increases with the square of the distance. 
	We compute the bias horizon for the three sources, and they are shown in Fig.~\ref{fig:horizon_bias_lines}.
	A cursory look at the \acp{SNR} required for a parameter to be biased reveals that biases could be present for much smaller \acp{SNR} than what is simulated here. Below, we will discuss the implications of these results in the context of various science applications. 
	The posteriors for events with low \ac{SNR} are not well approximated by a Normal distribution. 
	The simple scaling argument employed here fails for such cases. Therefore, for parameters for which the \acp{SNR} at the bias horizon is $<20$, we do not quote the exact value but rather denote it with an inequality sign. Similarly, we do not show the exact projected distances if they are larger than $25 \,\rm Gpc$ ($z\sim3$). Note that there is an implicit assumption in this scaling argument that the detector-frame masses are constant. For large distances, the source-frame masses would be materially smaller. Therefore, the binaries considered may no longer be appropriate for science objectives related to the source-frame mass. If, in turn, these binaries are placed at a larger distance, while keeping the source-frame quantities constant, they will have a larger detector-frame total mass and, consequently, fewer \ac{GW} cycles in the detectable band. Since the differences between various waveform models are greater closer to the merger, one expects that the bias will be larger for such systems compared to the scaled binaries. For instance, it can be seen from Figs.~\ref{fig:bh_DL} and ~\ref{fig:bh_DL_XG} that the bias horizon is generally greater for larger total detector-frame masses. In this sense, the reported scaled numbers can be considered conservative estimates.  	
	
	\begin{figure}
		\centering
		\includegraphics[width=0.95\columnwidth]{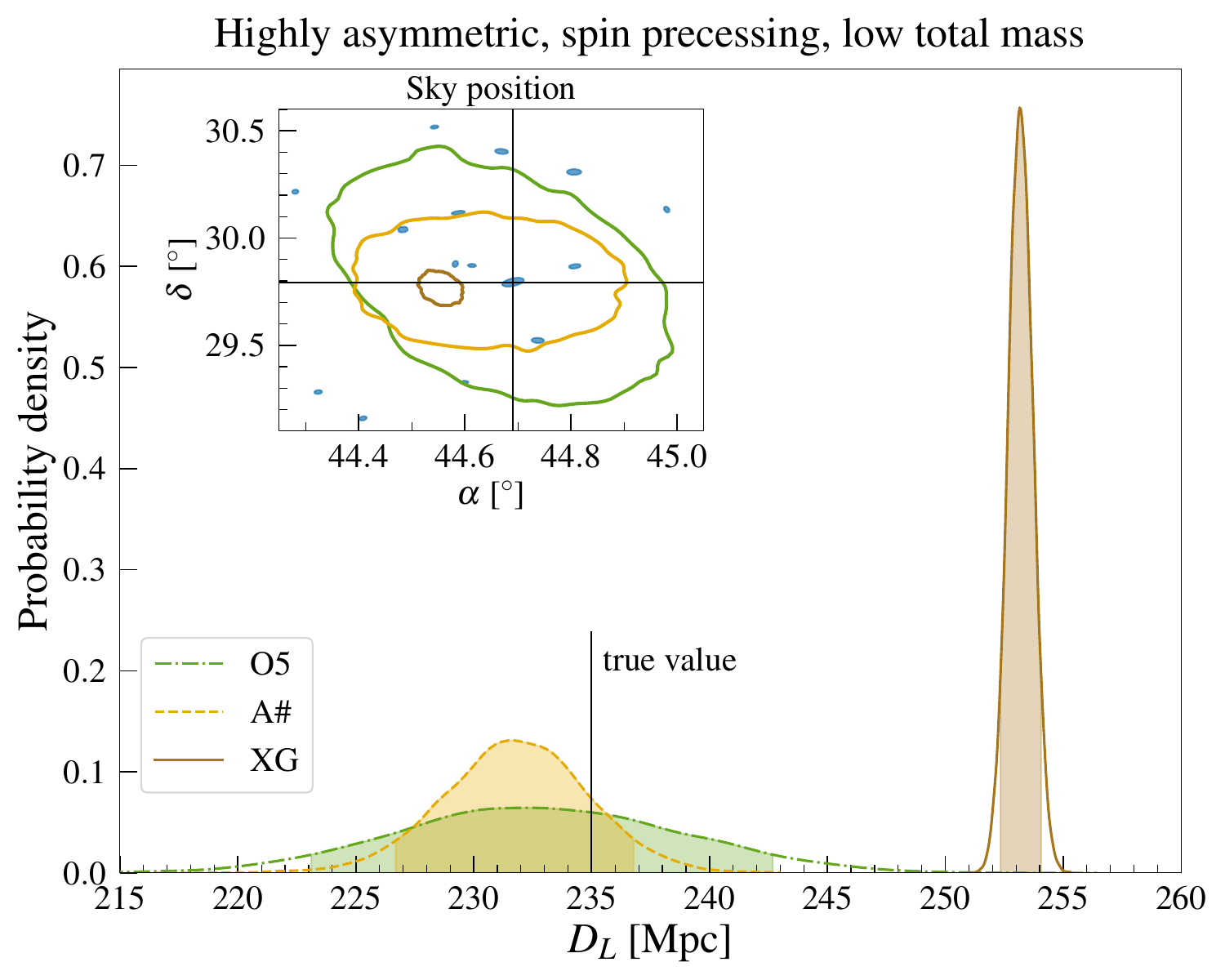}
		\includegraphics[width=0.95\columnwidth]{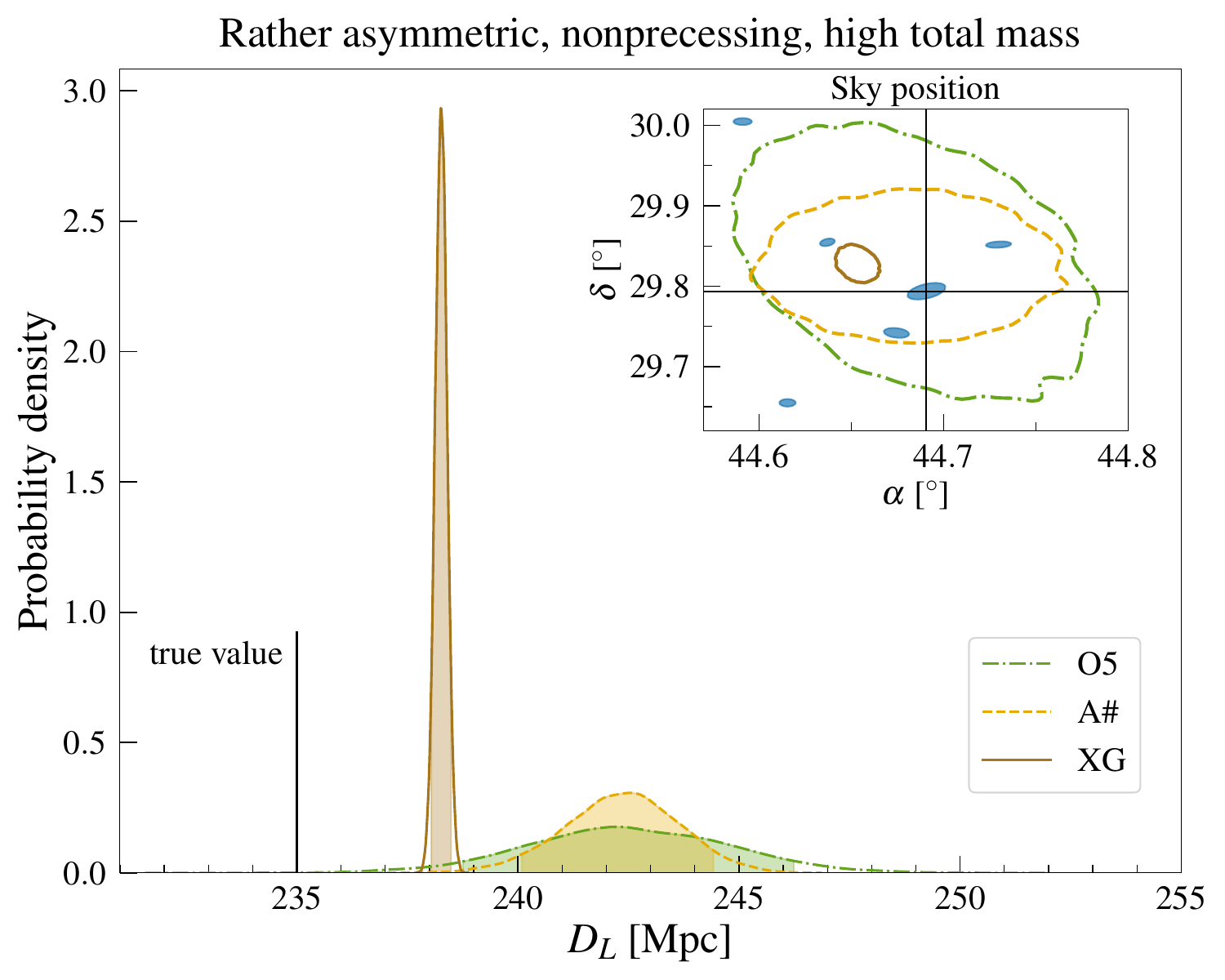}
		\includegraphics[width=0.95\columnwidth]{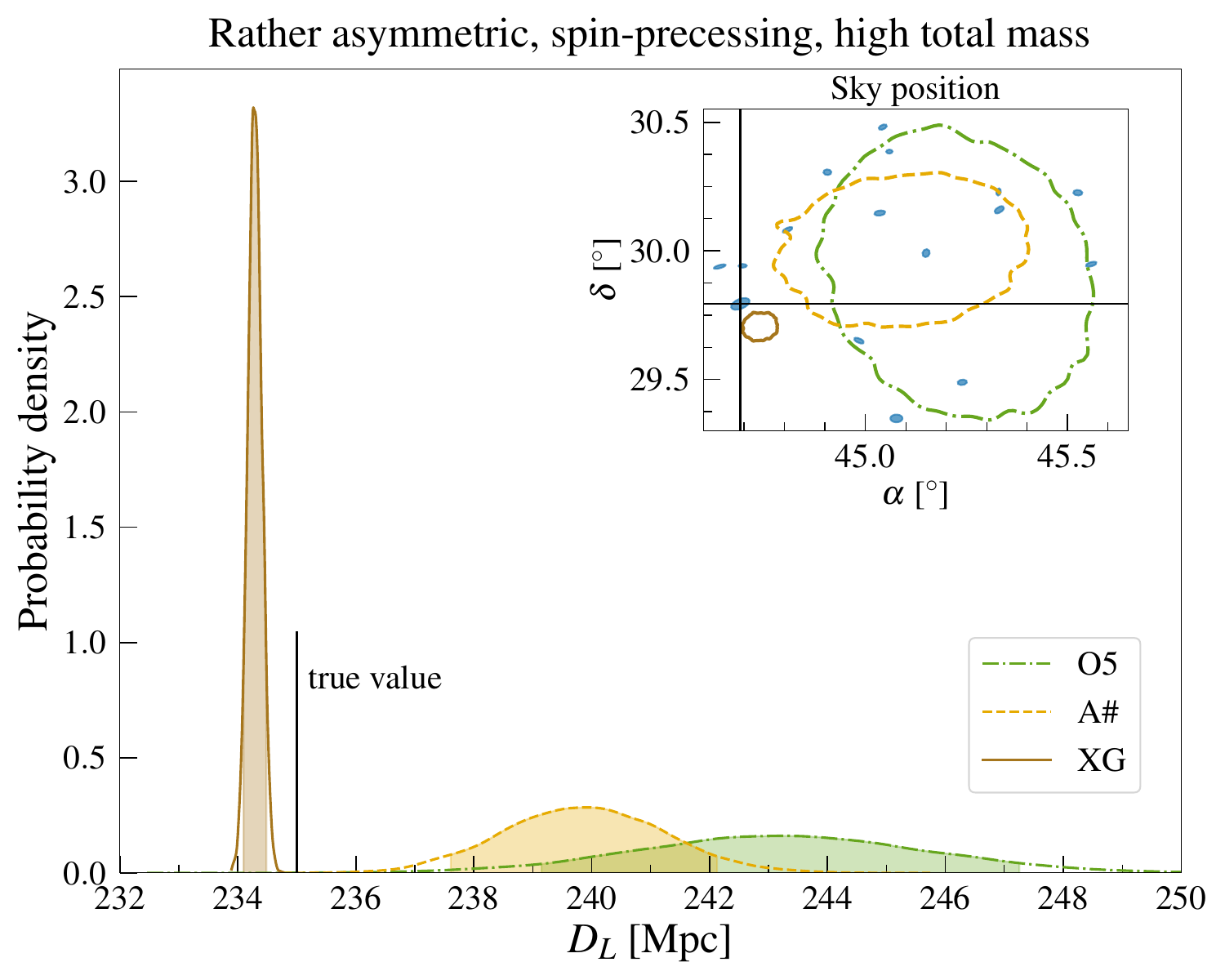}
		\caption{Distance posteriors for the \first, \second, and
			\third systems (from top to bottom) in the three detector
			networks considered in this study, The shaded region shows the
			90\% credible interval. The contours in the inset show the 90\%
			credible region for the corresponding sky position
			measurement. The true values are denoted by black
			lines.}
		\label{fig:distance_bias}
	\end{figure}

	\subsection{Impact on science objectives}
	In the rest of this section, we investigate the effect of systematic bias on various science objectives. Even though we isolate different science objectives 
to discuss them individually, this is an artificial separation since the science is interconnected and so are the biases. Therefore, some of the same results may be discussed in different sections in slightly different ways. For instance, a biased measurement of the maximum \ac{NS} mass not only has consequences for nuclear physics but also for cosmology. Similarly, different representations of the same parameter space may be helpful in highlighting different aspects of the science. For
instance, we refer to the magnitude and tilt of the spin vector when discussing lower mass-gap events, effective spin and spin-precession parameters when discussing astrophysical formation channels of heavy \acp{BBH}, and the relative orientations of the spin vectors when talking about spin morphologies. They are all different slices of the same spin space. However, the different parametrizations are useful for discussing and highlighting different aspects of the spin space. 
		
	\subsubsection{Cosmology}
	\label{subsubsec:cosmology}
	
	The luminosity distance and sky localization posteriors for the three simulated systems in the different detector networks are shown in Fig.~\ref{fig:distance_bias}. An illustration of an expected distribution of galaxies in a volume uncertainty region is depicted as blue ellipses. 
	The volume uncertainty includes galaxies with redshift between $z_{\rm min}=H_{0,{\rm min}}/D_{L,{\rm max}}$ and $z_{\rm max}=H_{0,{\rm max}}/D_{L,{\rm min}}$, where $H_{0,{\rm min}} = 35 \rm \, km \, s^{-1} Mpc^{-1}$ and $H_{0,{\rm max}} = 140 \rm \, km \, s^{-1} Mpc^{-1}$, and $D_{L,{\rm min}}$ and $D_{L,{\rm max}}$ are the edges of the 90\% credible interval of the $D_L$ posterior. The sky patch is the size of the inset panel that contains the sky location posterior. The areal density of luminous L* galaxies in the local Universe is taken to be $0.07\,\rm deg^{-2}$ at a distance of 100 Mpc~\cite{Singer:2016eax}.
	
	While the localization volume of the \first system (upper panel) is accurately measured for current detectors at design sensitivity and their next upgrade, both the distance and the sky position are biased for the \emph{XG} network. With a sky position measurement precision of $<0.1 \rm\, deg^2$ at 90\% credibility, at most a single galaxy is expected to be in a volumetric cone up to the distance to the event~\cite{Borhanian:2020vyr}. An inaccurate \ac{GW} measurement will completely miss the host galaxy in this specific example or result in the identification of the wrong host in general. From Fig.~\ref{fig:horizon_bias_lines}, it is clear that such a system will give a biased distance estimate for sources at distances up to $\sim35$ times farther, having an \ac{SNR} $\gtrsim30$ in the \ac{XG} network. 
	
	For the \second system (middle panel), while \emph{O5} and \emph{A\#} networks accurately recover the sky position again, the distance is biased for all three networks. A bias in the distance measurement towards larger values will cause the \hubble measurement to be systematically biased towards smaller values. As in the case of the previous system, the precision in the sky position measurement for the \emph{XG} network implies the existence of a single galaxy in the volume uncertainty region on average. Figure~\ref{fig:horizon_bias_lines} tells us that the distance bias is important in \emph{O5} and \emph{A\#} networks for \acp{SNR} $\sim70$ which implies distances of $\sim800$ Mpc and $\sim1300$ Mpc, respectively, while the distance at which $|\delta\vartheta/\Delta\vartheta|=1$ for the \emph{XG} network is $\sim5700$ Mpc. Even at a distance which is $\sim7$ ($\sim4$) times the maximum distances in \emph{O5} (\emph{A\#}), the event in \emph{XG} will have an \ac{SNR} of 130. 
	
	The measurement of both the distance and sky position is inaccurate for the \third system (lower panel) for all three detector networks under consideration. However, the distance bias, particularly for the \emph{XG} network, is smaller than the previous systems. On the other hand, the sky position is biased even for \emph{O5} and \emph{A\#} networks. 
	 Because of the smaller distance biases, we see from Fig.~\ref{fig:horizon_bias_lines} that the bias horizon for the \emph{XG} network is also smaller resulting in a very loud signal of \ac{SNR} $\sim400$. 
	 Meanwhile, since the \acp{SNR} in the three detector networks for this system are smaller than the \second system, an \ac{SNR} of $\sim35$ in the \emph{O5} network is sufficient for a material distance bias. 
	
	Given that a two percent measurement of the Hubble constant will resolve the Hubble tension, all three events in the \emph{XG} network can single-handedly resolve the tension while the \second and the \third events can do so even in an \emph{A\#} network, at a simulated distance of $D_L=235 \rm\, Mpc$. 
		
	\subsubsection{Lower mass gap}
	\label{subsec:lower_mass_gap}	
	
	\begin{figure}
		\centering
		\includegraphics[width=\columnwidth]{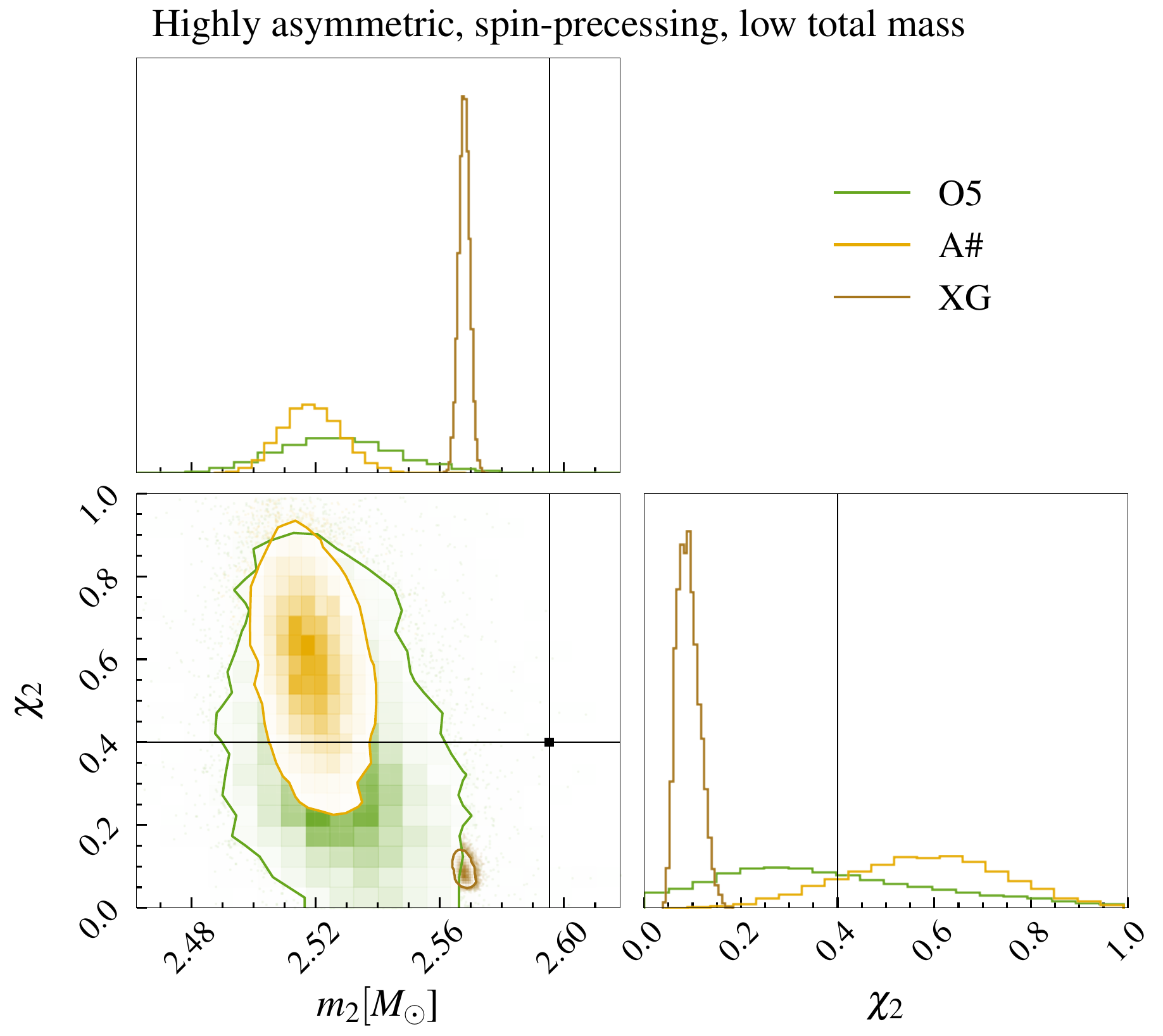}
		\caption{Corner plot of the posterior distributions for the mass and dimensionless spin magnitude of the secondary companion of the \first system in the three detector networks in which the system is simulated. The shaded region in the 1D posteriors and the contours in the 2D space denote the 90\% credible region. The black lines show the true injected value. A smaller minimum frequency for the \emph{XG} network ($f_{\rm low} = 5 \rm \, Hz$) compared to the \emph{O5} and \emph{A\#} networks ($f_{\rm low} = 10 \rm \, Hz$) results in a smaller bias for the mass.}
		\label{fig:mass_2}
	\end{figure}
	
	In Fig.~\ref{fig:mass_2}, the recovered distributions for the mass and dimensionless spin magnitude of the secondary companion of the \first system in the three detector networks are shown. Since this is a highly asymmetric merger, the spin of the secondary is poorly measured as is clear from the $\chi_2$ posteriors in the \emph{O5} and \emph{A\#} networks. Nevertheless, the measurement precision improves significantly in the \ac{XG} network, both due to its broad sensitivity improvement and
    a smaller minimum frequency. However, the measured value is materially inaccurate predicting a much lower value than the injection. This result would affect inference on hierarchical formation channels and accretion-induced mass growth of a \ac{NS} since both predict a large component spin. Instead, primordial \ac{BH} formation scenarios that predict small spins would be favored~\cite{Bianchi:2018ula}. The measurement bias in the component mass would also have impacts on constraints on the speed-of-sound relation in NSs~\cite{Tan:2020ics,Tan:2021ahl} and inference on the effect of rotation on \ac{NS} radii~\cite{Kruger:2023olj}.

	More directly, the upper edge in the mass distribution of astrophysical \acp{NS} can be used to determine the redshift and, thereby, estimate \hubble~\cite{Chernoff:1993th,Taylor:2012db,Farr:2019twy,Ezquiaga:2022zkx}. Hence, an inaccurate determination of the edge of the mass distribution could also bias cosmological parameter estimation. This effect can be quantified using some simple calculations. Assume that the \ac{NS} mass distribution is uniform and that this system lies at the edge of that distribution. It can be shown that the error in the determination of the upper edge of the mass distribution is given by $\Delta m_{\rm max} = {\rm max}\left[\sqrt{\sigma_m \, R_o/(N-1)}, R_o/(N-1)\right]$, where $R_o=(m_{{\rm max},o} - m_{{\rm min},o})$, $\sigma_m$ is the typical mass measurement uncertainty for systems near the upper edge, $N$ is the number of observations and $m_{{\rm max},o}$ and $m_{{\rm min},o}$ are the maximum and minimum observed \ac{NS} masses, respectively~\cite{https://doi.org/10.1111/j.2517-6161.1983.tb01268.x,Ezquiaga:2020tns}. 
	The posterior peak of the secondary companion in \emph{A\#} network is $\sim0.08 M_{\odot}$ away from the true value. It is easy to calculate from the above equation that with $N\sim20$ observations and $m_{\rm min} = 1 M_{\odot}$, the measurement error in determining the upper edge of the mass distribution is $\Delta m_{\rm max}\sim0.08 M_{\odot}$. 
	
	The merger rate of a \first system within a distance of 235\,Mpc is 1 every 3 years. Several studies have estimated that upcoming \ac{GW} observing runs are expected to detect tens of \ac{BNS} mergers per year~\cite{Borhanian:2022czq,Iacovelli:2022bbs}. Therefore, a systematic bias in the mass and spin measurement can bias the inference of \ac{NS} properties and cosmological parameters in the near future. 	
	
	\subsubsection{PISN mass-gap}
	\label{subsec:mass_gap}
	\begin{figure}
		\centering
		\includegraphics[width=\columnwidth]{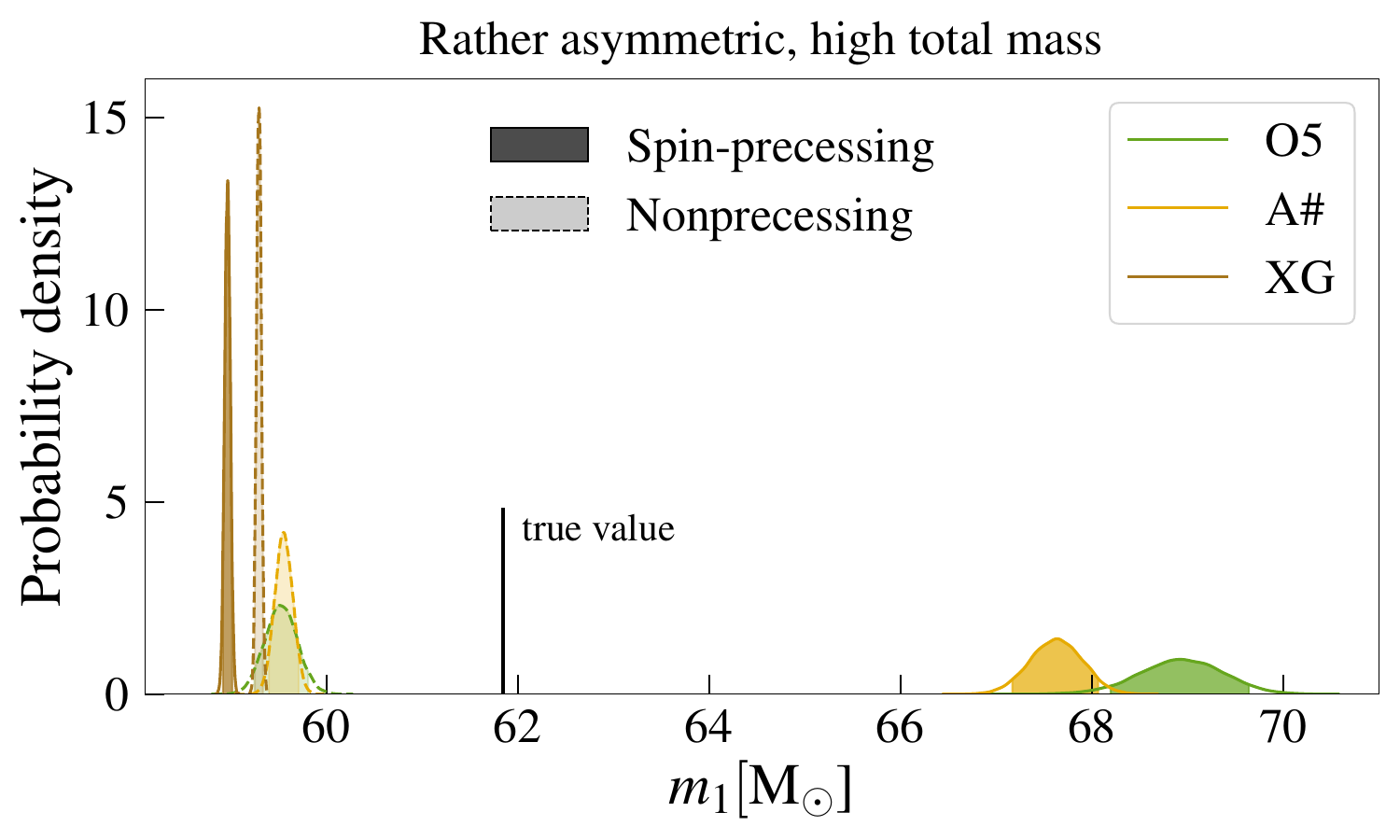}
		\caption{Posterior distributions of the primary mass for the \second (solid lines) and \third (dashed lines) systems in the three detector networks. The true injected value is shown by the vertical black line. The filled regions depict the 90\% credible interval.}
		\label{fig:mass_1_pisn}
	\end{figure}
	
	We report the measurements of the primary mass for the \second and \third systems in Fig.~\ref{fig:mass_1_pisn} in the three detector networks. While the biases in the aligned spin binary are less than in the precessing case, the measured values for both are significantly different from the injected value. 
	
	Furthermore, knowledge of the \ac{BH} mass spectrum, particularly the edge of the mass distribution, can be used for cosmological inference as with the \ac{NS} mass distribution. Following similar calculations and considering the simplified case where the \ac{BH} mass distribution follows a power law, $p(m)\propto m^{\alpha_m}$, with power law index $\alpha_m=-3.5$ and sharp cutoffs at $m_{\rm min}=10M_{\odot}$ and $m_{\rm max}=65M_{\odot}$, it can be shown that the error in the determination of the upper cutoff is given by, 
	\begin{equation}
		\Delta m_{\rm max} = \frac{(m_{{\rm max},o}^{\alpha_m+1}N-m_{{\rm min},o}^{\alpha_m+1})^{\frac{1}{\alpha_m+1}}}{(N-1)^{\frac{1}{\alpha_m+1}}} - m_{{\rm max},o},
	\end{equation}
	until $\Delta m_{\rm max} \sim \sigma_m$, the individual event mass-measurement uncertainty. 
	This uncertainty for $N=10^3$ events is then $\Delta m_{\rm max}\approx3\,M_{\odot}$. From Fig.~\ref{fig:mass_1_pisn}, it can be observed that the systematic error becomes dominant, even for 
	the \emph{O5} network which is expected to observe $\mathcal{O}(10^3)$ events every year.

	\subsubsection{Spin morphology}
	\label{subsubsec:spin_morphology}
	
	\begin{figure}
		\centering
		\includegraphics[width=\columnwidth]{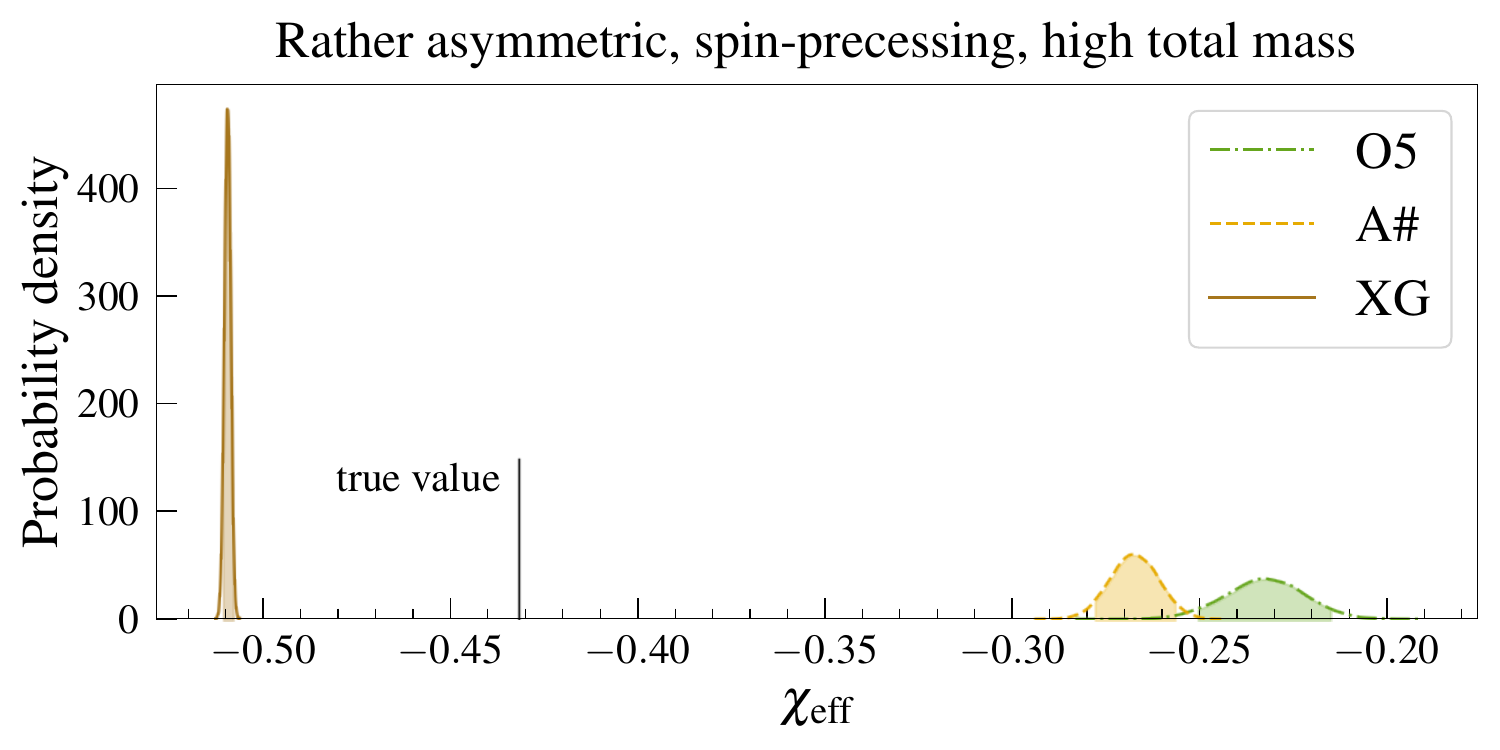}
		\includegraphics[width=\columnwidth]{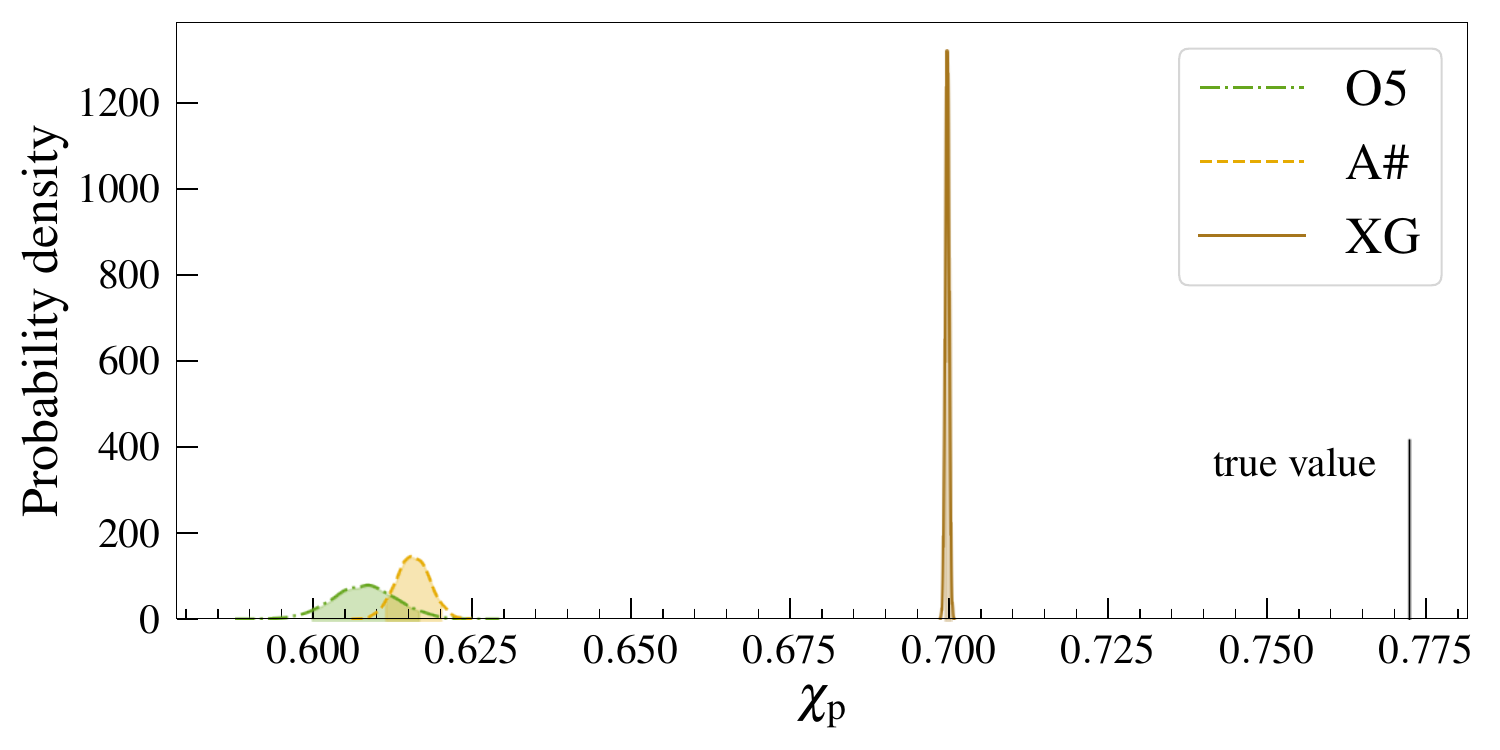}
		\caption{Posterior probability distribution of the $\chi_{\rm eff}$ (left) and $\chi_{\rm p}$ (right) parameters for the \third system in the three detector networks. The true value of the injection is shown by the black vertical line. The filled regions show the 90\% credible interval.}
		\label{fig:chieff_chip_bias}
	\end{figure}

	In Fig.~\ref{fig:chieff_chip_bias}, the posterior distributions of $\chi_{\rm eff}$ and $\chi_{\rm p}$ are depicted for the \third system in the three detector networks. It is immediately clear that the posterior distributions for both spin parameters and in all three detector networks are far from the true value. However, it is also noticeable that the absolute value of the bias is larger for the current networks and their upgrades compared to the future \emph{XG} network. Spin precession measurements are enabled by low-frequency sensitivity since the modulations in the \ac{GW} amplitude due to spin precession occur on these timescales. Moreover, different waveform models agree to a greater degree at lower frequencies because this regime is closely informed by \ac{PN} calculations in the various models. Therefore, a smaller minimum frequency and a better low-frequency sensitivity enable a more precise and accurate measurement of the $\chi_{\rm eff}$ and $\chi_{\rm p}$ parameters in the \emph{XG} network. Even so, the parameters are significantly biased. 
	
\begin{figure}
		\centering
		\includegraphics[width=\columnwidth]{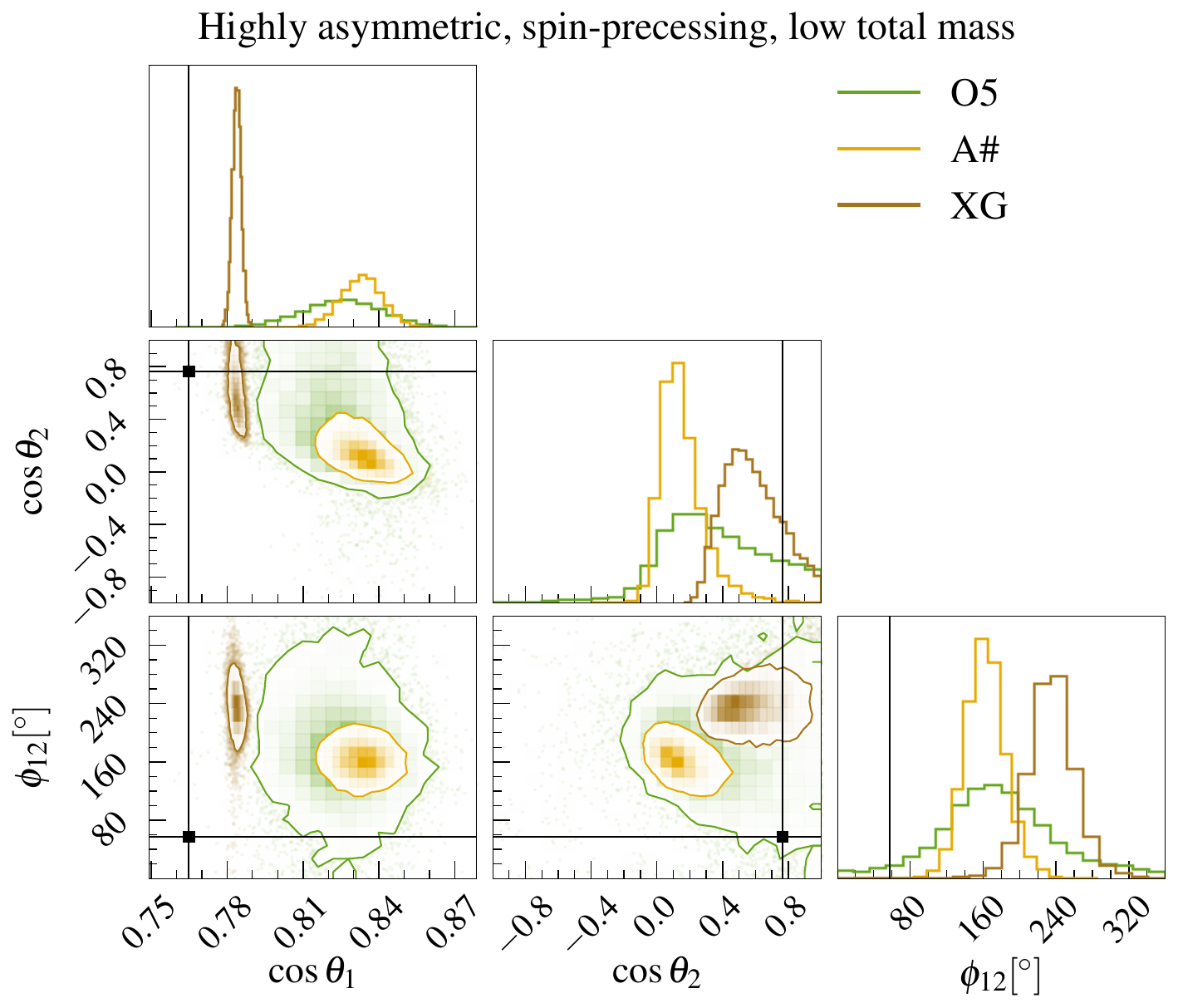}
		\caption{Posterior distributions of the cosine of the tilt angles, $\cos\theta_1$ and $\cos\theta_2$, and the relative in-plane spin angle, $\phi_{12}$, for the \first system. The black cross-hairs show the true injected value.}
		
		\label{fig:morphology_bias}
	\end{figure}

\begin{figure*}
		\centering
		\includegraphics[width=\columnwidth]{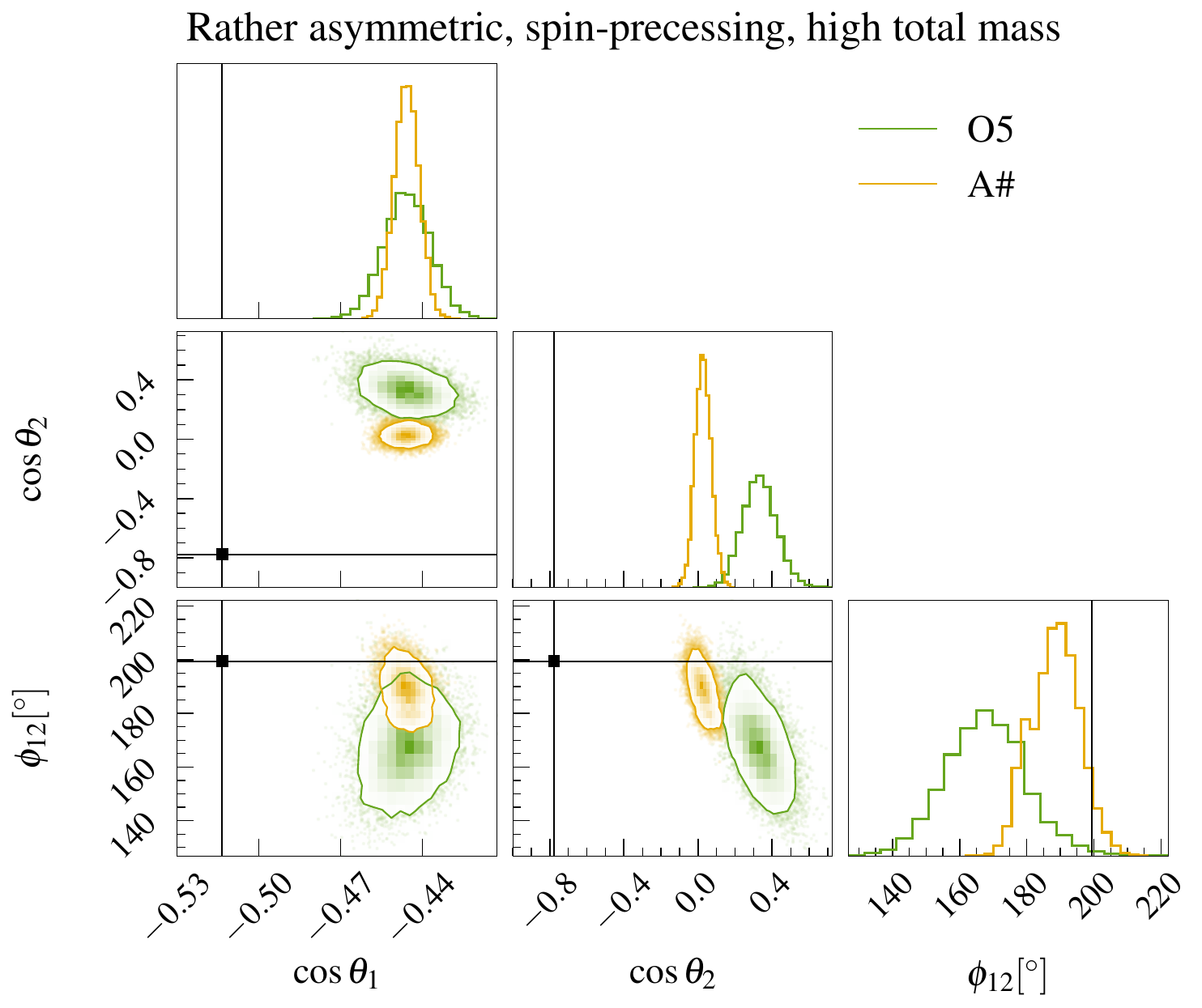}
		\includegraphics[width=\columnwidth]{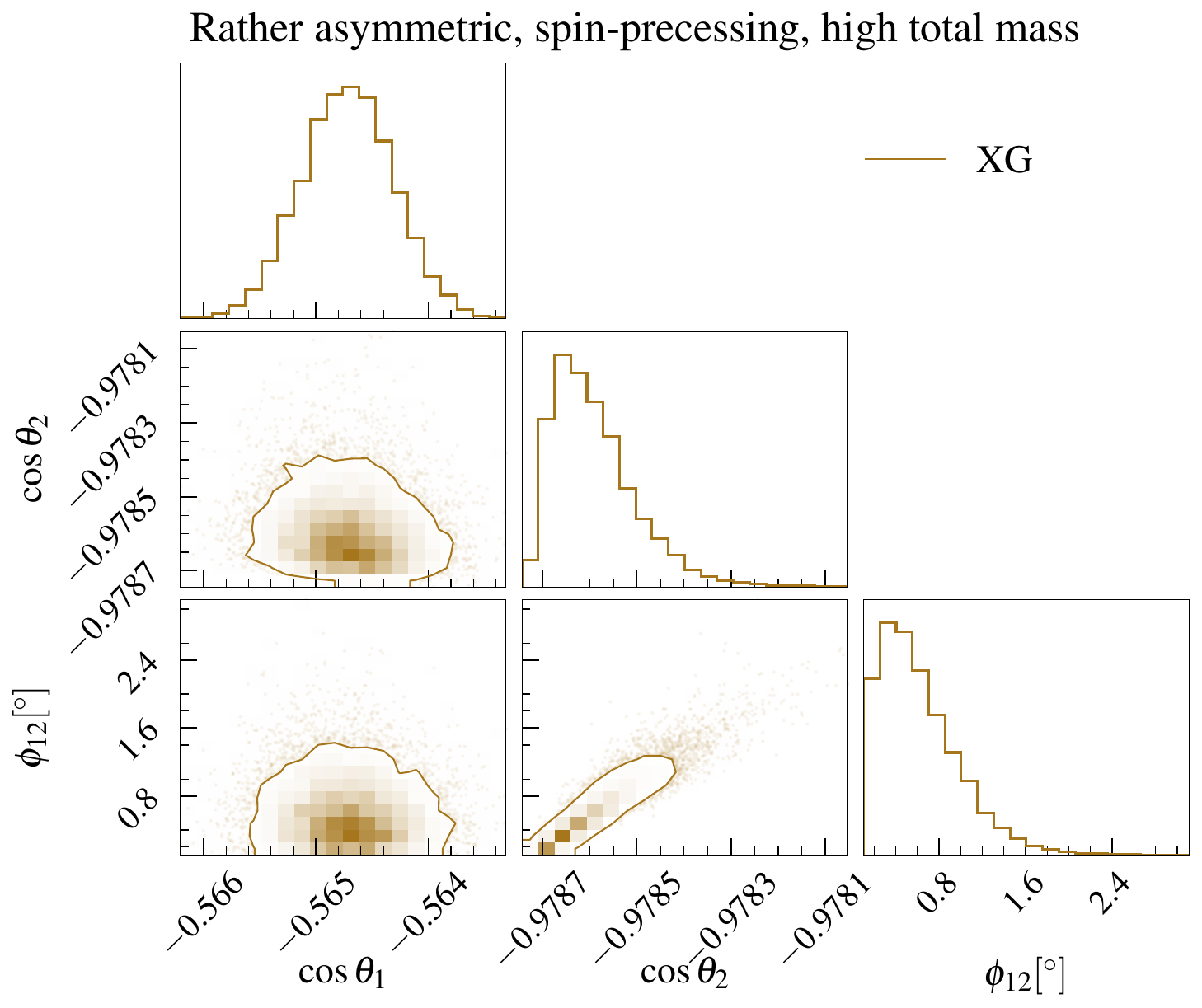}
		\caption{Posterior distributions of the cosine of the tilt angles, $\cos\theta_1$ and $\cos\theta_2$, and the relative in-plane spin angle, $\phi_{12}$, for the \third system in the \emph{O5} and \emph{A\#} networks (left) and the \emph{XG} network (right). The black cross-hairs show the true injected value. The \emph{XG} network is shown separately because it is significantly different from the other networks and incorporating them in the same figure distorts its appearance.}
		\label{fig:morphology_bias_XG}
	\end{figure*}

	The posterior distributions of the parameters characterizing the spin morphology for the \first and \third systems are reported in \cref{fig:morphology_bias,fig:morphology_bias_XG}. Let us analyze the \first system first. The tilt angles determine which morphology the binary falls into. We observe that both the tilt angles are recovered inaccurately. While the median bias for $\theta_2$ is greater than that for $\theta_1$, with the median value of $\theta_2$ being $1.5-2$ times the injected value, the poor measurement accuracy due to the highly asymmetric and low total mass nature of the binary results in the median value of $\theta_1$ being farther away from the true value, when expressed as a multiple of \emph{sigma} (the measurement error). The parameter $\phi_{12}$, which characterizes the spin morphology, is also heavily biased with the best-fit median values $2-3$ times the injected value depending on the detector network. 
	
	Now, considering the \third system, it is observed that the measurement accuracy increases due to the greater total mass and a resultant higher \ac{SNR}. Similar to the \first system, the systematic bias in the $\theta_1$ is smaller in absolute terms compared to $\theta_2$, whose median inferred value is $0.4-0.6$ the injected value. Interestingly, $\phi_{12}$ is less biased for the \emph{O5} and \emph{A$\#$} networks. The estimates in the \emph{XG} network are not only more biased but also completely different from the \emph{O5} and \emph{A$\#$} networks revealing the sensitivity of the measurement to the minimum frequency. Specifically, the measured value of $\phi_{12}$ in the \emph{XG} network is consistent with $0^{\circ}$ while the injected value is around $200^{\circ}$. 
	
	In this section, we do not explore the \second system in detail because it is a nonprecessing binary. Nevertheless, we report that the spin-precession parameter $\chi_{\rm p}$ is accurately recovered to be 0, and systematic biases do not lead to an inaccurate inference of an aligned-spin system as a spin-precessing system.
	
	\subsubsection{Remnant quantities}
	\label{subsec:remnant_quantities}

	\begin{figure*}
		\centering
		\includegraphics[width=2\columnwidth]{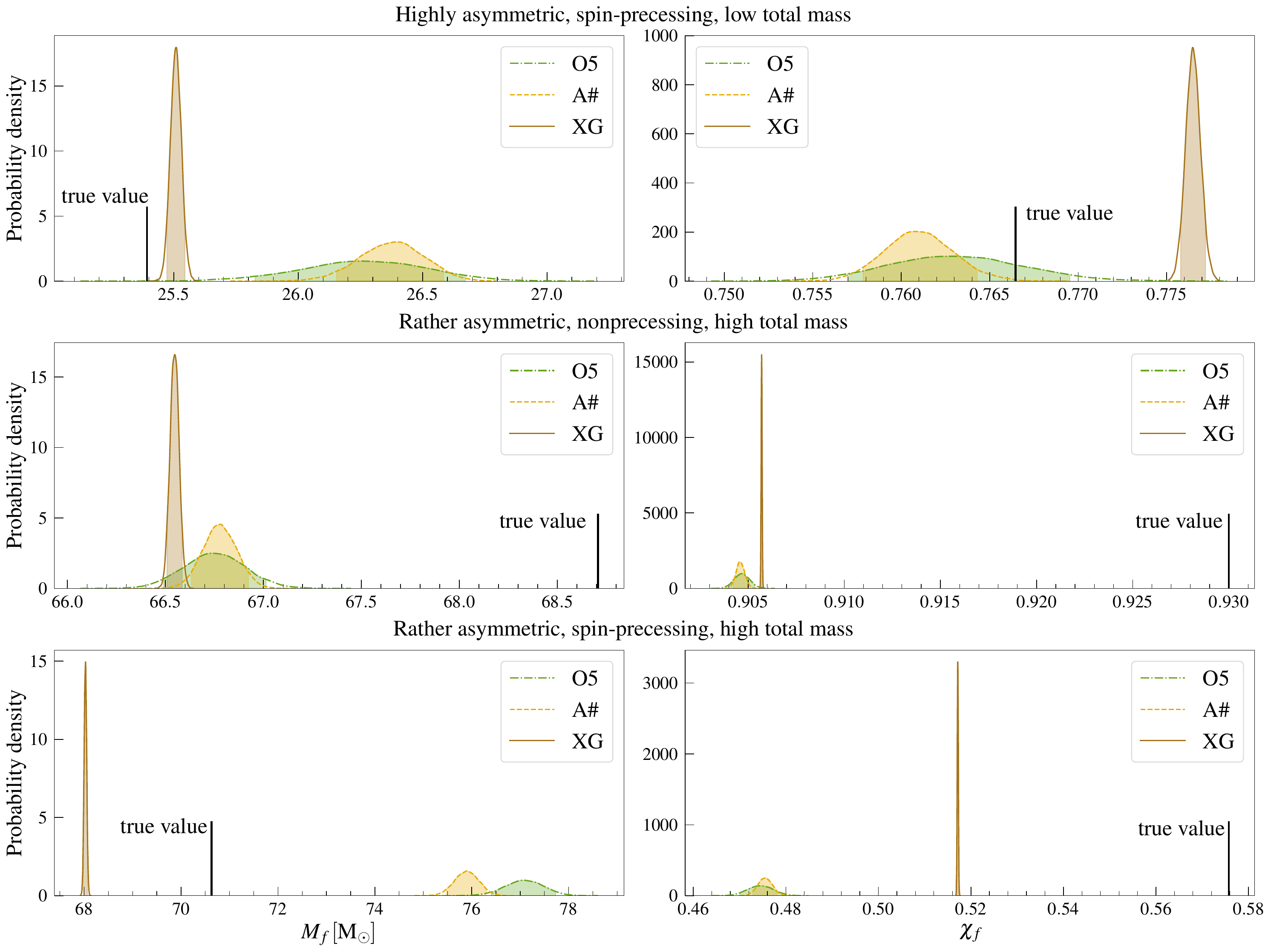}
		\caption{Final mass (left) and final spin (right) posteriors for the \first, \second, and \third systems (from top to bottom) in the three detector networks considered in this study. The true values are denoted by black lines.}
		\label{fig:remant_bias}
	\end{figure*}

	We report the inferred remnant quantities for the \first, \second, and \third systems (top to bottom) in the three detector networks in Fig.~\ref{fig:remant_bias}. Unsurprisingly, there is significant bias in most of the cases. Since we reported the biases in various mass and spin quantities in the preceding sections, the biases in the remnant quantities are expected. It can also be observed from the figure that the remnant properties of high-mass systems are better measured compared to the
    low-mass system. Consequently, the biases for \second and \third as a multiple of the measurement error are larger. 
	
	We illustrate the impact of these biases on inspiral-merger-ringdown consistency tests using GW150914 as a gauge. The 90\% confidence interval range on the oscillation frequency and damping time of the fundamental quasi-normal mode of GW150914, with a ringdown SNR of $\sim$8.5, is $\sim$29 Hz and $\sim$5.6 ms, respectively~\cite{LIGOScientific:2016lio}. In comparison, the biases in the fundamental quasi-normal mode calculated from the remnant quantities range between $\sim$0.2 Hz and $\sim$30 Hz for the oscillation frequency and $\sim$0.02 ms and $\sim$0.6 ms for the damping time across the different systems and networks considered here. Therefore, the consistency tests would fail for most of the cases considered here. 

	\subsubsection{Maximal \ac{BH} spin}
	\label{subsubsec:max_a2}
	\begin{figure*}
		\centering
		\includegraphics[width=2\columnwidth]{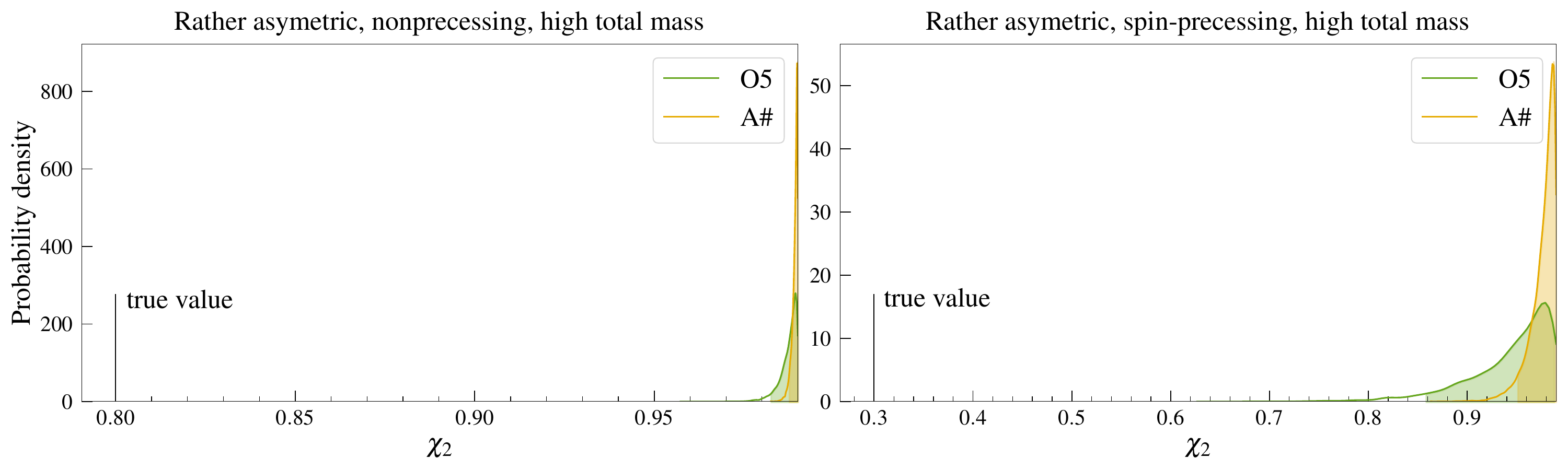}
		\caption{Probability distributions of the secondary dimensionless spin magnitude $\chi_2$ in the \second (left) and \third (right) systems in the \emph{O5} and \emph{A$\#$} networks. The injected signal is generated using \eob while the template model is \xphm. The estimates are biased for both the systems and in both the detector networks, but the more interesting feature is that the recovered $\chi_2$ distributions imply a maximally spinning secondary. We do not show the
        posteriors in the \emph{XG} network because they do not rail against the boundary, albeit still being biased.}
		\label{fig:a2_bias}
	\end{figure*}
	
	In Fig.~\ref{fig:a2_bias}, the posterior distributions on the secondary dimensionless spin magnitude $\chi_2$ are reported for the \second (left panel) and \third (right panel) systems. We observe that $\chi_2$ is biased for both systems. While the effect of $\chi_2$ on the \ac{GW} waveform is subdominant due to the high mass asymmetry and smaller magnitude compared to $\chi_1$, which results in a poorer measurement of this parameter compared to $\chi_1$, the significantly biased recovery
    hints at systematic differences in the modeling of the effects of the parameter on the \ac{GW} waveform. Perhaps, more interestingly, the secondary is measured to be maximally spinning, which is of great importance because such a measurement in a real event would be revolutionary. 
	We do not show the results for the \emph{XG} detectors here because $\chi_2$ is not inferred to be maximally spinning.
	
	As shown in Fig.~\ref{fig:bh_DL}, the measurement of $\chi_2$ will be biased even for low \ac{SNR}, albeit the prior will start becoming important at such \acp{SNR}.

	\section{Discussion and Conclusion}
	\label{sec:discussion_conclusion}
		
	In this work we have studied systematic biases arising in state-of-the-art \ac{BBH} waveform models used by the LVK Collaboration, in particular 
the quasi-circular, spin-precessing, multipolar \xphm and \eob models. With increasing detector sensitivities of the current LVK network, its future upgrades, and 
\ac{XG} detectors ahead, quantifying waveform systematics becomes central in achieving the promising science goals of \ac{GW} astronomy. Unbiased parameter estimation is crucial for a range of applications, from individual events to the entire \ac{BBH} population. Moreover, it is also vital for precision tests of \ac{GR} as the prevailing theory of gravity.

Throughout this work, we have assumed that the true signal can be represented by \eob, and we modeled it with \xphm. Although neither model represents exact solutions, using them in this way for injection-recovery studies allows one to explore a wide range of the \ac{BBH} parameter space, for which accurate \ac{NR} simulations are not yet available. 

To quantify the bias in parameter estimation, we utilized statistical tools from \ac{LSA} and full Bayesian analysis (see Sec.~\ref{sec:methods}). While \ac{LSA} is approximate and relies on large-\ac{SNR} events, it can forecast results for a large number of events, which is what we are interested in. Here, the two main methods are the \ac{FIM} to approximate measurement errors and the bias formula to predict the systematic errors. The full Bayesian analysis is computationally expensive but more reliable, which makes it suitable for understanding selected events in detail. With these methods, we studied biases and performed hierarchical inference on the \ac{BBH} population in Sec.~\ref{sec:LVK_bias}. We explored vast parts of the \ac{BBH} parameter space in Sec.~\ref{sec:parameterspace} and investigated in detail the impact of systematics on the science cases of individual events in Sec.~\ref{sec:science_case}. In the following, we summarize our main results.

Our first result is mainly on the ``use and abuse'' of the widely used bias formula based on the \ac{LSA}. By comparing with full Bayesian results, we explicitly demonstrated that the direct application of the bias formula without a ``waveform alignment'' procedure, as described in Eq.~\eqref{eq:mm}, can yield unreliable results. To our knowledge, this case has not been discussed in the literature yet, and was not explained in the main references for the bias
formula~\cite{Flanagan:1997kp,Cutler:2007mi}. We find good agreement with full Bayesian results only after performing alignment and proper bookkeeping of the adjusted parameters. When not correcting for it, one would predict biases in the \ac{BBH} population that could overestimate the actual bias by 2 orders of magnitude, which is particularly important when comparing waveforms of different families. Even after such caretaking, it is important to keep in mind that the \ac{LSA} is an approximate estimate of the biases and statistical errors. We include discussions on the validity of the estimates and possible shortcomings of the method. 

Regarding the LVK-like \ac{BBH} population studied in Sec.~\ref{sec:LVK_bias}, we find that the vast majority of events measured with the \emph{O5} and \emph{A$\#$} networks are not biased, around $2.5\%$ of the events, whereas, for \emph{XG} detectors, up to $25\%$ of the events are expected to show biases. This finding is most clearly reported in Fig.~\ref{fig:ratio}, which relates the mismatch of each event with the ratio of systematic-to-statistical error.  Although the bias relative to the
statistical error is larger for \emph{XG} detectors, the absolute bias per-se is smaller thanks to the detectors' improvement at low frequencies, which allows us to better observe the signal in the inspiral regime, where waveform models agree best. Thus, when combining all events through a hierarchical Bayesian analysis on the observed population, we find that the impact of systematic biases is more pronounced for \emph{A$\#$} than for \emph{XG} detectors. As illustrated in Fig.~\ref{fig:observed_pop}, the most biased parameter is the magnitude of the spin of the primary, with the recovered distribution peaking at smaller values, while exhibiting a tail at large spin magnitudes that is absent in the injected population. The latter feature, in particular, could erroneously guide astrophysical formation scenarios into explaining the existence of a sizable number of \ac{BH}s with large spins. We stress that the bias estimates obtained with the \ac{LSA} might be overly pessimistic, since it relies on the quadratic approximation to the likelihood, which holds only in the high \ac{SNR} regime. Thus, the impact on the population might be exaggerated for \emph{O5} and \emph{A$\#$} networks, but we expect our results for \emph{XG} to be more accurate.

When spanning the possible \ac{BBH} parameter space in Sec.~\ref{sec:parameterspace}, we sampled events with uniform distribution $\chi_\text{p} \in [0,1]$, uniform in total detector-frame mass $\in [10,200]\,M_\odot$ and uniform in inverse mass ratio $1/q \in [1,30]$. Although this sample does not represent the LVK \ac{BBH} population, exploring the challenging parts of the parameter space is important, since GW events in these regions may be discovered with more sensitive detectors in the
future. Using the \ac{LSA}, for each parameter, we computed the bias horizon, which describes the maximum distance up to which an event is systematically biased, after which the \ac{SNR} is low enough that the statistical error is larger than the systematic one. Since most of these binaries are difficult to model, analytically and numerically, not surprisingly the biases are more important than for the \ac{LVK}-like population of Sec.~\ref{sec:LVK_bias}. 

Furthermore, our detailed analysis of selected events summarized in Table~\ref{tab:params} in Sec.~\ref{sec:science_case} has several important findings depending on the science cases. 

Focusing on cosmological implications in Sec.~\ref{subsubsec:cosmology}, we found that distances and sky localization can be significantly biased. Here, one would be unable to infer the correct value of the Hubble-Lema\^itre parameter and thus would not resolve the Hubble-Lema\^itre tension. Biases in the sky position can be sufficiently large to prevent the correct identification of the host galaxy, which would be drastic, since single \emph{XG} events have the potential to determine $H_0$ to
a few percent. Furthermore, those biases may also affect the determination of $H_0$ from stacking GW events, requiring a dedicated future study.

When studying the lower mass gap in 
Sec.~\ref{subsec:lower_mass_gap}, we reported that the estimate of the secondary mass for the highly asymmetric, spin-precessing low total-mass system would be underestimated in all networks. The spin of the secondary would be significantly underestimated in the \emph{XG} network. This result could lead to wrong estimates of the upper edge of the \ac{NS} mass distribution, which would inflict further biases for studies of the equation of state and, again, $H_0$. %
The \ac{PISN} mass gap was investigated in Sec.~\ref{subsec:mass_gap}. Here, we showed that the estimate of the upper mass gap through the primary mass by the high total-mass binaries is strongly biased even for \emph{O5}. Although precession changes the primary mass posteriors of both binaries significantly, the injected value for $m_1$ is not recovered within the $90\,\%$ credible interval for any network. 
	
In Sec.~\ref{subsubsec:spin_morphology}, we looked into the spin morphology and found that \emph{XG} detectors are not always more prone to biases than \emph{O5} and \emph{A\#}. By extending the detectors' bandwidth to lower frequencies where the waveform models are more similar, the recovery of spin parameters can become closer to the injected values, at least for the low total-mass system.
%
The remnant quantities were studied in 
Sec.~\ref{subsec:remnant_quantities}, where we reported the final mass and final spin of the remnant for all three ``golden'' binaries. Both upper mass-gap events have significant measurement bias, excluding the injected values far outside the $90\%$ credible interval. For the highly asymmetric and precessing binary with low total mass, the overall biases are not as strong, besides the final spin in \emph{O5}, which is also outside the $90\%$ credible interval. By performing an independent
ringdown analysis, at least for \emph{XG}, one may conclude violations from the Kerr hypothesis and thus violations from \ac{GR} (not performed in this work). 
	
In Sec.~\ref{subsubsec:max_a2}, we observed that the two high total-mass events predict posteriors for $\chi_2$ that rail against the maximum spin of a BH for \emph{O5} and \emph{A\#}. If not identified as systematic bias, this prediction would certainly have important consequences for astrophysical formation channels in explaining such high spins in \ac{BBH} systems, as well as theoretical interest in extremal Kerr \acp{BH} and eventually exotic compact objects.

In summary, depending on the binary's parameters, biases can be present for the upcoming LVK O5 run and can affect crucial science. As expected, systematics become even more relevant with increasing detector sensitivity; thus they are important for future \emph{XG} detectors. The fact that many exciting science cases can be jeopardized by biases underlines the importance of improving existing waveform models. It also motivates the need to include modeling error estimates when performing
parameter estimation, even if it will inevitably broaden our posteriors. Lastly, much more work would be needed to more robustly qunatify the waveform systematics---for example by employing as signals NR and NRSur waveforms, where available, and extending the current study to binaries on generic orbits, notably \acp{BBH} on eccentric orbits. 

The main limitation of our work is the use of the \ac{LSA} for the population analysis. While it is the most readily available and feasible way to conduct a study like ours, one can be critical about its validity across the parameter space and view the ensuing conclusions with a grain of salt. In the future, we intend to carry out population-scale studies using modern data analysis tools, such as \texttt{DINGO}~\cite{Dax:2021tsq}, that allow for rapid evaluations of the posteriors.

Note added.---Recently, we became aware of a complementary study, Ref.~\cite{Kapil:2024zdn}, that focuses on assessing waveform systematics for \ac{XG} detectors using two quasi-circular aligned-spin models.

	\begin{acknowledgments}		
We wish to thank Antoni Ramos Buades, Nihar Gupte, Serguei Ossokine, and Michael P\"{u}rrer for collaboration during the early part of this project. We also thank Veome Kapil, Luca Reali, and Emanuele Berti for useful discussions. We thank Ish Mohan Gupta for his comments during the LIGO review.
        S.\,H.\,V. acknowledges funding from the Deutsche Forschungsgemeinschaft (DFG), Project No. 386119226.
The authors are grateful for computational resources provided by the \texttt{Hypatia} computer cluster at the Max Planck Institute for Gravitational Physics in Potsdam, the \texttt{Gwave} computing cluster at Penn State University, and the LIGO Laboratory which National Science Foundation Grants PHY-0757058 and PHY-0823459 support. 
Some of the results in this paper have been obtained using the \pesummary package~\cite{Hoy:2020vys}. 

	\end{acknowledgments}
	
	\appendix

	\section{Toy model}
	\label{sec:toy_model}
	We consider a simple toy model to illustrate the effect of nonuniform parameter definitions across
	waveform models on the systematic bias calculated under \ac{LSA} using Eq.~\eqref{eq:bias}. 
	For this example, we take the \xphm model as both the signal and the template. 
	However, we shift the signal by $\tau$, evaluating it at $t_c=\tau$, while the template is 
	evaluated at $t_c=0$. 
	Note that for a pair of arbitrary waveform models, we do not know of the shift $\tau$ \emph{a priori}
	and, hence, compute the bias at $t_c=0$. 
	In a Bayesian analysis, this would simply mean that the likelihood distribution for $t_c$ peaks at $t_c=\tau$,
	for a noiseless injection, without impacting any physical parameter.
	
	Let us now examine the predictions of the bias formula Eq.~\eqref{eq:bias}. 
	We consider a reduced two-dimensional parameter space, $\bm{\vartheta}=\{ D_L, t_c \}$.
	Here, the bias for the two parameters can be calculated analytically. 
	The distance bias is given by
	\begin{equation}
	\begin{aligned}
		\delta D_L &= \frac{D_L^2}{\left< h | h \right>} \left< -h/D_L | h(t_c=\tau) - h(t_c=0) \right> \\
				&= \frac{D_L^2}{\left< h | h \right>} \left< -h/D_L | h(t_c=0)e^{-2\pi if\tau} - h(t_c=0) \right> \\
				&= 0 + \mathcal{O}(\tau^2),
	\end{aligned}
	\label{eq:toy_dl_bias}
	\end{equation}
	while the $t_c$ bias reduces to
	\begin{equation}
	\begin{aligned}
		\delta t_c &= \frac{1}{\left< 2\pi ifh | 2\pi ifh \right>} \left< -2\pi ifh | h(t_c=\tau) - h(t_c=0) \right> \\
				&= \frac{-2\pi}{\left< 2\pi ifh | 2\pi ifh \right>} \left< ifh | h(t_c=0)e^{-2\pi if\tau} - h(t_c=0) \right> \\
				&= \tau + \mathcal{O}(\tau^2).
	\end{aligned}
	\label{eq:toy_tc_bias}
	\end{equation}

	\begin{figure}
		\centering
		\includegraphics[width=\columnwidth]{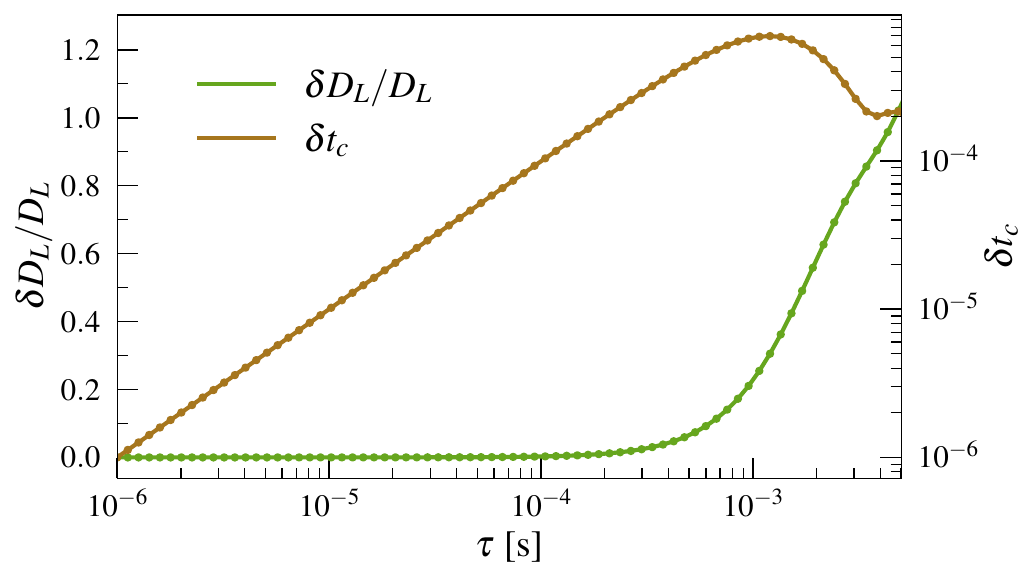}
		\caption{Bias in the luminosity distance and time of coalescence parameters as a function of a time shift in the 
				signal for the simple toy model considered in Sec.~\ref{sec:toy_model}. For very small values of the time 
				shift parameter $\tau$, the estimates of the bias formula are reliable, but they start deviating for $\tau$
				values that are still small.}
		\label{fig:toy_model}
	\end{figure}
	
	In Fig.~\ref{fig:toy_model}, we show the bias in the luminosity distance and time of coalescence calculated using Eq.~\eqref{eq:bias}. 
	We see that the estimates receive contributions from the quadratic and higher-order terms of Eq.~\eqref{eq:toy_dl_bias} and Eq.~\eqref{eq:toy_tc_bias}
	for small values of the time shift $\tau$. However, these corrections are not physical since $D_L$ should not be biased
	for simple time shifts of the signal and the bias in $t_c$ should simply correspond to the value of the time shift. 
	While the incorrect estimates for the $t_c$ bias are not of physical consequence in most cases, it is readily seen from the figure
	that the $D_L$ bias can be incorrectly estimated to very large values, impacting the outlook on science applications like cosmology
	where the $D_L$ parameter is crucial.

	\section{Effect of $f_{\rm low}$ for higher modes in SEOBNRv5PHM signal}
	\label{sec:eob_flo}
	\begin{figure*}
		\centering
		\includegraphics[width=2\columnwidth]{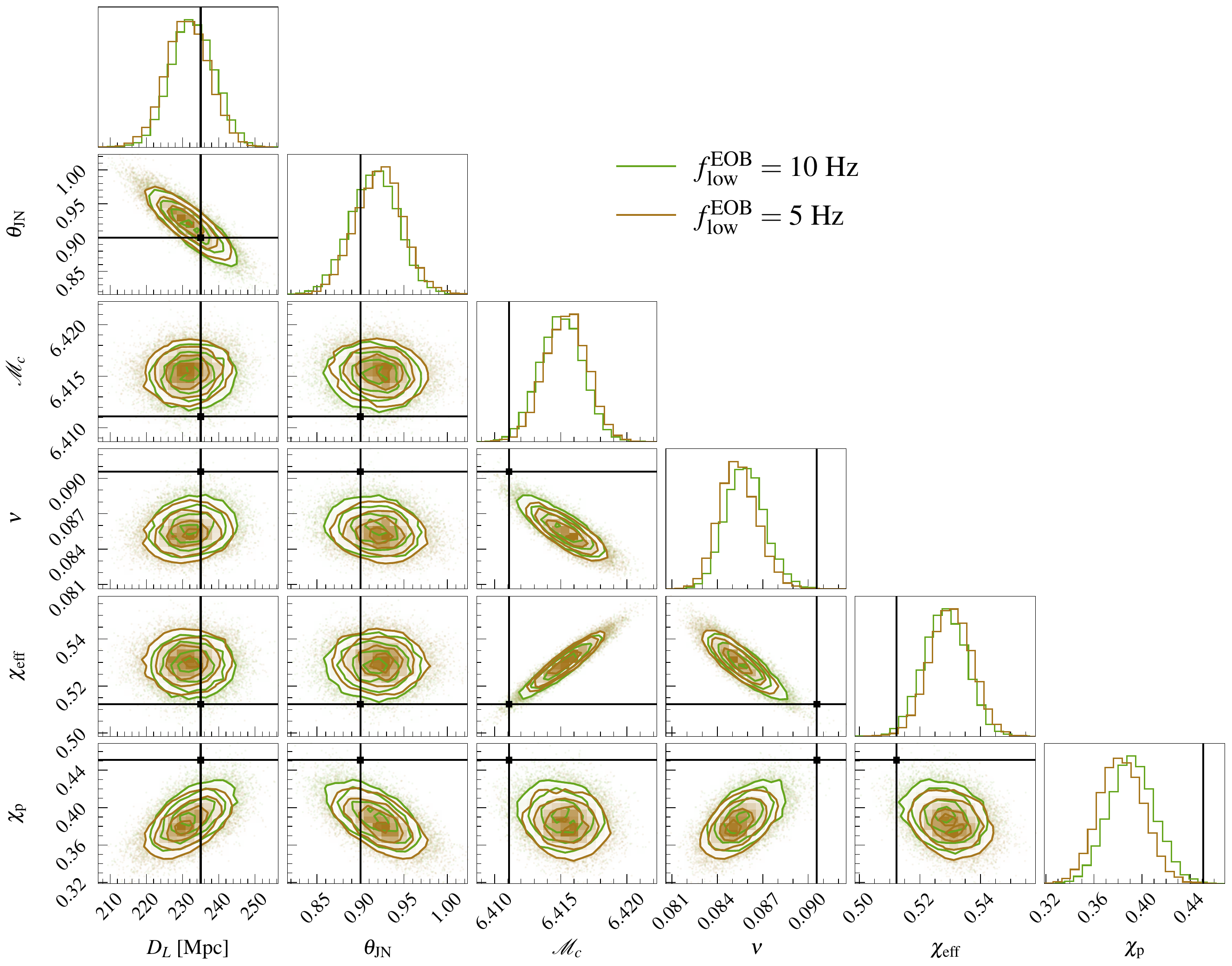}
		\caption{Posterior distributions of select parameters for the \first system in the \emph{O5} network with different minimum frequencies at which the \eob signal is generated. The starting frequency for the analysis is 10 Hz for both cases. The indistinguishability of the distributions in the two cases shows that the absence of certain subdominant harmonics in the signal at low frequencies for the case depicted in green is inconsequential to the analysis.}
		\label{fig:eob_fmin_HLV}
	\end{figure*}
	When generating an \eob waveform, the minimum frequency refers to the frequency of the $(l,m)=(2,2)$ harmonic. Higher harmonics in the given time segment occur at a higher frequency, which is a feature of all time-domain waveforms. Since phenomenological waveforms are constructed in the frequency domain, all the harmonics are present at any given frequency. Thus, in analysing a \ac{GW} signal generated using \eob with \xphm (with the same minimum frequency), the template contains the higher harmonics at frequencies lower than where the same is present in the signal. For instance, if $f_{\rm low}=10\rm\,Hz$ for both \eob and \xphm, the $(l,m)=(3,3), (4,4)$ harmonics for the \eob signal start at $15\rm\,Hz$ and $20\rm\,Hz$, respectively. 
	
	All analyses in the paper are performed by taking the same $f_{\rm low}$ for both \eob and \xphm waveforms. We verify in Fig.~\ref{fig:eob_fmin_HLV} that this approach does not affect any of the results of the paper. The figure shows the posterior distributions of select parameters of the \first system for two cases simulated in the \emph{O5} network. For the distributions plotted in orange, the \eob waveform is generated starting from $f_{\rm low}=5\rm\,Hz$ while the analysis is
    completed using \xphm setting $f_{\rm low}=10\rm\,Hz$. On the other hand, the \eob waveform is generated with $f_{\rm low}=10\rm\,Hz$ for the case in green and an identical analysis setting as the previous case. We remark that the posterior distributions in the two cases are indistinguishable.

	\section{Effect of $f_{\rm low}$ on parameter estimation and bias}
	\label{sec:flo_pe}
	\begin{figure}
		\centering
		\includegraphics[width=\columnwidth]{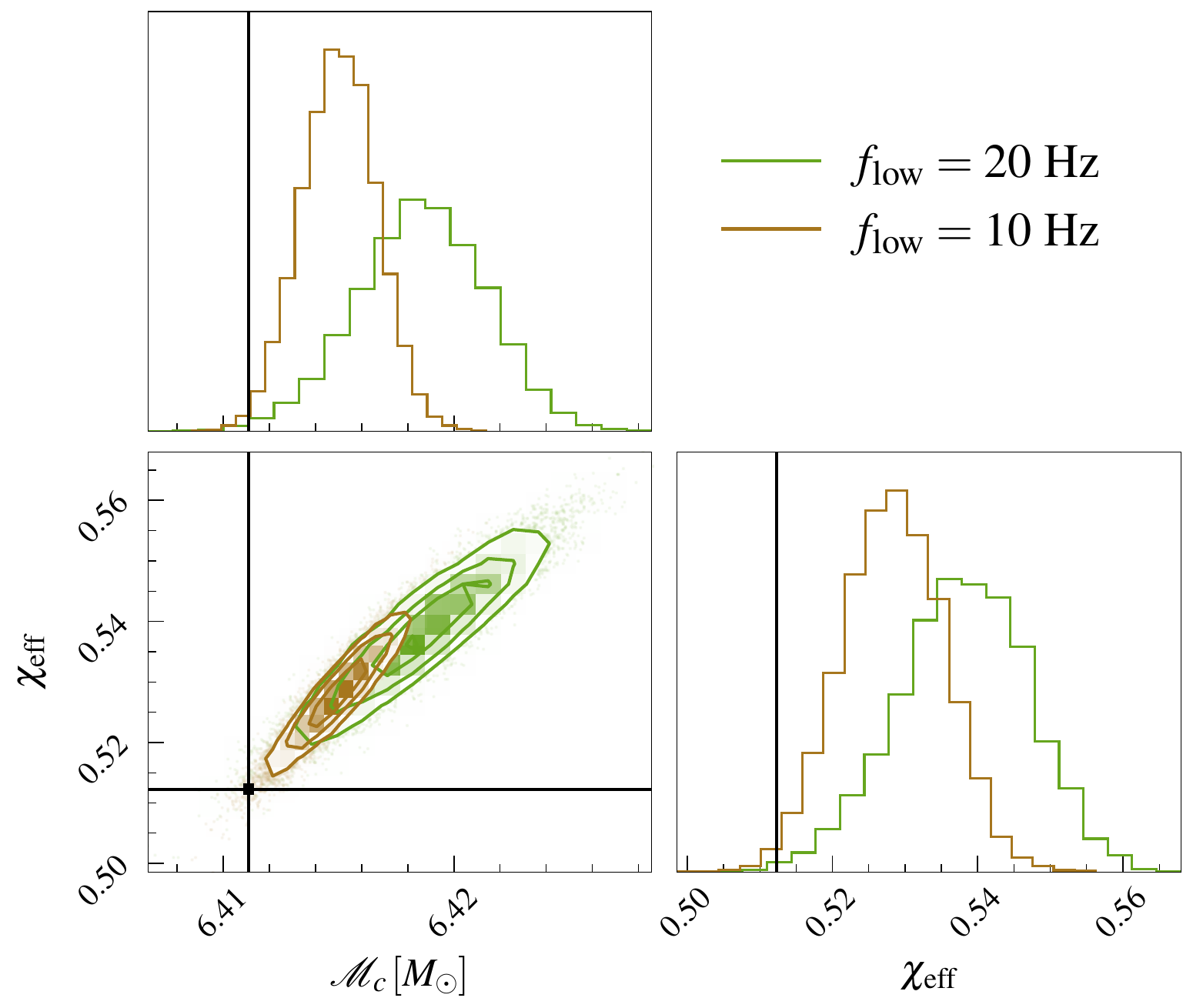}
		\caption{Posterior distributions in the $\mathcal{M}_c - \chi_{\rm eff}$ parameter space for the \first system in the \emph{O5} network with different minimum frequencies, $f_{\rm low}$. Note that a smaller minimum frequency results in a greater measurement precision and a smaller bias.}
		\label{fig:bias_fmin}
	\end{figure}
	We briefly discuss the impact of the minimum frequency $f_{\rm low}$ used in analyzing a \ac{GW} signal. This case is of particular importance given the excess low-frequency noise in the Advanced LIGO and Advanced Virgo detectors limiting the low-frequency cutoff to 20 Hz instead of the predicted 10 Hz. 
	
	We generate a signal corresponding to the \first system in the \emph{O5} network at two starting frequencies $f_{\rm low} = 10$ and $20 \rm\,Hz$ with the \eob model. This signal is then analyzed with the respective starting frequencies used in its generation with the \xphm model as the template. The posterior distributions in the $\mathcal{M}_c - \chi_{\rm eff}$ parameter space are reported in Fig.~\ref{fig:bias_fmin}. The distributions on the other nonderivative parameters are not reported since they are similar in both cases. It is observed that a lower minimum frequency leads to a smaller measurement error and bias for the chirp mass. Since $\chi_{\rm eff}$ has a large (positive) correlation with $\mathcal{M}_c$, as seen from the figure, it is also affected in a similar manner. 
	
	The number of cycles in a \ac{GW} waveform is inversely related to the minimum frequency and $\mathcal{M}_c$ is the leading-order contributor to this relation. Hence, the large number of \ac{GW} cycles between 10 Hz and 20 Hz leads to a better measurement precision for $\mathcal{M}_c$. It is also the part of the waveform where different models are in better agreement since all models have to reproduce the \ac{PN} limit leading to a more accurate measurement. Thus, the absolute magnitude of the $\mathcal{M}_c$ bias in the \emph{XG} network, which has $f_{\rm low}=5\rm\,Hz$, is smaller than the \emph{O5} and \emph{A\#} networks.

	\section{Dependence of $|\delta\vartheta/\Delta\vartheta|$ on the \ac{SNR}}
	\label{sec:ratio_snr}

	\begin{figure*}
		\includegraphics[width=2\columnwidth]{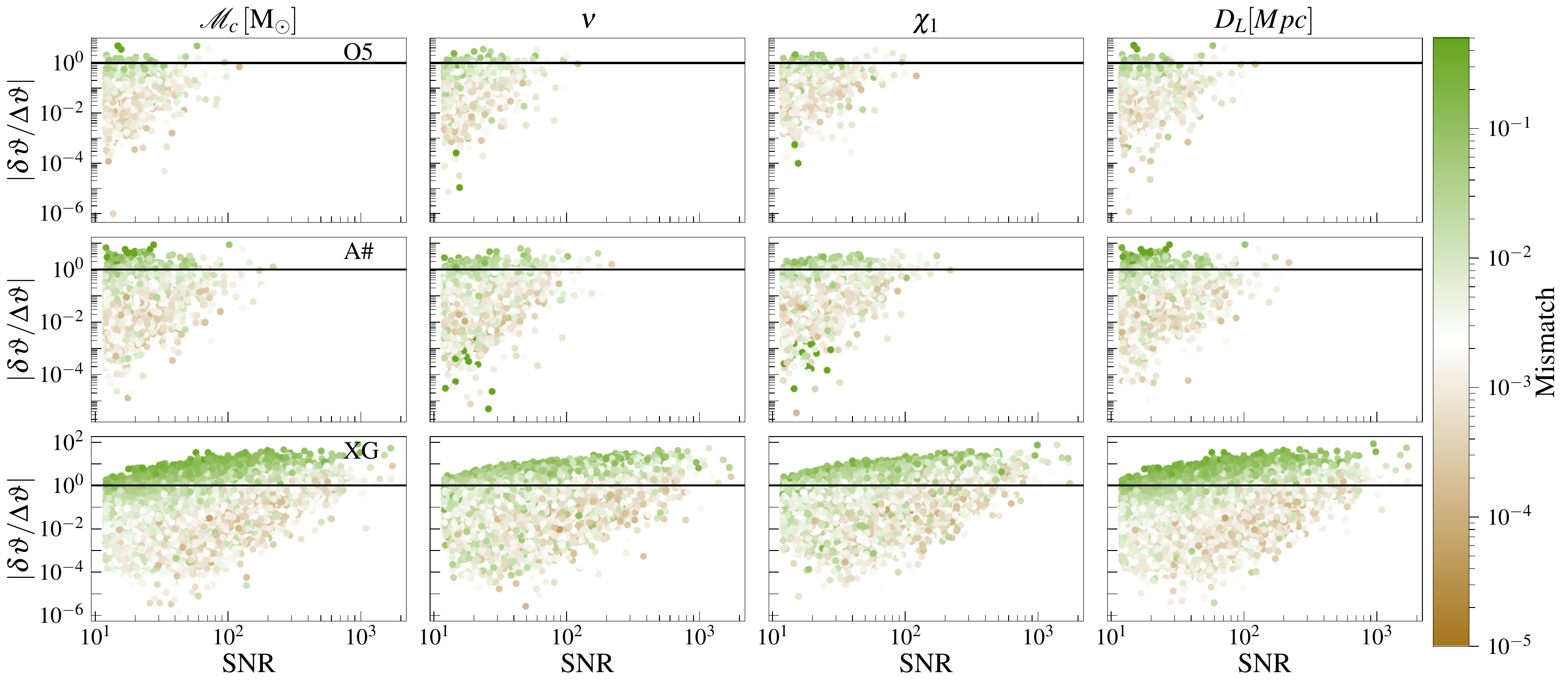}
		\caption{Ratio between the systematic bias and statistical errors for select parameters as a function of the \ac{SNR} for a population of \ac{BBH} mergers as observed by the \ac{LVK}. A network \ac{SNR} threshold of 12 was imposed on the $10^5$ binaries in the population resulting in $\sim1800$ and $\sim8100$ in the \emph{O5} (top) and \emph{A\#} (bottom) networks respectively. The color bar depicts the mismatch between \eob and \xphm waveform models.}
		\label{fig:ratio_snr}
	\end{figure*}
	
	In the following, we report complementary results for the same data as presented in Fig.~\ref{fig:ratio}. Figure~\ref{fig:ratio_snr} shows how the ratio $|\delta\vartheta/\Delta\vartheta|$ depends on the \ac{SNR} (instead of the mismatch), and the color bar now indicates the mismatch. The panels are structured similarly to those of Fig.~\ref{fig:ratio}. Overall, the cumulation of biased events (ratio larger than 1) depends strongly on the mismatch, although outliers exist for all networks.
    In all cases, the number of events with a ratio much smaller than 1 decreases as a function of the \ac{SNR}, although strongly biased events exist even for small \acp{SNR}. These results underline the importance of improving waveform modeling to reduce mismatches for all detector networks, not only for \emph{XG}. 
	
	\section{Distribution of population parameters}
	\label{sec:population_distribution}
	\begin{figure*}
		\includegraphics[width=2\columnwidth]{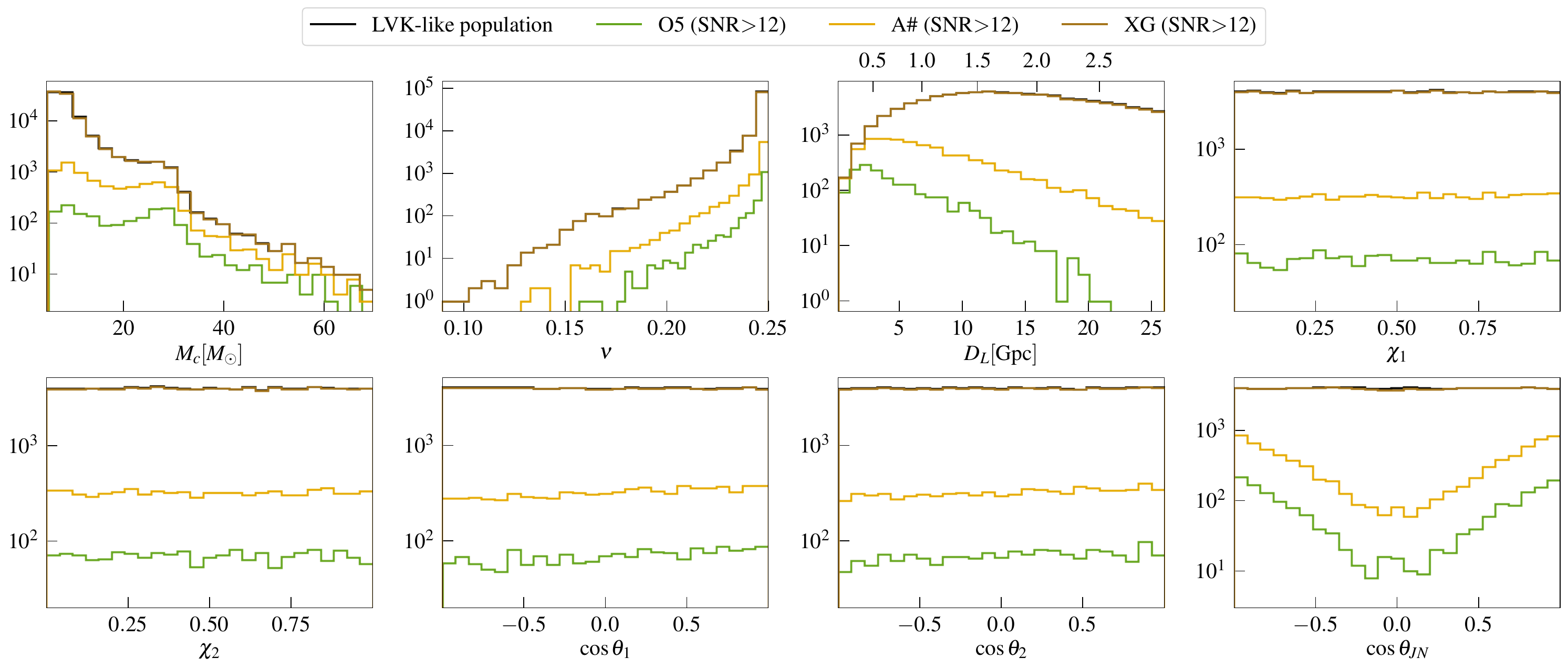}
		\caption{Distribution of select parameters of the LVK-like \ac{BBH} population and of the detectable population in the three detector networks. The black line overlaps almost entirely with the detectable population in the \emph{XG} network.}
		\label{fig:population_distribution}
	\end{figure*}
	
	In the panels of Fig.~\ref{fig:population_distribution}, we show the distribution of parameters of detectable events from the LVK-like BBH population for the different networks. The detectability criterion is \ac{SNR}$>12$, which acts as a strong filter for \emph{O5} and \emph{A\#}, but only very mildly impacts \emph{XG}. The parameters $a_1$, $a_2$, $\cos \theta_1$, and $\cos \theta_2$ are distributed uniformly for all networks, which agrees well with the population. However, the shape of
    the distributions describing the parameters $M_c$, $\nu$, $D_L$, and $\cos \theta_\text{JN}$ is more complicated and changes throughout the networks. This result implies a significant selection bias unless one uses \emph{XG}. Note that $\cos \theta_\text{JN}$ is sampled from a uniform distribution but shows strong selection bias for \emph{O5} and \emph{A\#}.

	\section{Bias in $\chi_1$ when scanning the parameter space}
	\label{sec:exploratory_bias}
	
	\begin{figure*}
		\includegraphics[width=\columnwidth]{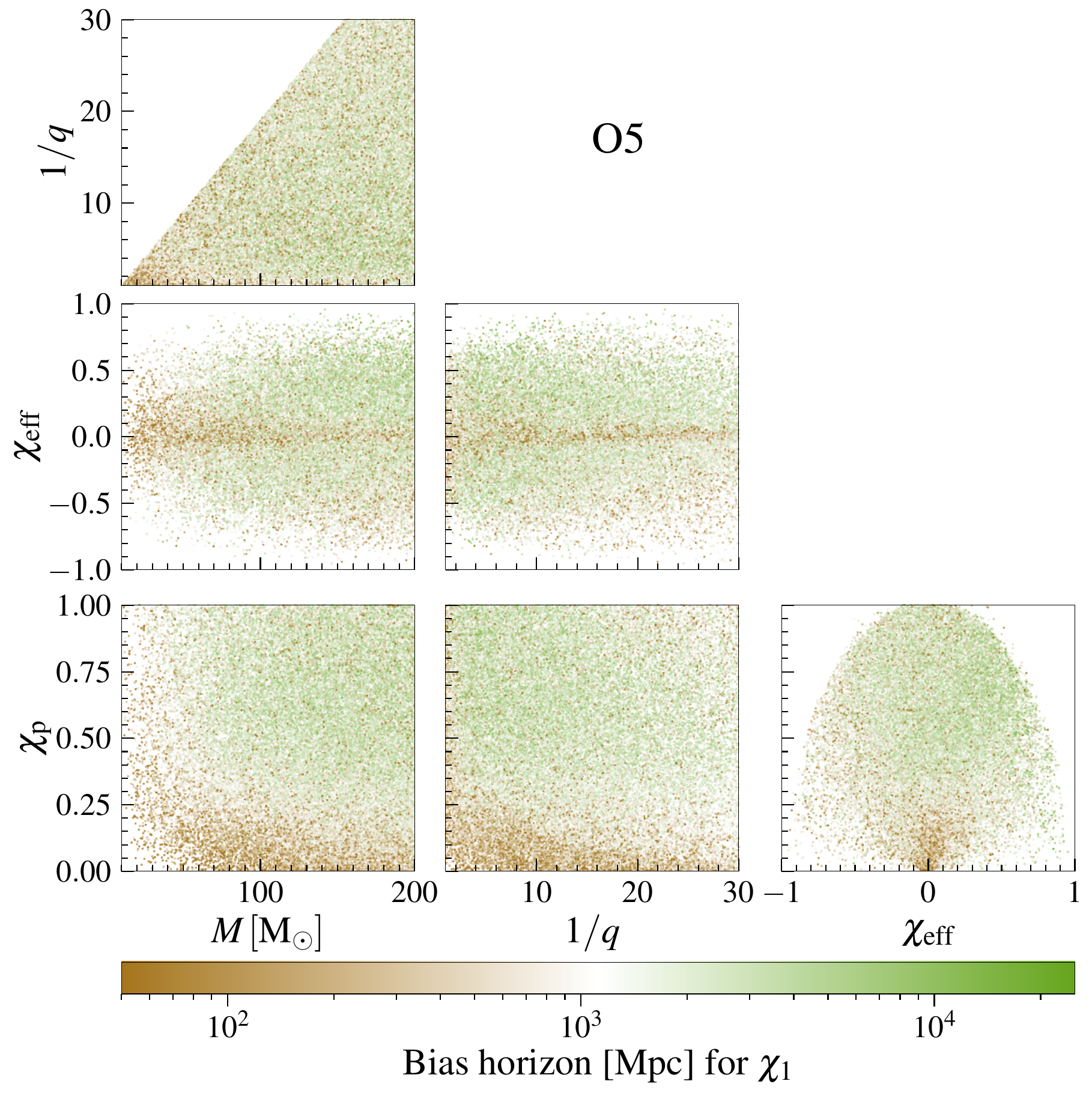}
		\includegraphics[width=\columnwidth]{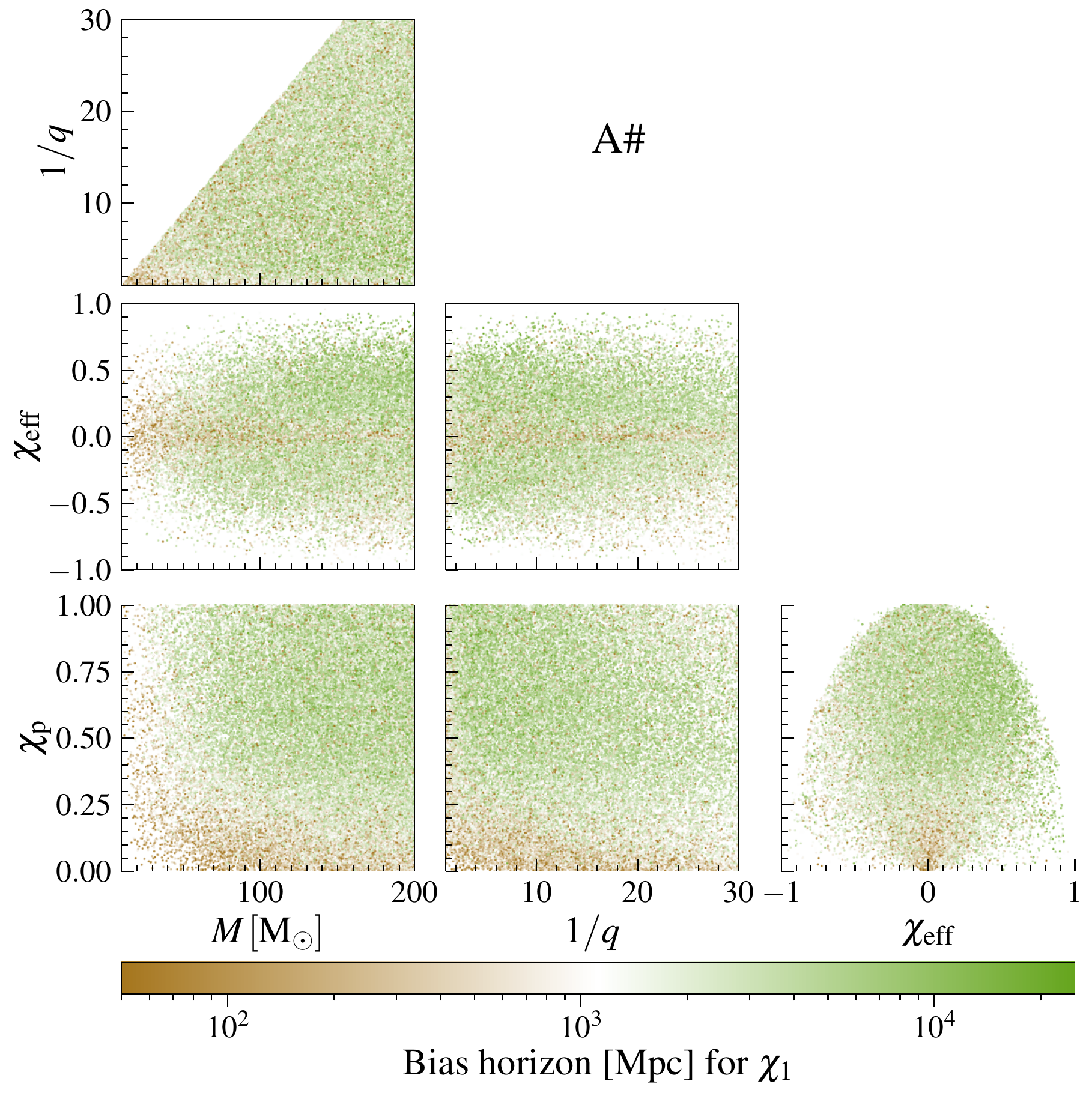}
		\caption{Distribution of the 50,000 binaries in the parameter space represented in Fig.~\ref{fig:lvk_agnostic}, with the color scale showing the distance to which the $\chi_1$ parameter is biased ($\delta \chi_1/\Delta \chi_1 \geq 1$) for the \emph{O5} (left) and \emph{A\#} (right) networks. Systematic biases become less important if a binary is at a larger distance since measurement precision decreases with distance. Therefore, a large bias horizon signifies that a given parameter
        ($\chi_1$ in this case) is measured well enough for systematic biases to be important even at such large distances. Notice that the binaries are biased up to a greater distance in the \emph{A\#} network compared to the \emph{O5} network due to its greater sensitivity, and the resultant improvement in measurement precision.}
		\label{fig:bh_a1}
	\end{figure*}
	
	\begin{figure}
		\includegraphics[width=\columnwidth]{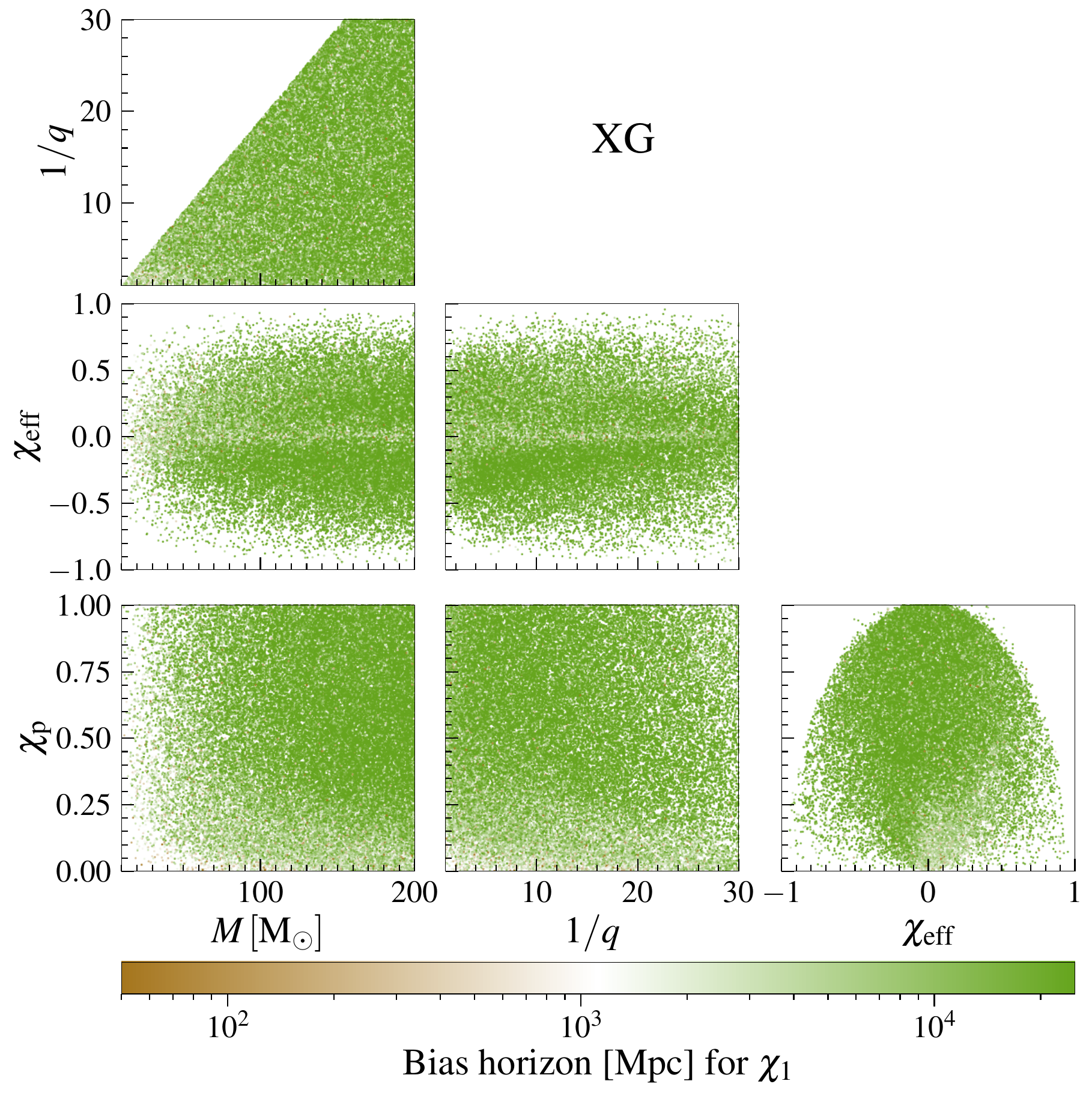}
		\caption{Same as Fig.~\ref{fig:bh_a1} for the \emph{XG} network. BBHs observed with the \emph{XG} network are biased up to a greater distance than those observed with either the \emph{O5} or \emph{A\#} network due to its greater sensitivity and the resultant improvement in measurement precision, with a majority of the binaries having a bias horizon $\geq 25 \, \rm Gpc$ ($z \approx 3$) beyond which stellar-origin \acp{BBH} are not expected to exist.}
		\label{fig:bh_a1_XG}
	\end{figure}
	
	In Fig.~\ref{fig:bh_a1} and Fig.~\ref{fig:bh_a1_XG}, we show the bias horizon for the parameter $\chi_1$. The binaries are distributed as described in Sec.~\ref{sec:parameterspace}.

	\clearpage
	\bibliography{references}

\end{document}